\documentstyle[12pt,bezier]{article}
\def\emline#1#2#3#4#5#6{%
       \put(#1,#2){\special{em:moveto}}%
       \put(#4,#5){\special{em:lineto}}}
\def\newpic#1{}

\newcommand{\tr}[1]{\,{\rm tr}\,#1\,}
\newcommand{\NP}[1]{Nucl. \ Phys.}
\newcommand{\PL}[1]{Phys. \ Lett.}
\newcommand{\p}[1]{\partial}
\newcommand{\PRL}[1]{Phys.\ Rev.\ Lett. }
\newcommand{\MPL}[1] { Mod. Phys. Lett. }
\newcommand{\IJMP}[1] { Int. J. Mod. Phys. }
\begin{document}

\title{
\begin{flushright}
{\small SMI-TH-15-96 }
\end{flushright}
\vspace{2cm}
Interaction Representation in Boltzmann Field Theory}
\author{
I.Ya.Aref'eva, \thanks{Steklov Mathematical Institute,
Vavilov 42, GSP-1, 117966, Moscow,
e-mail: arefeva@class.mian.su} \\
\\ and\\ A.P.Zubarev \thanks {Steklov Mathematical Institute,
Vavilov 42, GSP-1, 117966, Moscow, e-mail: zubarev@class.mian.su}}
\date{$~~$}
\maketitle
\begin {abstract}
We consider an interaction representation  in the Boltzmann field theory.
It describes  the master field for  a  subclass  of  planar
diagrams in matrix models,  so  called  half-planar  diagrams. 
This interaction  representation  was found in the 
previous paper by Accardi, Volovich and one of us (I.A.) and it 
has an unusual property  that  one deals with a rational function  of
the  interaction  Lagrangian instead of the ordinary exponential function.
Here we study the interaction representation in more details  and
show that under natural assumptions this representation is
in fact unique.  We demonstrate that corresponding  Schwinger-Dyson 
equations lead to a closed set of integral 
equations for two- and four-point correlation
functions. Renormalization of the model is performed
and renormalization group equations are obtained.
Some model examples with discrete number
of degrees of freedom are solved numerically. The solution for one degree of
freedom is compared with the planar approximation for one matrix model.
For large variety of
coupling constant it reproduces the planar approximation with good accuracy.
\end {abstract}

\newpage
\section{Introduction}
\setcounter{equation}{0}
There is an old problem in quantum field theory of finding the leading
asymptotics in $N \times N$ matrix models for large $N$ in
realistic space-time. Its solution may have
important applications to  hadron dynamics  \cite {tH}-\cite{AS}
since the large $N$ limit in QCD ($N$ is the number of colors)
enables us to understand some phenomenological features of strong 
interaction.

In the early 80-s it was suggested  \cite {Wit} that there exists a master
field $\Phi (x)$ such that  correlation functions of this  field
 are equal to the large $N$ limit of  invariant correlation functions
of a matrix field $M(x)$,
\begin{equation}
				      \label {mc}
\lim _{N\to\infty} \frac{1}{N^{1+n/2}}
\langle tr (M(x_{n})...M(x_{1}))\rangle
= <\Phi (x_{n}) ...\Phi (x_{1})>.
\end{equation}

The problem of the construction of the master field has been
discussed in many works, see for example \cite {Haan}-\cite {Doug1}.
Gopakumar and Gross \cite {GG} and Douglas \cite {Doug2}
have constructed the master field for an arbitrary matrix model
in terms of correlation functions by using methods of non-commutative
probability theory \cite{Voi}-\cite{Sin}.
There has been a problem of the construction an operator realization
for the master field without knowledge of correlation functions.
Recently this problem has been solved in \cite{AV}
and it was shown that the master fields satisfy to standard equations of
relativistic field theory but fields are quantized according to a new rule.
These fields have a realization in the free (Boltzmannian) Fock space.

Quantum field theory in the Boltzmannian Fock space has been 
considered in \cite{OWG,GG,AAV,AV,AZ,Ar96}.  Some special form of this 
theory realizes the master field for a subset of planar diagrams, for 
the so called half-planar diagrams.  This construction deals with  
the  master field in a  modified interaction representation in the 
free Fock space. This new interaction representation involves not the 
ordinary exponential function  of the interaction but a rational 
function of the interaction and correlation functions are given by 
the formula

$$
< \Phi (x_{m})...\Phi (x_{1})> =\langle 0|\phi (x_{m})...
\phi (x_{1})|\Omega\rangle ,
~ |\Omega\rangle =\Omega|0\rangle
$$
\begin {equation} 
							  \label {omega}
\Omega={1\over {1+S_{int}(\phi )}},
\end   {equation} 
where $\phi (x)$ is a field in the free
(Boltzmannian) Fock space,
\begin {equation} 
                                                          \label {Phi}
\phi (x)= \phi ^{-}(x)+\phi^{+} (x),
\end   {equation} 
satisfying the relation
\begin {equation} 
                                                          \label {Phi'}
\phi ^{-}(x)\phi^{+} (y)=\Delta (x-y),
\end   {equation} 
where $\Delta (x-y)$ is the free propagator and $\phi ^{-}|0>=0$.

In this  paper we study the Boltzmann field theory with correlation 
functions (\ref {omega}) and we show that the Boltzmann field theory 
gives an analytical summation of half-planar diagrams.

We prove in pure algebraic way that the correlation functions (\ref 
{omega}) satisfy a closed set of Schwinger-Dyson-like equations.  
Moreover, starting
with an arbitrary function of the interaction
$\Omega=\Omega(S_{int}(\phi))$ and
using natural assumptions we show that the form
(\ref {omega})  of $\Omega (S_{int})$
is in fact an unique one  which admits  Schwinger-Dyson-like equations.
We will call these equations the Boltzmannian Schwinger-Dyson
equations.

We will see that there are an essential simplifications
in the Boltzmannian Schwinger-Dyson equations.
Namely, among the Boltzmannian Schwinger-Dyson equations
for $n$-point Green's functions for $n$ greater then the degree of $S_{int}$
there are equations which relate
$n$-point Green's functions only with $k$-point Green's  functions for
$k\leq n$.  This is a distinguish feature of the Schwinger-Dyson equations
for the Boltzmann fields and it drastically simplifies the situation.
In particular in
the case of quartic interaction one has a closed set of equations  for two
 and four point correlation functions.  Let us stress that this
system of equations is exact one but not 
a truncation of the original system 
and it does not assume any approximation for
correlation functions (\ref {omega}).  The origin of this property of the
Boltzmannian Schwinger-Dyson equations is related with  
a specific features of the
Wick theorem for the free algebra (\ref{Phi}).  Note that
in the operator form the Boltzmannian Schwinger-Dyson equations
contain  projectors on the vacuum state in the interaction terms.

The Boltzmannian system  (\ref{omega}) can be considered as
a non-trivial approximation to the planar correlation functions.
Note in this context that in all previous attempts
of approximated treatment of the planar theory 
some non-perturbative  approximations 
were used \cite {Sl,IA,Fer}.  Topologically  diagrams  representing
the perturbative series of the Boltzmann correlation
functions (\ref {omega}) look as rainbow (or half-planar) graphs of 
the usual diagram technique for matrix models \cite{AAV}.   This is a 
raison to call the Boltzmann correlation functions  the half-planar 
correlation functions and the corresponding equations the half-planar 
Schwinger-Dyson equations.  Comparing three sets of the Schwinger-Dyson
equations, full, planar and half-planar ones, one can say that the
operator form of the full Schwinger-Dyson equations is the simplest one.
Indeed, the planar Schwinger-Dyson equations contain projectors in the
Schwinger terms and the half-planar Schwinger-Dyson equations  also
contain projectors in the interactions terms.
However just due to the presence of additional projectors
it turns out that
the half-planar   Schwinger-Dyson equations  admit
analytical investigations.

We will find explicitly the correlation
function (\ref{omega}) for   the
quartic interaction in the case of one degree of freedom.  We compare
numerically the two-, four- and six-point half-planar correlation functions
with the corresponding planar correlation functions for the one 
matrix model.  For large variety of the coupling constant the 
half-planar approximation reproduces the planar approximation with 
good accuracy. However the half-planar approximation does not 
reproduce asymptotics of correlation functions in the strong coupling 
regime. To reproduce the strong coupling regime one has to consider
more large class of graphs \cite{AZ96}.

For the case of finite number of degrees of freedom
and a special action with a diagonal quadratic term
it is also possible to get an analytical solutions of the Boltzmannian
Schwinger-Dyson equations.
For more complicated case of a non-diagonal quadratic action and quartic
interaction  the system of equations  for two- and four-point
correlations functions is reduced to an algebraic
system of equations that admits a numerical solution.
This system describes the half-planar approximation of the
two-matrix model which has been analyzed recently by  Douglas and Li
\cite{DL} using the language of functions of non-commutating variables.

An analytical investigation of the correlation functions
(\ref{omega}) is also possible for $D$-di\-men\-si\-o\-nal spa\-ce-time.
In the case of quartic interaction one has a closed set of integral
equations for two- and four-point correlation functions.  For four-point
correlation functions we get a Bethe-Salpeter-like equation.  A special
approximation reduces this system  of integral equations to a linear
integral equation which has been considered \cite{Rothe} in the rainbow
approximation in the usual field theory.

For  four-dimensional space-time  the perturbative expansion
of  correlation functions  (\ref {omega})
has ultraviolet divergences. To remove these divergencies we apply
R-operation \cite {BogS}. We show that an application of R-operation
is equivalent to an introduction of counterterms to the interaction
Lagrangian. As usual this fact permits to write down the
renormalization group equations. Because of special features
of the Boltzmann field theory the
renormalization group equations are different from the ordinary
renormalization group equations. There is also a difference in the 
explicit formula for the  beta-function in terms of renormalization 
constants.

The paper is organized as follows.
In Section 2 we deal with one degree of freedom,
in Section 3 we consider finite number of degrees of freedom
and the last section
is devoted to continuous fields in $D$ dimensions.
We start Section 2 by reminding  specific feature of free Fock space for
one degree of freedom. In subsection 2.2
we show in pure algebraic way that the correlation functions (\ref 
{omega}) satisfy a closed set of Schwinger-Dyson-like equations.  
Then we study a question of uniqness of
the form (\ref {omega}) for $\Omega$.
In 2.3 we present an operator form of the Boltzmannian Schwinger-Dyson
equations. In 2.4 we solve explicitly these equations.
In subsections 2.5, 2.6 and  2.7 we present a generating 
functional for correlation functions for different forms of interaction. 
In subsection 2.8 we  compare the half-planar
approximation with the exact
answer for the sum of the planar diagrams for the one
matrix model.
In Section 3 we consider the same questions for finite number of degrees
of freedom.
The  goal of  Section 4 is  to demonstrate that the infinite set of
 the Schwinger-Dyson
equations for Boltzmann correlation functions (\ref{omega})
in D-dimensions  allows a reduction to a finite  system of
integral equations for lower correlation functions.
In subsection 4.2 we consider the quartic interaction and present
a system of  integral equations for two- and four-point correlation
functions. In 4.3 we study the renormalization of the correlation 
functions (\ref {omega}) and derive the renormalization group 
equation for $D=4$.  
In Appendix  a calculation of combinatoric factors for 
planar correlation functions is presented.  
\section{One Degree of Freedom}

\setcounter{equation}{0}
\subsection{Boltzmannian Fock Space and Free Correlation Functions}

In this section we consider the free correlation functions in the 
Boltzmannian  Fock space in zero-dimensional space-time.  Let us 
start from the general definition of Boltzmannian Fock space, then we 
will restrict ourself to zero-dimensional case.

The free (or Boltzmannian)  Fock space ${\cal H}$ over the Hilbert space $H$
is just the tensor algebra
$${\cal H}=\oplus_{n=0}^{\infty}H^{\otimes n}.$$
Creation and annihilation operators are defined as
$$a^{+}(f) f_{1}\otimes...\otimes f_{n}=
f\otimes f_{1}\otimes...\otimes f_{n}$$
\begin{equation}
\label{cran}
a(f) f_{1}\otimes...\otimes f_{n}=<f,f_{1}>\otimes f_{2}\otimes...\otimes
f_{n}
\end{equation}
where $<f,g>$ is the inner product in $H$.
We have
\begin{equation}
\label{alcran}
a(f)a^+(g)=<f,g>.
\end{equation}

We shall consider the simplest case  $H=C$. One has the vacuum vector
$|0\rangle $,
\begin{equation}
\label{vacuum}
a|0\rangle =0
\end{equation}
and $n$-particle states
\begin{equation}
\label{n-part}
|n \rangle=(a^+)^n|0\rangle.
\end{equation}

The master field  in zero-dimensional case is
\begin{equation}
\label{3}
\phi =a + a^+,
\end{equation}
where $a$ and $a^+$ satisfy the following relation
\begin{equation}
\label{1}
 aa^+=1.
\end{equation}
This algebra has a realization in the free (or Boltzmannian) Fock space.

One can formulate the Boltzmannian Wick theorem which is an analog of
the ordinary Wick theorem.
For this purpose let us define the  normal product of operators
(\ref{3}) in the following way
\begin{equation}
\label{nprod}
:\phi ^n:
\equiv \sum _{m=0}^{n}(a^+)^m a^{n-m}.
\end{equation}
Note that (\ref{nprod}) is different from the
normal product of $n$-th power of ordinary  field $\varphi =
\alpha + \alpha ^+ $
$$
:\varphi ^n:  \equiv \sum _{m=0}^{n} \frac{n!}{(n-m)!m!}(\alpha ^+)^m
\alpha ^{n-m},
$$
where $[\alpha , \alpha ^+]=1.$
For an arbitrary product of creation and annihilation operators
we define the normal product  as
$$
:(a^+)^{n_1}(a)^{m_1}
(a^+)^{n_2}(a)^{m_2}
...(a^+)^{n_l}(a)^{m_l}:
\equiv
$$
\begin{equation}
\label{nprod1}
\sum _{k=1}^{l}
(a^+)^{N_k}
(a)^{\hat{M}_k}
\delta _{\hat{N}_k,0}
\delta _{M_k,0},
\end{equation}
where
$$
N_k=\sum _{i=1}^{k}n_i,~~
\hat{N}_k=\sum _{i=k}^{l}n_i,~~
M_k= \sum_{i=1}^{k-1}m_i, ~~
\hat{M}_k= \sum_{i=k+1}^{l}m_i
$$
and
\[ \delta _{n,0} = \left \{ \begin{array}{ll}
1,~n=0, \\
0,~n \ne 0.
\end{array}
\right. \]

Let us call a contraction
 of two operators  an admissible one if it relates  neighboring
operators.  The Boltzmannian Wick theorem states that the product of
operators (\ref{nprod1}) is equal to the  normal product of
these operators plus sum of  normal products of these
operators with all admissible contractions.
In particular,
\begin{equation}
\label{nprod2}
:\phi ^n : :\phi ^m :=
:\phi ^{n+m} :+
\sum _{k=1}^{min(n,m)} :\phi ^{n+m-2k}:.
\end{equation}

A free $n$-point Green's function is defined as the vacuum expectation
of  $n$-th power  of master field (\ref{3})
\begin{equation}
\label{4}
G^{(0)}_n=\langle 0|\phi ^n |0\rangle .
\end{equation}
As it is well-known, the Green's function (\ref{4}) is  given by a $n$-th
moment of Wigner's distribution  \cite{BIPZ,Voi}
\begin{equation}
\label{wig}
G^{(0)}_n=\frac{1}{2\pi }\int_{-2}^{2} \lambda ^n\sqrt{4-\lambda ^2}
d \lambda
\end{equation}
and $G_{2n}^{(0)}=c_n$, where $c_n$ are the Catalane numbers
\begin{equation}
\label{cat}
c_n=\frac{(2n)!}{n!(n+1)!}.
\end{equation}
In \cite{AAKV} 
an algebraic origin
of the representation (\ref {wig}) has been found. 
It is  related with the fact that the algebra (\ref{1})
is isomorphic to the quantum semigroup  $SU_q(2)$  with $q=0$
and (\ref{wig}) is the Haar measure on $SU_{q}(2)$.

The representation (\ref{wig})
can be also obtained as a solution of the Schwinger-Dyson equations.
To get these equations from  definitions (\ref{1})-(\ref{4})
let us
calculate (\ref{4}) using the Boltzmannian Wick theorem.
We present the result of an application of the Wick
theorem graphically (Fig. 1).
We draw the $n$-th power of $\phi$ as $n$ points
lying on a line and the contractions as upper half-ovals.
According to the relation $a a^+=1$ all points may be connected only by
non-overlapping upper half-ovals.
Considering all possible contractions
of the first point from the left and taking into account
that all odd correlation functions are equal to zero
we have
\begin{equation}
\label{8}
G^{(0)}_{2n}=\sum_{m=1}^{n}G^{(0)}_{2m-2}G^{(0)}_{2n-2m}.
\end{equation}

Equation (\ref{8}) leads to an
algebraic equation \cite {AAV} for a generation functional
\begin{equation}
\label{10}
Z^{(0)}(j)=\sum _{n=0}^{\infty}G^{(0)}_{n}j^{n}, ~~G^{(0)}_0=1.
\end{equation}
Indeed,
multiplying (\ref{8}) by $j^{2n}$ and making the summation
over $n$ we have
$$\sum_{n=1}^{\infty}G^{(0)}_{2n}j^{2n}=
\sum _{n=1}^{\infty}\sum_{m=1}^{n}
G^{(0)}_{2m-2}G^{(0)}_{2n-2m}j^{2n}.
$$
Using
$$
\sum _{n=1}^{\infty}\sum_{m=1}^{n}f(n,m)
=\sum _{m=1}^{\infty}\sum_{n=m}^{\infty}f(n,m)
$$
and making the change $m-1 \to m$, $n-m \to n$, we
get the equation
$$\sum_{n=1}^{\infty}G^{(0)}_{2n}j^{2n}=
\sum _{m=0}^{\infty}\sum_{n=0}^{\infty}
G^{(0)}_{2m}G^{(0)}_{2n}j^{2n+2m+2},
$$
which in terms of generation functional can be written in the form
\begin{equation}
\label{14}
Z^{(0)}(j)-1=j^2[Z^{(0)}(j)]^2.
\end{equation}
Under the initial condition
$Z^{(0)}(0)=1$
the solution of equation (\ref{14}) is
\begin{equation}
\label{15}
Z^{(0)}(j)=\frac{1-\sqrt{1-4j^2}}{2j^2}=
\sum_{n=0}^{\infty}\frac{(2n)!}{n!(n+1)!}j^{2n}.
\end{equation}
One has  the integral representation
for $Z^{(0)}(j)$
\begin{equation}
\label{int}
Z^{(0)}(j)=\frac{1}{2\pi}\int _{-2}^{2} \frac{1}{1-\lambda j}
\sqrt{4-\lambda ^2} d\lambda,
\end{equation}
that is in agreement with  (\ref{wig}).

Note that in the above formulae $j$ may be considered as an operator.
If one assumes that $j$ satisfies a non-standard differentiation rule
\begin{equation}
\label{dif}
\frac{d}{dj}j^n=j^{n-1},
\end{equation}
then one can deduce the following differential equation for $Z(j)$
\cite{Cvit}:
\begin{equation}
\label{de}
\frac{d^2}{dj^2}Z^{(0)}(j)=[Z^{(0)}(j)]^2.
\end{equation}
To reproduce (\ref{15}) one has to assume the following initial conditions
\begin{equation}
\label{ic}
Z^{(0)}(0)=1,~~(\frac{d}{dj}Z^{(0)})(0)=0.
\end{equation}

In what  follows we shall also use the following modified form of
equations (\ref{8})
\begin{equation}
\label{6}
G^{(0)}_{2n}=\sum_{m=0}^{k-1}G^{(0)}_{2k-2m-2}G^{(0)}_{2n+2m-2k}+
\sum_{m=k}^{n-1}G^{(0)}_{2m-2k}G^{(0)}_{2n+2k-2m-2}
\end{equation}
for  any $k=1,2,...,n$.
Equations (\ref{6}) are obtained by a simple change of summation index.
Indeed, making the change $m \to m'=k-m$  in the first sum of (\ref{6}),
 we get
$$
\sum _{m'=1}^{k}
G^{(0)}_{2m'-2}G^{(0)}_{2n-2m'}.
$$
Introducing the new summation index in the second term in right hand side
of (\ref{6}) as $m'=n+k-m$ we get
$$
\sum _{m'=k+1}^{n}
G^{(0)}_{2n-2m'}G^{(0)}_{2m'-2},
$$
so, (\ref{6}) is equivalent to (\ref{8}).
Equations (\ref{6})  are the Schwinger-Dyson equations
written for the case  when one traces down for
contractions  of
$2k$-th  operator in the vacuum expectation
$\langle \phi \phi ...\phi  \rangle $.
This is illustrated on Fig. 2.
The same equations are  obtained when one traces down for the
contraction of  $2k+1$-th operator (Fig. 3).

Equations (\ref{6}) are  0-dimensional analogues
of the  $D$-dimensional  Schwinger-Dyson
equations written for the case when
the operators
$(-\Delta _{x_{2k}} +m^2)$ or
$(-\Delta _{x_{2k+1}} +m^2)$
are applied
to a  Green's function
$G^{(0)}_{2n}(x_1,...,x_n)$.

\subsection{Boltzmannian Schwinger-Dyson Equations}

In the ordinary Euclidean formulation of
Bose or Fermi 0-dimensional quantum field theory
the interacting Green's functions have the form
\begin{equation}
\label{16bose}
G^{{\rm (Eucl.)}}_n=\langle 0|\varphi ^n 
e^{-S_{int}(\varphi )}|0\rangle , 
\end{equation} 
A natural generalization of this formula 
to the case of Boltzmann statistic is 
\begin{equation} 
\label{16} 
{\cal G}_n=\langle 0|\phi ^n \Omega (S_{int}(\phi) )|0\rangle , 
\end{equation}
 where $\Omega (S_{int}(\phi))$ is not necessary an exponent but some
 analytical function of the Lagrangian.
 First of all we will consider the function
 $\Omega (S_{int}(\phi))$
of special form \cite{AAV}
\begin{equation}
\label{first}
\Omega (S_{int})=\frac{1}{1+S_{int}}
\end{equation}
and we will derive the Schwinger-Dyson-like equation for interacting
Green's functions (\ref{16}).
Then under some natural assumptions discussed below we will prove
that the form (\ref{first}) is unique.

We will consider
the case of quartic interaction  $S_{int} =g\phi ^4$.  The generalization for an 
arbitrary polynomial interaction is straightforward.

Using the Taylor expansion for $\Omega
(g\phi ^4)$
\begin{equation}
\label{tay}
\Omega (g\phi ^4)=\sum
_{m=0}^{\infty}g^m\Omega _m \phi ^{4m},
\end{equation}
where $\Omega _m=(-1)^m $ for (\ref{first}),
we write the Green's
functions ${\cal G}_n$ in terms of the
free Green's functions ${\cal G}_n^{(0)}$:
\begin{equation}
\label{17}
{\cal G}_n=\sum _{m=0}^{\infty} g^m\Omega _m
\langle 0|\phi ^{n+4m}|0\rangle =
\sum _{m=0}^{\infty}g^m\Omega _m {\cal G}^{(0)}_{n+4m}.
\end{equation}
Using the free Schwinger-Dyson equations in the form
\begin{equation}
\label{55}
{\cal G}^{(0)}_n=\sum_{l=1}^{k-1}{\cal G}^{(0)}_{k-l-1}
{\cal G}^{(0)}_{n+l-k-1}+
\sum_{l=k+1}^{n}{\cal G}^{(0)}_{l-k-1}{\cal G}^{(0)}_{n+k-l-1},
~~k=1,...,n,~
\end{equation}
(that coincides with (\ref{6}) after setting
${\cal G}_n=0$ for odd $n$)  we get
\begin{equation}
\label{18}
{\cal G}_n=\sum _{m=0}^{\infty} g^m\Omega _m
[\sum _{l=1}^{k-1}{\cal G}^{(0)}_{k-l-1}{\cal G}^{(0)}_{l+n+4m-k-1}+
\sum _{l=k+1}^{n+4m}{\cal G}^{(0)}_{l-k-1}{\cal G}^{(0)}_{n+4m+k-l-1}].
\end{equation}
Writing the sum in the second term of (\ref{18}) as
\begin{equation}
\label{suu}
 \sum_{l=k+1}^{n+4m}=\sum _{l=k+1}^{n}+
 \sum_{l=n+1}^{n+4m}
\end{equation}
and making the change of the
summation index  $l-n \to l$ in the second sum of (\ref{suu}),
one obtains
$$
{\cal G}_n=\sum _{l=1}^{k-1}
{\cal G}^{(0)}_{k-l-1}{\cal G}_{l+n-k-1}+
\sum _{l=k+1}^{n}{\cal G}^{(0)}_{l-k-1}{\cal G}_{n+k-l-1}+$$
\begin{equation}
\label{19}
\sum _{m=0}^{\infty} g^m \Omega _m
\sum _{l=1}^{4m}{\cal G}^{(0)}_{n+l-k-1}{\cal G}^{(0)}_{4m+k-l-1}
=J_{I}+J_{II}+J_{III}.
\end{equation}
Using the formulas
$$
\sum _{l=1}^{4m}f(l)=
\sum_{p=0}^{m-1}[f(4p+1)+f(4p+2)+f(4p+3)+f(4p+4)],$$
\begin{equation}
\label{summ}
\sum _{m=0}^{\infty}\sum_{p=0}^{m-1}f(m,p)=\sum_{p=0}^{\infty}
\sum_{m=p+1}^{\infty}f(m,p)
\end{equation}
and making the change $m \to m+p+1$ in the second line of (\ref{summ}) one
can rewrite $J_{III}$ in the following way
$$J_{III}~=~\sum _{p=0}^{\infty}\sum_{m=0}^{\infty}
g^{m+p+1}\Omega _{m+p+1}
[{\cal G}^{(0)}_{n-k+4p}{\cal G}^{(0)}_{4m+k+2}+
{\cal G}^{(0)}_{n-k+4p+1}{\cal G}^{(0)}_{4m+k+1}+
$$
\begin{equation}
\label{j3}
+{\cal G}^{(0)}_{n-k+4p+2}{\cal G}^{(0)}_{4m+k}+
{\cal G}^{(0)}_{n-k+4p+3}{\cal G}^{(0)}_{4m+k-1}],
\end{equation}
so equation (\ref{19}) takes the form
$$
{\cal G}_n=\sum _{l=1}^{k-1}
{\cal G}^{(0)}_{k-l-1}{\cal G}_{l+n-k-1}+
\sum _{l=k+1}^{n}{\cal G}^{(0)}_{l-k-1}{\cal G}_{n+k-l-1}+$$
$$\sum _{p=0}^{\infty}\sum_{m=0}^{\infty}g^{m+p+1} \Omega _{m+p+1}
[{\cal G}^{(0)}_{n-k+4p}{\cal G}^{(0)}_{4m+k+2}+
{\cal G}^{(0)}_{n-k+4p+1}{\cal G}^{(0)}_{4m+k+1}+
$$
\begin{equation}
\label{123}
{\cal G}^{(0)}_{n-k+4p+2}{\cal G}^{(0)}_{4m+k}+
{\cal G}^{(0)}_{n-k+4p+3}{\cal G}^{(0)}_{4m+k-1}].
\end{equation}

Note that two first sums in (\ref{123}) are
nothing but Schwinger's terms.
Since the coefficients $\Omega _m$
satisfy  the relations
\begin{equation}
\label{co}
\Omega _{n+m}=\Omega _n \Omega _m,
\end{equation}
we have a possibility to represent the third sum as a combination of
 interacting Green's functions (\ref{16}).
With $\Omega $ in the form
(\ref{first}) equations (\ref{123}) take the form
$$
{\cal G}_n=
-g[{\cal G}_{n-k}{\cal G}_{k+2}+
{\cal G}_{n-k+1}{\cal G}_{k+1}+
{\cal G}_{n-k+2}{\cal G}_{k}+
{\cal G}_{n-k+3}{\cal G}_{+k-1}]+
$$
\begin{equation}
\label{23}
\sum _{l=1}^{k-1}
{\cal G}^{(0)}_{k-l-1}{\cal G}_{l+n-k-1}+
\sum _{l=k+1}^{n}{\cal G}^{(0)}_{l-k-1}{\cal G}_{n+k-l-1}.
\end{equation}
Equations
(\ref{23})
are the Schwinger-Dyson-like equations for 
zero-dimensional Boltzmann theory with quartic interaction.
We will  also call these equations the  Boltzmannian Schwinger-Dyson 
equations.

We see that the terms describing an interaction are quadratic on the
correlations functions.
The sum of indices in all these terms is constant and depends on
the power of the interaction.
One can prove that under the assumption of such structure for the interaction
terms in the Schwinger-Dyson-like equations the form of $\Omega$ is unique.
In fact, one assumes that the form of $\Omega$  does not depend
on the special choice of the interaction and $\Omega(0)=1$. These
assumptions mean that in
particular for the linear interaction
$S_{int}=g \phi $
one has the following
Schwinger-Dyson-like equations
\begin {equation} 
                                                          \label {lsd}
{\cal G}_{n}=-g{\cal G}_{n-1}{\cal G}_{0}+\mbox{Schwinger's~terms},~~ n>0.
\end   {equation} 
Let us derive that the form (\ref {first})  follows from (\ref{lsd}).
Expanding the both sides of (\ref {lsd}) on power coupling we have
\begin {equation} 
                                                          \label {pc}
\sum _{m=0}^{\infty}g^{m}\Omega _{m}{\cal G}^{(0)}_{n+m}=
-g\sum _{m=0}^{\infty}g^{m}\Omega _{m}{\cal G}^{(0)}_{n-1+m}\cdot
\sum _{k=0}^{\infty}g^{k}\Omega _{k}{\cal G}^{(0)}_{k}
+\mbox{Schwinger's~terms},
\end   {equation} 
Using equation (\ref {55}) and comparing the coefficients in front of 
the same powers of the coupling constant we obtain 
\begin {equation} 
\label {pex}
\sum
_{k=0}^{m}(\Omega _{k}\Omega _{m-k} +\Omega _{m+1})
{\cal G}^{(0)}_{n+k}{\cal G}^{(0)}_{m-k}=0
\end   {equation} 
For $m=0$ from (\ref {pex}) we have $\Omega _{0}^{2}+\Omega _{1}=0$
and taking into account $\Omega _{0}=1$ we get $\Omega _{1}=-1$.
Assuming
\begin {equation} 
                                                          \label {rfo}
\Omega _{m}=(-1)^{m}
\end   {equation} 
 for $m<m_{0}$ and using
equation (\ref {pex}) one can prove
that this relation  is also hold for
$m=m_{0}$, that gives by induction the prove of  (\ref {rfo}).

  From the above consideration we see that to get a closed set of
Schwinger-Dyson equations in the free Fock space for an interacting  case
one has to deal with a non-Heisenberg dynamics
and the Green's function in the interaction representation
are not  represented as
exponential function of the interaction but
are given by  formula (\ref{first}).
\subsection{An Operator Form of the Boltzmannian Schwinger-Dyson
Equations}

Let us write down the week operator form of equation (\ref {23}).
By the week operator form we mean the relation 
that is hold after averaging
with the function (\ref {first}). Let first consider the case
$k=n$. Then equations (\ref {23}) take the form
\begin {equation} 
                                                          \label {n=k}
{\cal G}_{n}=
-g[{\cal G}_{0}{\cal G}_{n+2}+
{\cal G}_{2}{\cal G}_{n}]+
\sum ^{n-2}_{l=0}{\cal G}^{(0)}_{l}{\cal G}_{n-l-2},
~~n>1.
\end   {equation} 
The left hand side of (\ref{n=k}) represents the term $\phi \phi ^{n-1}$.
The last term in the right hand side
of (\ref{n=k}) is the Schwinger term
and it can be represented by the following commutator
\begin {equation} 
                                                          \label {ost}
-i[\pi , \phi ^{n-1}],
\end   {equation} 
where $\pi$ is an element of the  following
algebra \cite{Haan,GH}
\begin {equation} 
                                                          \label {0pcr}
[\pi , \phi ]=i|0><0|.
\end   {equation} 
Indeed, after performing the average of (\ref {ost}) with $\Omega$
we get
\begin {equation} 
                                                          \label {oo1}
-i<0|[\pi , \phi ^{n-1}]\Omega|0>=
\sum ^{n-2}_{l=0}{\cal G}^{(0)}_{l}{\cal G}_{n-l-2}
\end   {equation} 
The sum in the square brackets in (\ref {n=k}) can be
represented as an
average of the product of
$\phi ^{n-1} $ with the following sum of operators
\begin {equation} 
                                                          \label {rr1}
\phi ^{3}|\Omega><0|+\phi ^{2}|\Omega><0|\phi+
\phi |\Omega><0|\phi ^{2} +|\Omega><0|\phi ^{3} .
\end   {equation} 
Note that these terms
can be represented as
$${\cal V}_{int} =
\frac{\delta S_{int}[\phi ]}{\delta \phi }|\Omega\rangle \langle 0|+
\frac{\delta ^{2}S_{int}[\phi ]}{\delta \phi ^{2}}
|\Omega\rangle \langle 0|\phi +...+
$$
\begin {equation} 
                                                          \label {erer}
\phi |\Omega\rangle \langle 0|\frac{\delta ^{2}S_{int}[\phi ]}
{\delta \phi ^{2}}+
|\Omega\rangle \langle 0|\frac{\delta S_{int}[\phi ]}{\delta \phi }
\end   {equation} 
if one assumes that non-commutative derivatives
\begin {equation} 
                                                          \label {ncd}
\frac{\delta \phi ^{n}}{\delta \phi}=\phi ^{n-1}.
\end   {equation} 
are implied  in (\ref {erer}).
Therefore, equations (\ref {n=k}) is the equation for the average
\begin {equation} 
                                                          \label {onk}
<0|{\cal E}\phi ^{n-1}|\Omega>=0,
\end   {equation} 
where the operator ${\cal E}$ can be represented as
\begin {equation} 
                                                          \label {er}
{\cal E}=
\frac{\delta S_{0}[\phi ]}{\delta \phi (x)}+
{\cal V}_{int} +{\cal S}
\end   {equation} 
with
\begin {equation} 
                                                          \label {ST}
{\cal S}=[\pi ,~\cdot ~].
\end   {equation} 
For $k<n$  equations (\ref {23}) can be also represented as the overage
\begin {equation} 
                                                          \label {ok}
<0|\phi ^{k}{\cal E}\phi ^{n-k-1}|\Omega>=0,
\end   {equation} 
where one has to use another form for ${\cal S}$.
\subsection{Solution of the Boltzmannian Schwinger-Dyson
Equations}

The distinguish feature of  equations (\ref{23}) is that for $n \ge
4$ and $2\le k\le n-1 $ the right hand side of (\ref{23}) does not contain
the  Green's functions ${\cal G}_m$ with $m > n$.
This fact permit us to write
down a closed system  of equations for any ${\cal G}_n$, $n \geq 4$.  Let 
us study this questions in more details.

Equations (\ref{23})
for $k=1$ and $k=2$
take the form
\begin{equation}
\label{24}
{\cal G}_{2n}=\sum _{l=1}^{n}
{\cal G}^{(0)}_{2l-2}{\cal G}_{2n-2l}-
g{\cal G}_{2n}{\cal G}_{2}-
g{\cal G}_{2n+2}{\cal G}_{0},
\end{equation}
\begin{equation}
\label{25}
{\cal G}_{2n}=\sum _{l=1}^{n-1}
{\cal G}^{(0)}_{2l-2}{\cal G}_{2n-2l}+ {\cal G}_{2n-2}-
g{\cal G}_{2n-2}{\cal G}_{4}-
g{\cal G}_{2n}{\cal G}_{2},
\end{equation}
where we use the fact that ${\cal G}_{2n+1}=0$.
Setting $n=1$ in (\ref{24}) and $n=2$ in (\ref{25}) we obtain
the system of two equations for three unknown functions
${\cal G}_0(g)$, ${\cal G}_2(g)$ and ${\cal G}_4(g)$:
$$
{\cal G}_2={\cal G}_0-g{\cal G}_2{\cal G}_2-g{\cal G}_4{\cal G}_0,
$$
\begin{equation}
\label{26}
{\cal G}_4=2{\cal G}_2-2g{\cal G}_2{\cal G}_4.
\end{equation}
To have the third equation let us note that there is an
additional set of equations for Green's functions
\begin{equation}
\label{id}
{\cal G}_{2n}={\cal G}^{(0)}_{2n}-g{\cal G}_{2n+4},
\end{equation}
that follow from the identity
\begin{equation}
\label{id1}
\langle 0| \frac{\phi ^{2n}}{1+g\phi ^4}|0\rangle
=\langle 0 |\phi ^{2n}-\frac{g\phi ^{2n+4}}{1+g\phi ^4}|0\rangle.
\end{equation}
For the vacuum Green's function
${\cal G}_0$ equation (\ref{id}) has the form
\begin{equation}
\label{26a}
{\cal G}_0=1-g{\cal G}_4.
\end{equation}
The system of equations (\ref{26}) and (\ref{26a}) lead to the following
equation for ${\cal G}_0$
\begin{equation}
\label{26c}
4g{\cal G}_0^4 + {\cal G}_0^2 - 1 = 0.
\end{equation}
Taking into account the condition ${\cal G}_0=1$  for $g=0$ we have
\begin{equation}
\label{26d}
{\cal G}_0=\sqrt{\frac{-1+\sqrt{1+16g}}{8g}}.
\end{equation}
The Green's functions ${\cal G}_2$ and
${\cal G}_4$ can be expressed in terms of ${\cal G}_0$:
\begin{equation}
\label{26e}
{\cal G}_2=\frac{1-{\cal G}_0}{2g{\cal G}_0},~~
{\cal G}_4=\frac{1-{\cal G}_0}{g}.
\end{equation}
High-point Green's functions can be found from (\ref{id})
after taking into account (\ref{26d}) and (\ref{26e}).
\subsection{Generating Functional }

A generating functional for  interacting Green's functions is
\begin{equation}
\label{31}
Z(j,g)=\sum _{n=0}^{\infty}{\cal G}_{2n}(g)j^{2n}.
\end{equation}
   From  (\ref{wig}) one has the following  integral representation for
$Z(j,g)$
\begin{equation}
\label{genf}
Z(j,g)= \frac{1}{2\pi}\int _{-2}^{2}
\frac{1}{1-\lambda j}
\frac{1}{1+g\lambda ^4}
\sqrt{4-\lambda ^2} d \lambda
\end{equation}
and the integral (\ref {genf}) can be calculated explicitly.
However it is instructive to get an explicit derivation of
$Z(j,g)$ in a  pure algebraic way,
since just this method can be generalized to
the $D$-dimensional space-time.

  From (\ref{24}) it  follows
\begin{equation}
\label{32}
j^2[Z(j)-{\cal G}_0]=j^4Z^{(0)}(j)Z(j)-gj^2{\cal G}_2[Z(j)-
{\cal G}_0]-g[Z(j)-{\cal G}_2j^2-{\cal G}_0]{\cal G}_0
\end{equation}
and
$$
Z(j)=\frac{j^2{\cal G}_0(1+2g{\cal G}_2)+
g{\cal G}_0^2}{g{\cal G}_0+j^2(1+g{\cal G}_2)-j^4Z^{(0)}(j)} =
$$
\begin{equation}
\label{32a}
\frac{2{\cal G}_0(j^2+g{\cal G}_0^2)}{2g{\cal G}_0^2+
j^2(1+{\cal G}_0)-2j^4{\cal G}_0 Z^{(0)}(j)}.
\end{equation}

Using (\ref{id}) one can write the equation for $Z(j)$
in an alternative form
\begin{equation}
\label{al}
j^4Z(j)=j^4Z^{(0)}(j)-g[Z(j)-{\cal G}_2j^2-{\cal G}_0],
\end{equation}
and therefore

$$
Z(j)=\frac{j^4Z^{(0)}(j)+g{\cal G}_2j^2+g{\cal G}_0}{j^4+g}=
$$
\begin{equation}
\label{al1}
\frac{2g{\cal G}_0Z^{(0)}(j)+
gj^2-g{\cal G}_0j^2+2g^2{\cal G}_0^2}{2g{\cal G}_0(j^4+g)}.
\end{equation}
Taking into account (\ref{14})
and (\ref{26c}) one can verify that (\ref{al1})
coincides with (\ref{32a}).
\subsection{Correlation Functions without Vacuum Insertions}

To calculate the Green's functions without vacuum insertions
$G_{n}$ note that they satisfy
the same Schwin\-ger-Dyson equations, but one
has to use another "initial" condition, namely $G_0=1$.
The system of equations for 2-, 4-, and 6-point Green's functions is
$$
G_2=1-gG_2G_2-gG_4,
$$
$$
G_4=2G_2-2gG_2G_4,
$$
\begin{equation}
\label{27}
G_6=G_2+2G_4-gG_4^2-gG_2G_6.
\end{equation}
   From the first two lines of (\ref{27})
one gets the following
equation for $G_2$:
\begin{equation}
\label{28}
2g^2G^3_2+3gG^2_2+G_2-1=0.
\end{equation}
Introducing a new variable $\tilde G_{2}$,
\begin{equation}
\label{cha}
 G_2=\frac{
{\tilde G}_2
-1}{2g},
\end{equation}
we have
\begin{equation}
\label{29}
{\tilde G}_2^3-
{\tilde G}_2
-4g=0.
\end{equation}
The solution of equations (\ref{29}) has a different form in 
dependence on values of $g$.

Using the Cordano formula we have
\begin{equation}
\label{30}
{\tilde G}_2=
\frac{2}{\sqrt{3}}{\rm Re}e^{i\frac{\pi}{6}}(\sqrt{1-108g^2}
-i6\sqrt{3}g)^{\frac{1}{3}}
\end{equation}
for $~g\le \frac{\sqrt{3}}{18}~$ and
\begin{equation}
\label{30n1}
{\tilde  G}_2=
\frac{1}{\sqrt{3}}[
(6\sqrt{3}g+\sqrt{108 g^2-1})^{\frac{1}{3}}+
(6\sqrt{3}g-\sqrt{108 g^2-1})^{\frac{1}{3}}]
\end{equation}
for $~g \ge \frac{\sqrt{3}}{18} ~$.
Therefore
$$G_2=  \frac{1}{2g}[
\frac{2}{\sqrt{3}}{\rm Re}e^{i\frac{\pi}{6}}(\sqrt{1-108g^2}
-i6\sqrt{3}g)^{\frac{1}{3}}-1]=
$$
\begin{equation}
\label{30n}
1-3g+16g^2-105g^3+768g^4...~
\end{equation}
for $~g\le \frac{\sqrt{3}}{18}~$ and
\begin{equation}
\label{30n2}
G_2=
\frac{1}{2\sqrt{3}g}[
(6\sqrt{3}g+\sqrt{108 g^2-1})^{\frac{1}{3}}+
(6\sqrt{3}g-\sqrt{108 g^2-1})^{\frac{1}{3}}-\sqrt{3}]
\end{equation}
for $~g \ge \frac{\sqrt{3}}{18}~$.

An equation for a generating functional for Green's functions without
vacuum insertions ${\cal Z}(j)$ follows from (\ref{32}) by setting
$G_0=1$:
\begin{equation}
\label{33}
j^2[{\cal Z}(j)-1]=j^4Z^{(0)}(j){\cal Z}(j)-
gj^2G_2[{\cal Z}(j)-1]-g[{\cal Z}(j)-G_2j^2-1].
\end{equation}
  From (\ref{33}) we find
\begin{equation}
\label{34}
{\cal Z}(j)=
\frac{g+j^2(2gG_2+1)}{g+j^2(gG_2+1)-j^4Z^{(0)}(j)},
\end{equation}
where $Z^{(0)}(j)$ and $G_2$ are given by (\ref{15}) and (\ref{30n}),
(\ref{30n2}), respectively.

The generating functionals
$Z(j,g)$
and
${\cal Z}(j,g)$
are related
by the formula
\begin{equation}
\label{rel}
Z(j,g)=G_0(g)
{\cal Z}(j,gG_0(g)),
\end{equation}
which follows from the relation
\begin{equation}
\label{relg}
G_n(g)=G_0(g)G_n(gG_0(g)).
\end{equation}
The last relation is true due to the
Boltzmannian Wick theorem. It is easy to check
that
$ Z(j,g)$ and
${\cal Z}(j,g)$
given by (\ref{32a}) and (\ref{34}) satisfy the relation (\ref{rel}).
Indeed, from (\ref{34}) it  follows that

$$
G_0(g){\cal Z}(j,gG_0(g))=
\frac{gG_0^2+j^2(1+2gG_0G_2(gG_0))}
{gG_0+j^2(1+gG_0G_2(gG_0)-j^4Z^{(0)}(j)}.
$$
Replacing in the right hand side of this relation $G_0G_2(gG_0)$
by $G_2$ and taking into account (\ref{32}) we get
the expression for $Z(j,g)$.
\subsection{Correlation Functions for Normal Ordered Interaction}

In  this subsection we will consider the Boltzmann theory
with a normal ordered interaction $:\phi ^4:$. 
Let us denotes by $\bar{G} _n$ the $n$-point Green's function without
vacuum insertions in such theory.
According to the definition of Boltzmannian normal product we have
\begin{equation}
\label{nophi4}
:\phi ^4:=
\phi ^4
-3\phi ^2-2
\end{equation}
and so
\begin{equation}
\label{no}
\bar{G}_{2n}=\langle 0 |\phi ^{2n}
\frac{1}{1+g(\phi ^4-3\phi ^2-2})|0\rangle.
\end{equation}

Let us write down the Schwinger-Dyson equations for Green's functions
$\bar{G} _n$.
Repeating  all steps as for derivation of (\ref{24}) and (\ref{25})
we get
\begin{equation}
\label{no1}
\bar{G}_{2n}=\sum _{l=1}^{n}
\bar{G}^{(0)}_{2l-2}\bar{G}_{2n-2l}-
g\bar{G}_{2n}\bar{G}_{2}-
g\bar{G}_{2n+2}\bar{G}_{0}+3g\bar{G} _{2n},
\end{equation}
\begin{equation}
\label{no2}
\bar{G}_{2n}=\sum _{l=1}^{n-1}
\bar{G}^{(0)}_{2l-2}\bar{G}_{2n-2l}+ \bar{G}_{2n-2}-
g\bar{G}_{2n-2}\bar{G}_{4}-
g\bar{G}_{2n}\bar{G}_{2}+3g\bar{G}_{2n-2}\bar{G} _2.
\end{equation}

Putting $n=1$ in (\ref{no1}), $n=2$ and $n=3$ in (\ref{no2}) we obtain
the  following equations for $\bar{G} _2$, $\bar{G} _4$ and $\bar{G} _6$ 
$$
\bar{G}_2=1-g\bar{G}_2\bar{G}_2-g\bar{G}_4+3g\bar{G}_2,
$$
$$
\bar{G}_4=2\bar{G}_2-2g\bar{G}_2\bar{G}_4 +3g\bar{G} _2\bar{G} _2.
$$
\begin{equation}
\label{no3}
\bar{G} _6=\bar{G} _2 + 2 \bar{G} _4 -g \bar{G} _4^2
-g\bar{G} _2 \bar{G} _6 + 3 g \bar{G}_ 2 \bar{G} _4.
\end{equation}
   From the first two lines of (\ref{no3})
one can derive
the equation for $\bar{G} _2$
\begin{equation}
\label{no4}
2g^2 \bar{G}_2^3+3g(1-g)\bar{G}_2^2+(1-3g)\bar{G} _2 -1=0.
\end{equation}
We restrict ourself by writing the perturbative solution of (\ref{no4}) up
to order $g^4$:
\begin{equation}
\label{no4a}
\bar{G}_2=1+g^2-3g^3+9g^4+...~.
\end{equation}
\subsection{The Half-Planar Approximation
and the Two-Field  Boltzmann Theory}

In the previous subsections we have investigated the interacting Boltzmann
theory itself. In this subsection we are going to compare the Boltzmann
theory and the planar approximation for the one-matrix model.
Green's functions for the one-matrix model in the planar approximation
are defined as
\begin{equation}
\label{mm}
\Pi _{2n}(g) =\lim _{N\to \infty }\frac{1}{N^{1+n}}
\frac{1}{{\cal Z}}\int dM \tr (M^{2n})\exp[-\frac{1}{2}\tr (M^2)-
\frac{g}{4N}\tr (M^4)],
\end{equation}
$${\cal Z}=\int DM exp[-\frac{1}{2}\tr (M^2)-
\frac{g}{4N}\tr(M^4)],$$
The integration in (\ref {mm}) is over $N\times N$ hermitian matrices.
According the 't Hooft diagram technique the perturbative expansion
in the coupling constant of the correlation functions (\ref {mm})
is given by a sum of all planar double-line graphs  \cite {tH}.
Due to a normalization factor $N^{-(1+n)}$ external lines corresponding to
$\tr M^{2n}$ can be treated as lines of a generalized vertex.
 Note that all vertices of these graphs
have the equal orientation (left or right
in dependence on an adopted convention). We shall call two
double-line  planar graphs
topologically equivalent  if one of them can be transformed into the
other by a continuous deformation on the plane.

A  planar non-vacuum graph is  an
half-planar graph if it can  be drawn  
so that all its vertices lie on some plane
line  in the right of the
generalized vertex $\tr M^{2n}$ and all propagators lie in the upper-half
 plane without
overlapping  \cite {AAV}.
Also we shall call a planar graph  
the half-planar irreducible graph if it can be
represented as an half-planar graph in an unique way.  A simple analysis
shows that  an half-planar graph is the half-planar-irreducible
one if it does  not contain any tadpole subgraphs.  By a graph with a 
tadpole, as usual, we mean a graph  with a subgraph which contains two 
lines coming from the same vertex so that after removing of these two 
lines the remaining subgraph becomes disconnected with the rest of the 
graph.

In \cite{AAV} it has been shown  that if one considers
some special approximation
for the planar theory, namely so called half-planar approximation, then
Green's functions of the one-matrix model in this approximation
coincide with correlation functions in a corresponding  Boltzmann
theory,
\begin {eqnarray} 
\Pi ^{HP} _{2n}(g)=<\phi ^{2n}(1+g\phi ^{4})^{-1}>.
							  \label {ahp}
\end   {eqnarray} 
In both hand sides of (\ref{ahp}) we omit all graphs with tadpoles.
The raison to omit  the graphs with tadpoles  is  that otherwise we
have a double-counting for graphs with tadpoles for the Boltzmann theory.
We call this approximation the half-planar approximation.

A pure combinatorial proof of equation (\ref {ahp})
is based on  the following two statements.
The first one states that  the planar correlation functions
without vacuum insertions in all order of perturbation theory are
represented by
a sum of all topologically non-equivalent graphs without
any combinatoric factors. The proof of this statement is presented in
Appendix B.
According the second one in the Boltzmann theory all graphs contribute
into correlations functions  without any combinatoric factors.

Let us compare
the half-planar approximation with
the planar approximation
in first orders of perturbative expansion in the coupling constant.
We collect corresponding diagrams on Figures 5-7.

The explicit formula for an arbitrary planar
Green's function without vacuum
insertions is \cite{BIPZ,KNN}
\begin{equation}
\label{Pi}
\Pi _{2n}(g)=\frac{(2n)!}{n!(n+2)!}a^{2n}(2n+2-na^2),
\end{equation}
where
\begin{equation}
\label{ag}
a^2=\frac{1}{6g}[\sqrt{1+12g}-1].
\end{equation}
For $n=1,~2,~3$ we have
$$
\Pi _2=\frac{1}{3}a^2(4-a^2)=1-2g+9g^2-54g^3+378g^4-...~,
$$
$$
\Pi _4=a^4(3-a^2)=2-9g+54g^2-378g^3+2920g^4-...~,
$$
\begin{equation}
\label{Pi1}
\Pi _6=a^6(8-3a^2)=5-36g+270g^2-2164g^3+...~.
\end{equation}

On  Table 1
 the results of numerical calculations of half-planar Green's functions
${\cal G}_2$, ${\cal G}_4$, ${\cal G}_6$ and
the planar Green's function $\Pi _2$, $\Pi _4$, $\Pi _6$
for some values of the coupling constant $g$ are presented
$$~$$

Table 1

\begin{tabular}{llllllll}
\hline
$g$ & $10^{-3}$ & $10^{-2}$ & $10^{-1}$ & $1$ & $10$ & $10^2$&$10^3$
\\
\hline
$\Pi _2$
& $9.98\cdot  10^{-1}$
& $9.81 \cdot 10^{-1}$
& $8.6 \cdot 10^{-1}$
& $5.2 \cdot 10^{-1}$
& $2.1 \cdot 10^{-1}$
& $7.4 \cdot 10^{-2}$
& $2.4 \cdot 10^{-2}$
\\
${\cal G}_2$
& $9.97\cdot  10^{-1}$
& $9.71 \cdot 10^{-1}$
& $8.0 \cdot 10^{-1}$
& $4.0 \cdot 10^{-1}$
& $1.3 \cdot 10^{-1}$
& $3.2 \cdot 10^{-2}$
& $7.4 \cdot 10^{-3}$
\\
\hline
$\Pi _4$
& $1.99$
& $1.96$
& $1.42$
& $4.84 \cdot 10^{-1}$
& $7.87 \cdot 10^{-2}$
& $9.26 \cdot 10^{-3}$
& $9.76 \cdot 10^{-4}$
\\
${\cal G}_4$
& $1.99$
& $1.91$
& $1.38$
& $4.43 \cdot 10^{-1}$
& $7.16 \cdot 10^{-2}$
& $8.65 \cdot 10^{-3}$
& $9.37 \cdot 10^{-4}$
\\
\hline
$\Pi _6$
& $4.96$
& $4.67$
& $2.92$
& $5.48 \cdot 10^{-1}$
& $3.47 \cdot 10^{-2}$
& $1.38 \cdot 10^{-3}$
& $4.70 \cdot 10^{-5}$
\\
${\cal G}_6$
& $4.97$
& $4.70$
& $3.11$
& $7.78 \cdot 10^{-1}$
& $9.64 \cdot 10^{-2}$
& $1.00 \cdot 10^{-2}$
& $1.00 \cdot 10^{-3}$
\\
\hline
\end{tabular}
\vspace{10mm}

One can see that the answers for half-plane Green's functions
${\cal G}_2$, ${\cal G}_4$ practically saturate the planar Green's
functions $\Pi _2$, $\Pi _4$ in board interval of the values of $g$.
${\cal G}_6$ is in the good accordance with $\Pi _6$ for $g$
only for small $g$.

Let us make also a comparison of these two theories for the case 
when all tadpole graphs are neglected.
The matrix theory  the rejection of tadpole graphs is equivalent
to adding to the Lagrangian the term 
\begin{equation}
\label{mp2}
-g {\tilde \Pi}_2 \tr(M^2),
\end{equation}
where    ${\tilde \Pi}_2$ is tadpole-free two-point planar Green's
function. So the tadpole-free planar Green's functions
are defined as
\begin{equation}
\label{mp3}
{\tilde \Pi}_{2n}=\lim _{N \to \infty}
\frac{1}{N^{1+n}}\int DM
tr(M^{2n})\exp[-\frac{1}{2}
(1-2g{\tilde \Pi}_2 )tr(M^2)- \frac{g}{4N}tr(M^4)].
\end{equation}
The perturbative expansion in the theory (\ref{mp3})
is valid only for  
\begin{equation}
\label{restr}
1-2 g {\tilde \Pi}_2 >0.
\end{equation}
For this case
making the change  of variable
$$M=(1-2g{\tilde \Pi}_2)^{-\frac{1}{2}}{\tilde M}$$
in the (\ref{mp3}) one gets
\begin{equation}
\label{mp4}
{\tilde \Pi}_{2n}
=\int D{\tilde M}
(1-2g{\tilde \Pi}_2)^{-n}
tr({\tilde M}^{2n})
exp[-\frac{1}{2}tr({\tilde M}^2)-
\frac{{\tilde g}}{4N}tr({\tilde M}^4)],
\end{equation}
where
$${\tilde g}=
\frac{g}{(1-2g{\tilde \Pi}_2)^2}.$$
So the Green's functions
${\tilde \Pi}_{2n}(g)$
in that all tadpole contribution are dropped out
can be expressed in terms of full Green's functions
$ \Pi _{2n}(g) $  
\begin{equation}
\label{mp5}
{\tilde \Pi}_{2n}(g)=
(1-2g{\tilde \Pi}_2)^{-n}
\Pi _{2n}(
\frac{g}{(1-2g{\tilde \Pi}_2)^2}).
\end{equation}

Now let us reformulate the Boltzmann theory in such way
that it  reproduces only tadpole-free half-planar graphs.
One possible way to do this is to consider the
formulation of Boltzmann theory using two fields.

For this purpose let us defined the fields $\psi$ and $\phi$ as follows
\begin{equation}
\label{2f.1}
\phi = a+a^+,~~
\psi = b + b^+,
\end{equation}
where $a$, $a^+$, $b$ and $b^+$ satisfy the following relations
$$
a a^+=1,~~ b b^+=1,
$$
\begin{equation}
\label{2f.2}
a b^+=0,~~b a^+=0.
\end{equation}
This algebra has the realization in the Boltzmannian Fock space.
For two degrees of freedom
the vacuum  $|0\rangle $ satisfies
\begin{equation}
\label{2f.3}
a|0\rangle=0,~~
b|0\rangle=0.
\end{equation}

Now let us define an interaction Green's functions by the formula
\begin{equation}
\label{2f.5}
{\cal G}^{2-field}_{k_1 l_1,...,k_m,l_m} \equiv \langle 0 | \psi ^{k_1}
\phi ^{l_1}...  \psi ^{k_m} \phi ^{l_m} \frac{1}{1+S_{int}}|0\rangle,
\end{equation}
where the interaction $S_{int}$ has the form
\begin{equation}
\label{2f.4}
S_{int}=
g\psi :\phi \phi : \psi=
g\psi \phi \phi  \psi -
g\psi \psi.
\end{equation}
One can see that  the connected tadpole-free parts of Green's
functions in Boltzmann theory described in subsection 2.2
coincide with  the connected parts of Green's functions
${\cal G}^{2-field}_{1,n-2,1}$
\begin{equation}
\label{Fn}
 F_n  \equiv
{\cal G}^{2-field,~conn.}_{1,n-2,1}
=\langle 0|\psi :\phi ^{n-2}:\psi
\frac{1}{1+S_{int}} |0\rangle .
\end{equation}

Let us write down the system of Schwinger-Dyson equations for
the Green's functions $F_2$ and $F_4$
$$
F_2=1-gF_4,
$$
\begin{equation}
\label{2f.8}
F_4=-g F_2 F_4 -g F_2^2.
\end{equation}

The results of numerical  calculation of $F_2$ and $F_4$ related to
a perturbative branch of solution of (\ref{2f.8})
are presented on Table 2.  This solution has a phase transition point
$g_0=0$. When $g$ goes to $g_0$ then $F_2$ and $F_4$ go to infinity.
Also on Table 2 we present the numerical values for ${\tilde \Pi}_2$
and $ {\tilde \Pi}_4^{connected}$ $=$
$ {\tilde \Pi}_4-2 {\tilde \Pi}_2^2 $ for $g<g_0=0.421875$,
where  $g_0$ is determined from the following equation
\begin{equation}
\label{g0}
1-2 g {\tilde \Pi}_2 =0.
\end{equation}

Table 2

\begin{tabular}{llllllll}
\hline
$g$ & $10^{-3}$ & $10^{-2}$ & $10^{-1}$ & $0.4$ 
& $0.421870$ & $0.9$ &$0.99$
\\
\hline
$F _2$
& $1.00$
& $1.00$
& $1.01$
& $1.14$
& $1.16$
& $2.82$
& $9.56$
\\
${\tilde \Pi}_2$
& $1.00$
& $1.00$
& $1.01$
& $1.16$
& $1.18$
& $~~~-$
& $~~~-$
\\
\hline
$F _4$
& $-9.99 \cdot 10 ^{-4}$
& $-9.90 \cdot 10^{-3}$
& $-0.0925$
& $-0.359$
& $-0.382$
& $-2.03$
& $-8.64$
\\
${\tilde \Pi}_4^{connected}$
& $-9.98 \cdot 10 ^{-4}$
& $-9.81 \cdot 10^{-3}$
& $-0.0872$
& $-0.394$
& $-0.439$
& $~~~-$
& $~~~-$
\\
\hline
\end{tabular}
\vspace{5mm}

Perturbation series for $F_2$, $F_4$,
${\tilde \Pi}_2$  and
${\tilde \Pi}_4^{connected}$ have the forms
$$
F_2=\frac{-1+g +\sqrt{1+2g-3g^2}}{2g(1-g)}=
1+g^2-g^3+3g^4-6 g^5+...~,
$$
\begin{equation}
F_4=\frac{1+g -2g^2-\sqrt{1+2g-3g^2}}{2g^2(1-g)}=
-g+2g^2-5g^3+6 g^4 -15 g^5 +...~,
\label{2f.10}
\end{equation}

$$
{\tilde \Pi}_2=
1+g^2-2 g^3+10 g^4-42 g^5+...~,
$$
\begin{equation}
{\tilde \Pi}_4^{connected}=
-g+2g^2-10 g^3+42 g^4 -209 g^5 +...~.
\label{2f.11}
\end{equation}
\section{Boltzmann Theory for Finite
Number of Degree of Freedom}
\setcounter{equation}{0}
\subsection{The  Schwinger-Dyson Equations}

In this section we will study the  Boltzmannian
  Schwinger-Dyson equations for the theory
with finite number of degrees of freedom.

First of all let us derive the Boltzmannian Schwinger-Dyson equations.
This can be done similarly to the zero-dimensional case. 
We adopt the following notations.
Let $ a,b,c,...$ are multi-indices  which label
degrees of freedom,
$\phi (a)=\phi ^+(a)+\phi ^-(a)$,
where $\phi ^+(a)$ and $\phi ^-(a)$
satisfy the relation
\begin{equation}
\label{z1}
\phi ^-(a) \phi ^+(b)=G^{(0)} (a,b)
\end{equation}
and $G^{(0)} (a,b)$ is a matrix which has the inverse $K(a,b)$:
\begin{equation}
\label{inv}
\sum _{c} K(a,c)G(c,b)=\delta (a,b).
\end{equation}

The algebra (\ref{z1}) has the realization in the
Bolzmannian Fock space  generated by the vacuum $|0\rangle$,
$\phi ^-(a)|0\rangle =0 $,
and $n$-particle states
\begin{equation}
\label{z2}
|a_1,...,a_n\rangle =\phi ^+(a_1)...\phi ^+(a_n)|0\rangle .
\end{equation}
Note that states (\ref{z2}) have no any symmetry properties
under permutation of particles because there is no relations between
 $\phi ^+(a)$ $\phi ^+(b)$  and
 $\phi ^+(b)$ $\phi ^+(a)$.

Free correlation functions
$$G^{(0)}(a_1,...,a_n)=
\langle 0|\phi (a_1)...\phi (a_n)|0\rangle $$
in the Bolzmannian Fock space satisfy the
relations
$$
\sum _{c}K(a_k,c)G^{(0)}(a_1,...,a_{k-1},c,a_{k+1},a_n)=$$
$$=\sum _{p=k+1}^{n}\delta(a_k,a_p)
G^{(0)}(a_{k+1},...,a_{p-1})
G^{(0)}(a_{1},...,a_{k-1},a_{p+1},...,a_n)+$$
\begin{equation}
\label{z6}
\sum _{p=1}^{k-1}\delta (a_p,a_k)
G^{(0)}(a_{p+1},...,a_{k-1})
G^{(0)}(a_{1},...,a_{p-1},a_{k+1},...,a_n),
\end{equation}
that follow from  the Wick
theorem in the Boltzmannian Fock space. We also call
equations (\ref{z6}) the free Schwinger-Dyson equations.

Interacting Green's functions
${\cal G}(a_1,...,a_n)$  are defined
by an ana\-logy with the 0-di\-men\-si\-onal mo\-del:
\begin{equation}
\label{z3}
 {\cal G}(a_1,...,a_n)=
\langle 0|\phi (a_1)...\phi (a_n)
\Omega (\phi)
|0\rangle ,~~\Omega (\phi )=
\frac{1}{1+V_{int}(\phi)}.
\end{equation}
Let us note that the Green's functions (\ref{z3}) have a symmetry under 
the permutation of arguments in inverse order
\begin{equation}
\label{z3a}
 {\cal G}(a_1,...,a_n)=
 {\cal G}(a_n,...,a_1).
\end{equation}
There are no another 
symmetry properties.

The form of $V_{int}(\phi)$ depends on the model under consideration.
We are restricted ourself to the case with $V_{int}$ in the form
\begin{equation}
\label{z4}
V_{int}(\phi)=g\sum _{a}(\phi (a))^r.
\end{equation}

To derive the Scwinger-Dyson equations
for correlation functions (\ref{z3})
let us rewrite
the Green's function (\ref{z3}) in terms of the free Green's 
functions, $${\cal G}(a_1,...a_{n})= \sum _{m=0}^{\infty}\sum 
_{b_1,...b_m} (- g)^m \langle 0|\phi (a_1)...\phi (a_{n}) (\phi 
(b_1))^r...(\phi (b_m))^r|0\rangle =$$ \begin{equation} \label{z5} 
\sum _{m=0}^{\infty}\sum _{b_1,...b_m}
{G}^{(0)}(a_1,...,a_n,\underbrace
{b_1,...,b_1}_{r},...,
\underbrace{b_m,...,b_m}_{r})
\end{equation}
Using the free Schwinger-Dyson equations (\ref{z6})
we have
$$
\sum _{c}K(a_k,c){{\cal G}}(a_1,...,a_{k-1},c,a_{k+1},...,a_n)=$$
$$
\sum _{c}K(a_k,c)
\sum _{m=0}^{\infty}\sum_{b_1,...,b_m}
(-g)^m
{G}^{(0)}(a_1,...,a_{k-1},c,a_{k+1},a_n,\underbrace{b_1,...,b_1}_{r},...,
\underbrace{b_m,...,b_m}_{r})=$$
\begin{equation}
\label{j}
=J_{I}+J_{II}+J_{III},
\end{equation}
where
$$
J_{I}=\sum _{m=0}^{\infty}\sum_{b_1,...,b_m}
\sum_{p=1}^{k-1} (-g)^m
\delta(a_p,a_k){G}^{(0)}(a_1,...,a_{p-1},a_{k+1},...,a_n,
\underbrace
{b_1,...,b_1}_{r},...,
\underbrace
{b_m,...,b_m}_{r}) \times$$
\begin{equation}
\label{j1}
\times {G}^{(0)}(a_{p+1},...,a_{k-1})
\end{equation}

$$
J_{II}=\sum _{m=0}^{\infty}\sum_{b_1,...,b_m}
\sum_{p=k+1}^{n} (-g)^m
\delta(a_k,a_p){G}^{(0)}(a_1,...,a_{k-1},a_{p+1},...,a_n,
\underbrace {b_1,...,b_1}_{r},...,
\underbrace {b_m,...,b_m}_{r}) \times $$
\begin{equation}
\label{j2}
\times {G}^{(0)}(a_{k+1},...,a_{p-1})
\end{equation}

$$
J_{III}=\sum_{b_1,...,b_m}
\sum_{l=1}^{m}\sum_{s=1}^{r}(-g)^m\delta(a_k,b_l)
{G}^{(0)}(
a_1,...,a_{k-1},
\underbrace{b_1,...,b_1}_{r},...,
\underbrace{b_{l-1},...,b_{l-1}}_{r},
\underbrace{b_l,...,b_l}_{s-1}
)
\times $$
\begin{equation}
\label{J3}
\times {G}^{(0)}(
\underbrace {b_l,...,b_l}_{r-s},
a_{k+1},...,a_{n},
\underbrace {b_{l+1},...,b_{l+1}}_{r},...,
\underbrace {b_{m},...,b_{m}}_{r}).
\end{equation}
Using the  relation
$$\sum_{m=0}^{\infty}\sum_{l=1}^{m}f(m,l)=
\sum_{l=1}^{\infty}\sum_{m=l}^{\infty}f(m,l)$$
and making the summation over $b_l$ and then changing the summation 
index $$m \to m'=m-l,~$$ one can represent the term $J_{III}$ in the 
form

$$ J_{III} =
\sum_{l=1}^{\infty}\sum_{m'=0}^{\infty}\sum_{b_i}
 \sum_{s=1}^{r}(-g)^m {G}^{(0)}(
a_1,...,a_{k-1}, \underbrace {b_1,...,b_1}_{r},...,
\underbrace {b_{l-1},...,b_{l-1}}_{r},
 \underbrace {a_k,...,a_k}_{s-1},
)
\times $$
$$
\times {G}^{(0)}(
\underbrace{a_k,...,a_k}_{r-s},
a_{k+1},...,a_{n},
\underbrace{b_{l+1},...,b_{l+1}}_{r},...,
\underbrace{b_{m'+l},...,b_{m'+l}}_{r})=$$
\begin{equation}
\label{z8}
-g\sum_{s=1}^{r}{{\cal G}}(
a_1,...,a_{k-1},
\underbrace{a_k,...,a_k}_{s-1},
)
{{\cal G}}(
a_{k+1},...,a_n
)
\end{equation}

Putting all together we obtain the half-plane Schwinger-Dyson equations
$$\sum _{c}K(a_k,c){{\cal G}}(a_1,...,a_{k-1},c,a_{k+1},...,a_n)=$$
$$
\sum_{p=1}^{k-1}
\delta(a_p,a_k){{\cal G}}(a_1,...,a_{p-1},a_{k+1},...,a_n)
{G}^{(0)}(a_{p+1},...,a_{k-1})+$$
$$
\sum_{p=k+1}^{n}
\delta(a_k,a_p){{\cal G}}(a_1,...,a_{k-1},a_{p+1},...,a_n)
{G}^{(0)}(a_{k+1},...,a_{p-1})-$$
\begin{equation}
\label{z9}
g\sum_{s=1}^{r}{{\cal G}}(
a_1,...,a_{k-1},
\underbrace{a_k,...,a_k}_{s-1}
)
{{\cal G}}(
\underbrace{a_k,...,a_k}_{r-s},
a_{k+1},...,a_n
).
\end{equation}

Also we have the equations
\begin{equation}
\label{bbb}
{{\cal G}}(a_1,...,a_n)={G}^{(0)}(a_1,...,a_n)
-g\sum_{b}{{\cal G}}(a_1,...,a_n,\underbrace{b,...,b}_{r})
\end{equation}
which follow from the average the operator identity
\begin{equation}
\label{bbb1}
\frac{1}{1+V_{int}}=1-\frac{V_{int}}{1+V_{int}}.
\end{equation}

Let us consider the interaction
 $g\sum _{a} [\phi (a)]^4$,
and write down the Scwinger-Dyson equations (\ref{z9}) corresponding to
 $(n,k)$ $=$
$(2,1),(4,2) $.
We have
\begin{equation}
\label{z19}
\sum _{e}K(a,e){\cal G}(e,b)
=\delta(a,b){\cal G}-g{\cal G}(a,b)
{\cal G}(a,a)-g{\cal G}(a,a,a,b){\cal G},
\end{equation}

$$\sum_{e}K(b,e){\cal G}(a,e,c,d)
=\delta (b,c){\cal G}(a,d)+\delta(a,b){\cal G}(c,d)-
$$
\begin{equation}
\label{z20}
g{\cal G}(c,d){\cal G}(a,b,b,b)-g{\cal G}(a,b){\cal G}(b,b,c,d),
\end{equation}
where ${\cal G}$ is vacuum Green's function.
Equation (\ref{z19}) and (\ref{z20})
together with the equation for vacuum Green's function
that follows from (\ref{bbb})
\begin{equation}
\label{vac}
{\cal G}=1-g\sum _{a}{\cal G}(a,a,a,a)
\end{equation}
form the closed system for ${\cal G}$, 
${\cal G}(a,b)$ and ${\cal G}(a,b,c,d)$.
\subsection{Diagonal Kinetic Operator}

In this subsection we will study the theory with $K$
degrees of freedom and diagonal kinetic operator.
We assume the following form of Lagrangian
\begin {equation} 
\label {ll1}
{\cal L}=\frac{1}{2}\sum_{a=1}^{K} \Phi _a^2
+g\sum_{a=1}^{K} \Phi _a^4.
\end   {equation} 
The latin indices $a,b,c,...$   run the values $1,2,...,K$.
The operators $\Phi _a$ are the sum of Boltzmann creation and annihilation
operators $\Phi _a =\Phi _a^+ + \Phi _a^- $
which satisfy the  relation
\begin {equation} 
\label {ll2}
\Phi _a^- \Phi _b^+ =\delta _{a,b}.
\end   {equation} 
The Green's functions are defined by
\begin {equation} 
\label {ll3}
{\cal G}(a,b,...,c) \equiv {\cal G}_{ab...c} =
\langle 0|
\Phi _a \Phi _b ... \Phi _c \frac{1}{1+g\sum _{a}\Phi _a^4}
 |0 \rangle .
\end   {equation} 
It is evident that
\begin{equation}
\label{y2}
{{\cal G}}_{\underbrace{a...a}_{n}}=
{{\cal G}}_{\underbrace{b...b}_{n}}
\end{equation}
for any $a$ and $b$ and
\begin{equation}
\label{y3}
{{\cal G}}_{abb...b}=0
\end{equation}
for $a \ne b$.

Let us write down the
 equations (\ref{z19}), (\ref{z20}) and (\ref{vac})
for this case:
\begin{equation}
\label{y4}
{\cal G}_{aa} ={\cal G}-g{\cal G}_{aa} ^2-g{\cal G}{\cal G}_{aaaa},
\end{equation}
\begin{equation}
\label{y5}
{\cal G}_{aaaa}=2{\cal G}_{aa} -2g{\cal G}_{aa} {\cal G}_{aaaa},
\end{equation}
\begin{equation}
\label{y6}
{\cal G}=1-Kg{\cal G}_{aaaa}.
\end{equation}
  From these equations one can write the equations for ${\cal G}$:
\begin{equation}
\label{y7}
4g{\cal G}^4+12(K-1)g{\cal G}^3+[12(K-1)^2g+K]{\cal G}^2+[2K(K-1)+
4g(K-1)^3]{\cal G}-K(2K-1)=0.
\end{equation}
This equation can be solved and the Green's functions
 ${\cal G}_{aa}$ and ${\cal G}_{aaaa} $ are expressed 
in terms of ${\cal G}$.

\subsection{Two Degree of Freedom and Non-Diagonal Quadratic Kinetic
Operator}

In this subsection we shall consider the theory with two degree of
freedom and non-diagonal  kinetic operator.
The Lagrangian has the form
\begin {equation} 
\label {la1}
{\cal L}=\frac{1}{2}\sum_{a,b} K_{a,b}\Phi _a \Phi _b
+g\sum_{a} \Phi _a^4.
\end   {equation} 
The indices $a,b,c,...$ will take the values 1 and 2.
We adopt the following parametrization of the
kinetic operator
 $K_{ab}$
\begin{equation}
\label{z26a}
K_{11}=K_{22}=\alpha,~~K_{12}=K_{21}=-\beta.
\end{equation}
In this case we have
\begin{equation}
\label{z26}
\Phi _1^- \Phi _1^+
=\Phi _2^- \Phi _2^+
=\frac{\alpha}{\alpha ^2
-\beta ^2},
~~\Phi _1^- \Phi _2^+
=\Phi _2^- \Phi _1^+
=\frac{\beta}{\alpha ^2-\beta ^2}.
\end{equation}

The Lagrangian is invariant under the transformations
$\Phi _1 \to \Phi _2$,
$\Phi _2 \to \Phi _1$.
Therefore all Green's functions are invariant under the change
of indices
$1 \to 2$, $2 \to 1$. In particular, we have
\begin{equation}
\label{ch1}
{\cal G}_{ab}={\cal G}_{ba},
\end{equation}
\begin{equation}
\label{ch2}
{\cal G}_{\underbrace{a...a}_{n}}=
{\cal G}_{\underbrace{b...b}_{n}}
\end{equation}
for any $a$,  $b$ and $n$.

Let us write (\ref{vac}),
(\ref{z19}) for $(a,b)=(1,1),~(1,2)$
  and (\ref{z20}) for $(a,b,c,d)=$
$(1,1,1,1)$, $(2,1,1,1)$, $(1,2,1,1)$, $(2,2,1,1)$:
\begin{equation}
\label{GV}
{\cal G}=1-g{\cal G}_{1111}-g{\cal G}_{2222},
\end{equation}

\begin{equation}
\label{z21}
\alpha {\cal G}_{11}-
\beta {\cal G}_{21}={\cal G}-g{\cal G}^2_{11}-g{\cal G}_{1111}{\cal G},
\end{equation}

\begin{equation}
\label{z22}
\alpha {\cal G}_{12}-\beta {\cal G}_{22}=
-g{\cal G}_{12}{\cal G}_{11}-g{\cal G}_{1112}{\cal G},
\end{equation}

\begin{equation}
\label{z23}
\alpha {\cal G}_{1111}-\beta {\cal G}_{1211}=
2{\cal G}_{11}-2g{\cal G}_{11}{\cal G}_{1111},
\end{equation}

\begin{equation}
\label{z24}
\alpha {\cal G}_{2111}-\beta {\cal G}_{2211}={\cal G}_{21}-
g{\cal G}_{11}{\cal G}_{2111}-g{\cal G}_{21}{{\cal G}}_{1111},
\end{equation}

\begin{equation}
\label{z25}
-\beta {\cal G}_{1111}+\alpha {\cal G}_{1211}=
-g{\cal G}_{11}{\cal G}_{1222}-
g{\cal G}_{12}{\cal G}_{2211},
\end{equation}

\begin{equation}
\label{z25a}
-\beta {\cal G}_{2111}+\alpha {\cal G}_{2211}=
{\cal G}_{11}-
g{\cal G}_{11}{\cal G}_{2222}-
g{\cal G}_{22}{\cal G}_{2211}.
\end{equation}

The following  symmetry properties of Green's functions are evident:
$$
{\cal G}_{11}={\cal G}_{22},~~{\cal G}_{12}={\cal G}_{21} ,
~~ {\cal G}_{1111}={\cal G}_{2222},$$

\begin{equation}
\label{z27}
{\cal G}_{1211}={\cal G}_{2122},
~~~  {\cal G}_{1122}={\cal G}_{2211},~~~
{\cal G}_{1112}={\cal G}_{2221}={\cal G}_{1222}={\cal G}_{2111}.
\end{equation}
So the system (\ref{GV})-(\ref{z25a}) contains $7$ independent equations
and $7$ unknowns variables
${\cal G}$, ${\cal G}_{11}$,
${\cal G}_{12}$,
${\cal G}_{1111}$,
${\cal G}_{1112}$,
${\cal G}_{1211}$,
${\cal G}_{1122}$ and therefore can be solved.

To get the equations for Green's functions without vacuum insertions
one has to set ${\cal G}=1$ in (\ref{z21}) - (\ref{z25a}).
Then we have a system of 6 equations for 6 variables.
The numerical solutions of this system for $\alpha =1$
and some values of $\beta$ and $g$ are presented on  Tables 3-5.

\vspace{10mm}

Table 3. $\beta=0.01$

\begin{tabular}{llllllll}
\hline
$g$ & $10^{-3}$ & $10^{-2}$ & $10^{-1}$ & $1$ & $10$ & $10^2$&$10^3$
\\
\hline
$G_{11}$
& $9.97\cdot  10^{-1}$
& $9.72 \cdot 10^{-1}$
& $7.99 \cdot 10^{-1}$
& $3.98 \cdot 10^{-1}$
& $1.26 \cdot 10^{-1}$
& $3.2 \cdot 10^{-2}$
& $7.45 \cdot 10^{-3}$
\\
$G_{12}$
& $9.94 \cdot  10^{-3}$
& $9.43 \cdot 10^{-3}$
& $6.38 \cdot 10^{-3}$
& $1.58 \cdot 10^{-3}$
& $1.58 \cdot 10^{-4}$
& $1.03 \cdot 10^{-5}$
& $1.51 \cdot 10^{-6}$
\\
 $G_{1111}$
& $1.99$
& $1.91$
& $1.38$
& $4.43 \cdot 10^{-1}$
& $7.16 \cdot 10^{-2}$
& $8.65 \cdot 10^{-3}$
& $9.37 \cdot 10^{-4}$
\\
$G_{1112}$
& $1.98 \cdot 10^{-2}$
& $1.85 \cdot 10^{-2}$
& $1.10 \cdot 10^{-2}$
& $1.77 \cdot 10^{-3}$
& $9.01 \cdot 10^{-5}$
& $2.77 \cdot 10^{-6}$
& $6.98 \cdot 10^{-8}$
\\
$G_{1211}$
& $1.99 \cdot  10^{-2}$
& $1.88 \cdot 10^{-2}$
& $1.25 \cdot 10^{-2}$
& $3.48 \cdot 10^{-3}$
& $5.77 \cdot 10^{-4}$
& $7.66 \cdot 10^{-5}$
& $8.82 \cdot 10^{-6}$
\\
$G_{1122}$
& $9.94\cdot  10^{-1}$
& $9.44 \cdot 10^{-1}$
& $6.38 \cdot 10^{-1}$
& $1.59 \cdot 10^{-1}$
& $1.58 \cdot 10^{-2}$
& $1.03 \cdot 10^{-3}$
& $5.55 \cdot 10^{-5}$
\\
\hline
\end{tabular}
\vspace{10mm}

Table 4. $\beta =0.1$

\begin{tabular}{llllllll}
\hline
$g$ & $10^{-3}$ & $10^{-2}$ & $10^{-1}$ & $1$ & $10$ & $10^2$&$10^3$
\\
\hline
$G_{11}$
& $1.01$
& $9.80 \cdot 10^{-1}$
& $8.03 \cdot 10^{-1}$
& $3.98 \cdot 10^{-1}$
& $1.26 \cdot 10^{-1}$
& $3.20 \cdot 10^{-2}$
& $7.45 \cdot 10^{-3}$
\\
$G_{12}$
& $1.00 \cdot  10^{-1}$
& $9.52 \cdot 10^{-2}$
& $6.4 \cdot 10^{-2}$
& $1.58 \cdot 10^{-2}$
& $1.58 \cdot 10^{-3}$
& $1.03 \cdot 10^{-4}$
& $1.51 \cdot 10^{-5}$
\\
 $G_{1111}$
& $2.03$
& $1.94$
& $1.39$
& $4.45 \cdot 10^{-1}$
& $7.17 \cdot 10^{-2}$
& $8.66 \cdot 10^{-3}$
& $9.37 \cdot 10^{-4}$
\\
$G_{1112}$
& $2.02 \cdot 10^{-1}$
& $1.89 \cdot 10^{-1}$
& $1.11 \cdot 10^{-1}$
& $1.77 \cdot 10^{-2}$
& $9.00 \cdot 10^{-4}$
& $2.77 \cdot 10^{-5}$
& $6.96 \cdot 10^{-7}$
\\
$G_{1211}$
& $2.03 \cdot  10^{-1}$
& $1.91 \cdot 10^{-1}$
& $1.26 \cdot 10^{-1}$
& $3.50 \cdot 10^{-2}$
& $5.79 \cdot 10^{-3}$
& $7.67 \cdot 10^{-4}$
& $8.83 \cdot 10^{-5}$
\\
$G_{1122}$
& $1.02$
& $9.71 \cdot 10^{-1}$
& $6.50 \cdot 10^{-1}$
& $1.59 \cdot 10^{-1}$
& $1.58 \cdot 10^{-2}$
& $1.02 \cdot 10^{-3}$
& $5.51 \cdot 10^{-5}$
\\
\hline
\end{tabular}
\vspace{10mm}

Table 5. $\beta =0.9$

\begin{tabular}{llllllll}
\hline
$g$ & $10^{-3}$ & $10^{-2}$ & $10^{-1}$ & $1$ & $10$ & $10^2$&$10^3$
\\
\hline
$G_{11}$
& $4.63$
& $2.78$
& $1.03$
& $3.05 \cdot 10^{-1}$
& $7.94 \cdot 10^{-2}$
& $1.91 \cdot 10^{-2}$
& $4.34 \cdot 10^{-3}$
\\
$G_{12}$
& $4.11$
& $2.28$
& $6.37 \cdot 10^{-1}$
& $1.03 \cdot 10^{-1}$
& $1.16 \cdot 10^{-2}$
& $9.86 \cdot 10^{-4}$
& $6.60 \cdot 10^{-5}$
\\
 $G_{1111}$
& $44.9$
& $19.7$
& $4.33$
& $6.95 \cdot 10^{-1}$
& $8.68 \cdot 10^{-2}$
& $9.46 \cdot 10^{-3}$
& $9.77 \cdot 10^{-4}$
\\
$G_{1112}$
& $39.8$
& $15.8$
& $2.27$
& $1.40 \cdot 10^{-1}$
& $5.06 \cdot 10^{-3}$
& $1.43 \cdot 10^{-4}$
& $3.55 \cdot 10^{-6}$
\\
$G_{1211}$
& $40.1$
& $16.9$
& $3.51$
& $5.66 \cdot 10^{-1}$
& $7.31 \cdot 10^{-2}$
& $8.20 \cdot 10^{-3}$
& $8.62 \cdot 10^{-4}$
\\
$G_{1122}$
& $40.1$
& $16.0$
& $2.38 $
& $1.68 \cdot 10^{-1}$
& $8.39 \cdot 10^{-3}$
& $4.02 \cdot 10^{-4}$
& $1.94 \cdot 10^{-5}$
\\
\hline
\end{tabular}
\vspace{10mm}

The dependence of the 2-point Green's functions $G_{11}$ and
$G_{12}$ on the coupling constant $g$ for different values of $\beta$
is also presented on  Fig.8.

The limit $\beta \to 0$ corresponds to decoupling of the degrees of freedom.
In this limit the mixed Green's function $G_{12}$
goes to zero
and we reproduce the case of one degree of freedom.
For $\beta=0.9$ the mixed Green's function is of the same order as
the diagonal Green's function.

\section{Boltzmann Correlation Functions in $D$-Di\-men\-si\-o\-nal 
Spa\-ce-Time}
\setcounter{equation}{0}
\subsection{The  Schwinger-Dyson Equations}

In this section we will derive the
  Schwinger-Dyson equations
for Boltzmann correlation functions in $D$-dimensional Euclidean space.

We adopt the following notations.
Let
\begin {equation} 
							  \label {phi}
\phi (x)=\phi ^+(x)+\phi ^-(x)
\end   {equation} 
be the Bolzmann field with creation and annihilation operators
satisfying the relation
\begin{equation}
\label{z0}
\phi ^-(x) \phi ^+(y)=D(x,y),
\end{equation}
where
\begin{equation}
\label{pro}
D(x,y)=\frac{1}{(2\pi)^D}\int d ^Dp\frac{e^{-ip(x-y)}}
{p^2+m^2}
\end{equation}
is $D$-dimensional Euclidean propagator.
In what follows we  will assume that $\phi (x)$ is a scalar field.
A generalization to the case when $\phi $ carries
 isotopic or tensor indices is straightforward.

By analogy with the case of finite number  degrees of freedom we
consider the following correlation functions
\begin {equation} 
							  \label {cor.fun}
 F_{n}(x_n,...,x_1)=
\langle 0|\phi (x_n)...\phi (x_1)
 \frac{1}{1+\int d^{D}x {\cal L}_{int}(\phi(x))}
|0\rangle ,
\end   {equation} 
where
${\cal L}_{int}(\phi(x))$
is a local invariant polynomial in the field and its derivatives.
As before we  restrict ourself by the case of the quartic interaction
${\cal L}_{int}(\phi)=g \phi ^4$.

Using the results of the previous section one can immediately write down
the  Boltzmannian Schwinger-Dyson equations for correlation functions
(\ref{cor.fun}). We have

$$(-\bigtriangleup +m^{2})_{x_{2l}}F_{2n}(x_{2n},...x_{1})=
-g(F_{2l}(x_{2l},x_{2l-1},...x_{1})F_{2n-2l+2}
(x_{2n},...x_{2l+1},x_{2l},x_{2l})
+$$
$$
F_{2l+2}(x_{2l},x_{2l},x_{2l},x_{2l-1},...x_{1})
F_{2n-2l}(x_{2n},...x_{2l+1})
)+$$
$$
\sum _{i <  2l}\delta (x_{2l}-x_{i})F_{2n-2l+i-1}(x_{2n},...
x_{2l+1},x_{i-1},...x_{1})
F^{(0)}_{2l-i-1}(x_{2l-1},...x_{i+1})+$$
\begin {equation} 
							  \label {ebf}
\sum _{2l <  i}\delta
(x_{2l}-x_{i})F_{2n+2l-i-1}(x_{2n},...x_{i+1},x_{2l-1},...
x_{1})F^{(0)}_{i-2l-1}(x_{i-1},...x_{2l+1}),~~~l\leq n,
\end   {equation} 
and the similar equations for $x_{2l+1}$
$$(-\bigtriangleup +m^{2})_{x_{2l+1}}F_{2n}(x_{2n},...x_{1})=
$$
$$
-g(F_{2l}(x_{2l},x_{2l-1},...x_{1})F_{2n-2l+2}(x_{2n},...x_{2l+2},x_{2l+1},
x_{2l+1},x_{2l+1})+
$$
$$
F_{2l+2}(x_{2l+1},x_{2l+1},x_{2l},...x_{1})F_{2n-2l}(x_{2n},...x_{2l+2},
x_{2l+1})
)+$$
$$
\sum _{i <  2l+1}\delta (x_{2l}-x_{i})F_{2n-2l+i-2}(x_{2n},...
x_{2l+2},x_{i-1},...x_{1})
F^{(0)}_{2l-i}(x_{2l},...x_{i+1})+$$
\begin {equation} 
							  \label {ebf'}
\sum _{i > 2l+1}\delta
(x_{2l}-x_{i})F_{2n+2l-i}(x_{2n},...x_{i+1},x_{2l},...
x_{1})F^{(0)}_{i-2l-2}(x_{i-1},...x_{2l+2}), ~~~
l\leq n.
\end   {equation} 

Let us compare this set of equations with the usual Schwinger-Dyson
equations and with the planar Schwinger-Dyson equations.
The usual Schwinger-Dyson equations for scalar field
with $\varphi ^4$ interaction have the form
$$
(-\bigtriangleup  +m^{2})_{x_{n}}G_{n}(x_{n},...x_{1})=
-gG_{n+2}(x_{n},x_{n},x_{n},x_{n-1},...,x_{1})
+$$
\begin {equation} 
							  \label {usual}
\sum_{i=1}^{n-1}\delta (x_{n}-x_{i})G_{n-2}(x_{n-1},...x_{i-1},
x_{i+1},...,x_{1})
\end   {equation} 
and $G_{n}(x_{n},...x_{1})$ is a symmetric function of its arguments.

To write down  the planar Schwinger-Dyson equations one has to modify in
(\ref{usual}) only the Schwinger's terms:
$$
(-\bigtriangleup  +m^{2})_{x_{n}}\Pi_{n}(x_{n},x_{n},x_{n},...,x_{1})=
-g\Pi_{n+2}(x_{n},x_{n},x_{n},x_{n-1},...,x_{1})+
$$
\begin {equation} 
							  \label {pl}
+\sum _{i=1}^{n-1}\delta (x_{n}-x_{i})
\Pi_{i-1}(x_{i-1},...,x_{1})
\Pi_{n-i-1}(x_{n-1},...,x_{i+1}),
\end   {equation} 
where
\begin {equation} 
							  \label {G}
\Pi_{n}(x_{n},...x_{1})=
\lim _{N\to\infty} \frac{1}{N^{1+n/2}}<\tr (M(x_{n})...M(x_{1}))>
\end   {equation} 
and $\Pi_{n}(x_{n},...x_{1})$ is
invariant under cyclic permutations of its arguments.
The Schwinger terms for the
Boltzmannian Schwinger-Dyson equations  differ from the Schwinger 
terms for the planar Schwinger-Dyson equations.  From (\ref {ebf}) we 
see that the Schwinger terms are linear with respect to the 
interacting  Boltzmann correlation functions meanwhile the Schwinger 
terms in the planar equations are quadratic with respect to the 
planar correlation functions.  There is also a modification in the 
term representing the interaction.  Instead of the linear term for 
full or planar equations we have quadratic terms in the Boltzmannian 
Schwinger-Dyson equations.  The usual  Schwinger-Dyson  equations may 
be written in the operator form as follows 
\begin {equation} 
\label {SD}
\frac{\delta (S_{0}+S_{int})[\varphi ]}{\delta \varphi (x)}+
~{\rm Schwinger's ~terms}=0.
\end   {equation} 

The planar Schwinger-Dyson equations in the operator form
  can be presented in  terms of
the master field $\Phi $ and the master momentum $\Pi $ \cite
{Haan,GG,Doug1,Doug2}
\begin {equation}
\label {omf}
[i \frac{\delta
(S_{0}+S_{int})[\Phi ]}{\delta \Phi (x)}+2\Pi (x)]|0\rangle =0,
\end   {equation}
where
$S$ is an action and  $\Phi $ and $\Pi$ satisfy \cite {Haan,HalSc}
\begin{equation}
\label {pcr}
[\Pi (x),\Phi (y)] =i\delta ^{(D)}(x-y)|0\rangle
\langle 0|.
\end{equation}

The Boltzmannian Schwinger-Dyson equations can be written down
schematically  as
$$\frac{\delta S_{0}[\Phi ]}{\delta \Phi (x)}+
\frac{\delta S_{int}[\Phi ]}{\delta \Phi (x)}|\Omega\rangle \langle 0|+
\frac{\delta ^{2}S_{int}[\Phi ]}{\delta \Phi (x)^{2}}
|\Omega\rangle \langle 0|\Phi (x)+...+
$$
\begin {equation} 
							  \label {rsd}
\Phi (x)|\Omega\rangle \langle 0|\frac{\delta ^{2}S_{int}[\Phi ]}
{\delta \Phi (x)^{2}}+
|\Omega\rangle \langle 0|\frac{\delta S_{int}[\Phi ]}{\delta \Phi (x)}+
~{\rm Schwinger's}~ {\rm terms} =0.
\end   {equation} 
We see that the Boltzmannian Schwinger-Dyson equations in the operator form
are more complicated than the usual Schwinger-Dyson equations since
the interaction terms contain projectors on the vacuum state.
But as we will see in the next subsection
just due to these projectors it will be possible to get a closed set of
integral equations for two- and four-point Green's functions.

\subsection{A Closed System of Equations for Two- and Four-Point
Green's Functions}

Let us write down  equations (\ref{ebf}) and (\ref{ebf'}) for
the two-point correlation function
 \begin {equation} 
							  \label {2p1}
(-\bigtriangleup +m^{2})_{x}F_{2}(x,y)=
-gF_{2}(x,y) F_{2}(x,x) -gF_{4}(x,x,x,y)+\delta (x-y) ,
\end   {equation} 

\begin {equation} 
							  \label {2p2}
(-\bigtriangleup +m^{2})_{y}F_{2}(x,y)=
-gF_{2}(y,y) F_{2}(x,y) -gF_{4}(x,y,y,y)+\delta (x-y),
\end   {equation} 
and the four-point correlation function
$$(-\bigtriangleup +m^{2})_{x}F_{4}(x,y,z,t)=
-gF_{4}(x,y,z,t) F_{2}(x,x) -gF_{6}(x,x,x,y,z,t)+$$
\begin {equation} 
							  \label {4p1}
\delta (x-y)F_{2}(z,t) +\delta (x-t)F^{(0)}_{2}(y,z),
\end   {equation} 

$$(-\bigtriangleup +m^{2})_{y}F_{4}(x,y,z,t)=
-gF_{2}(z,t)F_{4}(x,y,y,y) -gF_{4}(y,y,z,t)F_{2}(x,y)+$$
\begin {equation} 
							  \label {4p2}
\delta (x-y)F_{2}(z,t) +\delta (y-z)F_{2}(x,t),
\end   {equation} 

$$(-\bigtriangleup +m^{2})_{z}F_{4}(x,y,z,t)=
-gF_{2}(z,t)F_{4}(x,y,z,z)  -gF_{4}(z,z,z,t)F_{2}(x,y)+$$
\begin {equation} 
							  \label {4p3}
\delta (z-t)F_{2}(x,y) +\delta (y-z)F_{2}(x,t),
\end   {equation} 

$$(-\bigtriangleup +m^{2})_{t}F_{4}(x,y,z,t)=
-gF_{6}(x,y,z,t,t,t)-gF_{2}(t,t)F_{4}(x,y,z,t) )+$$
\begin {equation} 
							  \label {4p4}
\delta (z-t)F_{2}(x,y) +\delta (x-t)F^{(0)}_{2}(y,z).
\end   {equation} 
 Here we assume that all vacuum insertions are dropped out.
Note that not all these equations are  independent. 
Indeed,
having in mind the symmetry 
of half-plane Green's functions under
the permutation of their arguments in the inverse order
(\ref{z3a}) one can reproduce  equations
(\ref{2p2}), (\ref{4p2}) and
(\ref{4p4}) from  equations (\ref{2p1}),
(\ref{4p3}) and \ref{4p1}), respectively.

We see that equations
(\ref{4p3}) and (\ref{4p2}) in accordance with the general
formula (\ref {rsd}) do not
contain six-point correlation functions. Let us show
that from (\ref{4p2}) and (\ref{2p1})
one can get a closed set of  integral equations which is
sufficient to find
$F_{2}$ and $F_{4}$. To this purpose let us rewrite (\ref{4p2})
 in the integral form
$$F_{4}(x,y,z,t)=
-g\int D(y,u)[F_{2}(z,t)F_{4}(x,u,u,u)  +F_{4}(u,u,z,t)
F_{2}(x,u)]du+$$
\begin {equation} 
							  \label {4pi3}
D(x,y)F_{2}(z,t) +D(y,z)F_{2}(x,t)
\end   {equation} 
(see Fig. 10). The integral representation for the Schwinger-Dyson 
equation (\ref{2p1}) is drawn on Fig. 9. As has been mentioned above 
the Boltzmann correlation functions do not possess any symmetry 
property except the invariance (\ref{z3a}).  Therefore 
 drawing correlations functions one has to mark the order of the 
external lines. The possibility to do this is to draw arguments of 
 the correlation function on a line as it is done on Figs. 9 and 10. 

 For the two-point correlation function we have
\begin {equation}
\label {2pi1}
F_{2}(x,y)=
-g\int [D(x,u)F_{2}(u,u)F_{2}(u,y)+
D(x,u)F_{4}(u,u,u,y)]du+D(x,y),
\end   {equation}
(see Fig.9).

  From (\ref {4pi3}) and (\ref {2pi1}) one gets the  relation between
${\cal F}_{4}(x,y,z,t) $, the connected
part of $F_{4}(x,y,z,t)$, and $F_{2}(x,y)$.
To write down this relation note that  
\begin {equation} 
							  \label {42r}
F_{4}(x,y,z,t)={\cal F}_{4}(x,y,z,t)+ F_{2}(x,y)F_{2}(z,t)+
F_{2}(x,t)D(y,z),
\end   {equation} 
(see Fig.11).
Finally we get
\begin {equation} 
							  \label {42i}
{\cal F}_{4}(x,y,z,t)=
-g\int F_{2}(x,u)D(y,u)[D(u,z)F_{2}(u,t)  +
{\cal F}_{4}(u,u,z,t)]du
\end   {equation} 

\begin {equation} 
							  \label {24i}
F_{2}(x,y)= D(x,y)
-g\int D(x,u)[ D(u,u)F_{2}(u,y) +2F_{2}(u,u)F_{2}(u,y)+
{\cal F}_{4}(u,u,u,y)]du.
\end   {equation} 

Let us write down this system in a more suitable form.
Note that there are tadpole contributions in the right hand side of equation
(\ref {24i}). It is convenient to include these terms
into a mass renormalization.
To this purpose let us substitute the representation
(\ref{42r}) in (\ref{2p2}). We have
\begin {equation} 
							  \label {2p2r}
(-\bigtriangleup +m^{2})_{y}F_{2}(x,y)=
-g(2c_{1}+c_{0}) F_{2}(x,y) -g{\cal F}_{4}(x,y,y,y)+\delta (x-y),
\end   {equation} 
where $c_{1}=F_{2}(x,x)$, $c_{0}=D(x,x)$.
Including these constants in a new mass $M$ we get
\begin {equation} 
\label {b2}
F_2(x,y)=
D_{M}(x-y)
-g\int du{\cal F}_4(x,u,u,u)D_{M}(u,y).
\end   {equation} 

Instead of equation (\ref {42i}) it is more convenient to deal with an
equation for a {\it partially} one-particle irreducible  (1PI) 4-point 
function $\Gamma_{4}(x,y,z,t)$ which is defined as 
\begin {equation} 
\Gamma_{4}=F^{-1}_{2}D^{-1}D^{-1}F^{-1}_{2}{\cal F}_{4},
\end   {equation} 
or more precisely
\begin {equation} 
\label {ppp}
\Gamma_{4}(x,y,z,t)=\int F^{-1}_{2}(x,x^{\prime})D^{-1}(y,y^{\prime})
D^{-1}(z,z^{\prime}))F^{-1}_{2}(t,t^{\prime}){\cal F}_{4}(x^{\prime},
y^{\prime},z^{\prime},t^{\prime})dx^{\prime}dy^{\prime}dz^{\prime}
dt^{\prime}.
\end   {equation} 
Note that in the right hand side of
(\ref {ppp}) we multiply ${\cal F}_{4}$ only by two full 2-point
Green's functions while in the usual case to get 1PI Green's 
function one
multiplies  an $n$-point
Green's function by $n$ full 2-point functions. 
From (\ref {42i}) and
(\ref {ppp}) we get
\begin {equation} 
\label {bsl}
\Gamma_{4}(p,k,r)= -g-g\int dk^{\prime}
F_{2}(p+k-k^{\prime})D(k^{\prime})\Gamma_{4}(p+k-k^{\prime},
k^{\prime},r).
\end   {equation} 
Equation  (\ref{b2}) can be also rewritten in terms of
$\Gamma _4$
\begin {equation} 
\label {bsl1}
F_2(p)= D{p}-g\int 
dk^{\prime}
dk^{\prime \prime}
D(k^{\prime})
D(k^{\prime \prime})
F_{2}(p
-k^{\prime}
-k^{\prime \prime})
\Gamma_{4}(p
-k^{\prime}
-k^{\prime \prime},
k^{\prime },
k^{\prime \prime}).
\end   {equation} 
Equations (\ref{bsl}) and (\ref{bsl1}) are presented graphically on 
Fig. 12.

Equation (\ref {bsl}) is   the Bethe-Salpeter-like equation
with the kernel which contains an unknown  function $F_{2}$.
As in the usual case we can write down $F_{2}$ in term of self-energy
$\Sigma _{2}$,
\begin {equation} 
			 \label {si}
F_{2}=\frac{1}{p^{2}+M^{2}+\Sigma _{2}}
\end   {equation} 
and write down equation (\ref {b2}) as an equation for $\Sigma _{2}$,
i.e.
\begin {equation} 
							  \label {si4}
\Sigma _{2} (p)=\int dkdq F_{2}(k)D(q)D(p-k-q)\Gamma _{4}(p,k,q).
\end   {equation} 
Equation (\ref {si4}) is similar to a usual
relation between the self-energy function $\Sigma _{2}$ and 4-point vertex
function for $\varphi ^{4}$ field theory. Note that
a closed form of the representation for
4-point vertex function (\ref {bsl}) is a distinctive feature of the
Boltzmann field theory.

One can choose an approximation in the system of equations (\ref {si4}),
(\ref {bsl}) which gives a so-called rainbow approximation.
To get this approximation
one  takes $\Gamma _{4}\approx -g$ and restricts
by the first two terms in the relation (\ref {si}),
\begin {equation} 
							  \label {apsi}
F ^{appr}_{2}=\frac{1}{p^{2}+M^{2}}-
\frac{1}{p^{2}+M^{2}}\Sigma _{2}\frac{1}{p^{2}+M^{2}}
\end   {equation} 
Within this approximation
the solution $\Sigma ^{appr}_{2}\equiv \Pi $  of equations (\ref {si4})
is reduced to the solution of the following  equation for $\Pi $
\begin {equation} 
							  \label {insi}
\Pi (q)=\Pi _{0}(q)+g^{2}\int \frac{d^{4}q'}{(2\pi)^{4}}
\frac{I(q-q')}{(q^{2}+M^{2})^{2}}\Pi (q'),
\end   {equation} 
where
$\Pi _{0}(q)$
denotes the  second order contribution to the self-energy
\begin {equation} 
							  \label {up}
\Pi _{0}(q)=g^{2}\int \frac{d^{4}k_{1}}{(2\pi)^{4}}
\int \frac{d^{4}k_{2}}{(2\pi)^{4}}
\frac{1}{(k_{1}^{2}+m^{2})(k_{2}^{2}+m^{2})
((q-k_{1}+k_{2})^{2}+M^{2})},
\end   {equation} 
and $I(q)$ is a "potential"
\begin {equation} 
							   \label {pot}
I(p)= \int \frac{d^{4}q}{(2\pi)^{4}}
\frac{1}{(q^{2}+m^{2})((p-q)^{2}+m^{2})}.
\end   {equation} 
Equation (\ref {insi}) is known as the rainbow equation
and it was studied by Rothe \cite{Rothe}.
The integrals (\ref {pot}) and (\ref {up}) contain divergences.
Performing a subtraction one gets the following expression for
the renormalized "potential"  (for simplicity we assume that $m=0$)
\begin {equation} 
							  \label {rI}
I(p)=\frac{1}{(4\pi)^{2}}\ln (\frac{p^{2}}{\kappa ^{2}})
\end   {equation} 
with $\kappa$ being an arbitrary subtraction point.
The renormalized second order self-energy   has the form
(eq. (A.9) from \cite{Rothe})
\begin {equation} 
							  \label {rp}
\Pi _{0}(q) =\frac{M^{2}g^{2}}{2(4\pi)^{2}}
\{A+Bq^{2}+(\frac{M^{2}}{q^{2}}-\frac{q^{2}}{M^{2}})
\ln (1-\frac{q^{2}}{M^{2}})+2\int ^{1}_{0}\frac{d\chi}{\chi}
\ln (1-\chi\frac{q^{2}}{M^{2}}) \},
\end   {equation} 
where $A$ and $B$ are arbitrary renormalization constants.

For the case of exact equations
(\ref {bsl}), (\ref {si}) and (\ref {si4}) one also has to perform appropriate
renormalizations. There are two possibilities to discuss renormalizations.
One can do an appropriate  subtractions directly in the integral
equations or one can do them order by order in the perturbative expansion.
Both ways are equivalent but the second one is more closed to
renormalizations of usual field theories.
We are going to discuss it in the next
subsection.

\subsection{Renormalization of the Boltzmann Field Theory}

In this subsection we will describe the renormalization of the
Boltzmann field theory.
To make a formal expression for  correlation functions
\begin {equation} 
                                                          \label {5}
 F_{n}(x_1,...,x_n)=
\langle 0|\phi (x_1)...\phi (x_n) \frac{1}{1+\int d^{D}xL_{int}(\phi(x))}
|0\rangle ,
\end   {equation} 
finite we apply R-operation \cite{BogS}. In (\ref{5})
$\phi (x)$ has the decomposition (\ref {phi}) with
$\phi ^{+}(x)$ and $\phi ^{-}(x)$  satisfying (\ref {z0}) and 
the interaction Lagrangian is an  invariant polynomial of $\phi ^+$,
$\phi ^-$ and
its derivatives.
The perturbative expansion of Boltzmann correlation functions (\ref {5.0})
is represented by the sum of half-planar  Feynman graphs
$\{ H \}$.
We write $F_{H}$ for the unrenormalized value of graph $H$ 
\begin {equation} 
                                                          \label {5.1}
 F_{H}(p_{1},...p_{n})=\int dk_{1}...dk_{l}
  I_{H}(k_{1},...,k_{l};p_{1},...,p_{n}),
\end   {equation} 
where $p_{1},...,p_{n}$ are external momenta and the integration
is over internal momenta $k_{1},...,k_{l}$.

Let us apply the standard 
$R$-operation  to Boltzmannian Green's functions
(\ref{5}).
If $H$ is not a renormalization part  ($H$ has no overall divergence)
we set
\begin {equation} 
                                                          \label {5.2}
R_{H}=\bar {R}_{H},
\end   {equation} 
and if $H$ is a renormalization part ($H$  is 1PI and has
an  overall divergence)
\begin {equation} 
                                                          \label {5.3}
R_{H}=(1-T_{H})\bar {R}_{H},
\end   {equation} 
where $T_{H}$ is a subtracting operation which fixes the renormalization
scheme. The function $\bar {R}_{H}$ is defined recursively by
\begin {equation} 
                                                          \label {5.4}
\bar {R}_{H}=I_{H}+\sum _{\{\gamma _{1},...,\gamma _{s}\}}
I_{H/\{\gamma _{1},...,\gamma _{s}\}}
\prod _{i=1}^{s}(-T_{\gamma_{i}}\bar {R}_{\gamma_{i}}).
\end   {equation} 
The sum in (\ref{5.4}) is extends over all sets $\{\gamma _{1},...,
\gamma _{s}\}$ of renormalization
parts of $H$ which are mutually disjoint and different from $H$.
The reduced diagram $H/\{\gamma _{1},...,\gamma _{s}\}$ is defined
by contracting
each $\gamma _{i}$ to a point.
The explicit formula for the
renormalized  integrand  is given by the forest formula,
\begin {equation} 
                                                          \label {5.5}
R_{H}=(1+\sum _{U} \prod _{\gamma \in U}(-T_{\gamma}))I_{H}
\end   {equation} 
with the sum going over all $H$-forests.
Up to this point a specific character of the Boltzmann field theory
has not been taken into account.

To prove the renormalizability for some given ${\cal L}_{int}$ we 
have to prove that all subtractions can be collected in corresponding 
counterterms.  At this point we meet with a specific character of the 
Boltzmann field theory.  There is the following subtlety   
in this step.  Half-planar graphs  contain  
renormalization parts which are half-planar as well as 
renormalization parts which are not belong to  the set of half-planar 
graphs.  With an half-planar renormalization part we have to 
associate the counterterm 
\begin{equation} 
\label{re6a} 
{\cal 
L}_{n}(H_n)=(-T_{H_n}\bar{R}_{H_n})\phi ^n .
\end{equation} 
Note that 
there is no the combinatoric factor.  We will see that with  
non-half-planar renormalization parts we have to associate 
counterterms which are not  functions of $\phi $ but are functions of 
$\phi ^{\pm}$.

To specify our discussion let us consider
correlation functions (\ref{5.0}) for the $\phi ^4$ interaction in $D=4$.
To avoid  problems with tadpoles we consider the two-field formulation
\begin {equation} 
                                                          \label {2-f}
 F_{n}(x_1,...,x_n)=
\langle 0|\psi (x_1) \phi(x_2) \phi(x_3)
...\phi (x_{n-1}) \psi (x_n) \frac{1}{1+g\int
d^{D}x
\psi  :\phi \phi :
\psi }
|0\rangle ,
\end   {equation} 
where
$$\psi (x)=\psi ^-(x) + \psi ^+ (x),~~
\phi (x)=\phi ^-(x) + \phi ^+ (x),$$
$$\psi ^-(x)
\psi ^+(y)= D_M(x,y),~~
\phi ^-(x)
\phi ^+(y)= D_m(x,y),$$
$$
\psi ^-(x)\phi ^+(y)=
\phi ^-(x)\psi ^+(y)=0,
$$
\begin {equation} 
D_{{\cal M}}(x,y) =
\int \frac{d^{D}k}{(2 \pi)^D}\frac{1}{p^2+
{\cal M}^2}e ^{i k (x-y)},~~~
{\cal M}=(M,m).
                                                         \label {re104'}
\end   {equation} 
We draw $\phi $-propagator by thin lines and $\psi $-propagator by
thick lines. Only two- and four-point graphs may be divergent.
Any two-point subgraph
of the half-planar graph is also from the set of half-planar
graphs and performing contractions of the two-point subgraphs
of the half-planar graph one gets again a half-planar graph.
But four-point subgraphs of a half-planar graph may be of half-planar type
and may be not.
We refer to the later case as to the case of subgraphs of $\Pi $-type.
Particular examples of divergent
parts of $\Pi $-type of the two-point half-planar
graph  as well as an
 examples of    half-planar divergent subgraphs are presented on 
 Fig. 13.

At the first sight a  four-point
divergent part of $\Pi $-type requires a counterterms  $\psi ^4$.
But a careful examination of half-planar
diagram shows that in fact only
the counterterms
\begin {equation} 
                                                     \label {counter}
:\psi \psi :  \psi ^-\psi ^-
\end   {equation} 
are necessary. Indeed, let us consider a graph with a four point
divergent part of $\Pi $-type. 
There are two possibilities to make
a contraction of this graph to a point, namely, making a contraction to the
left  vertex $A$, or making a contraction to the right vertex $B$
(see Fig.14). In the both cases
after the contractions we are left with half-planar graphs with the insertions
\begin {equation} 
                                                          \label {psict}
:\psi \phi :  \psi ^-\psi ^-
\end   {equation} 
and
\begin {equation} 
                                                          \label {pcict+}
\psi ^+ \phi ^+ : \psi \psi : ,
\end   {equation} 
respectively. One has to choose one of these possibilities.
Let us take the first one.

The above consideration shows that
reduced graphs $H/\{\gamma_{1},...\gamma_{s}\}$
are always from the set of half-planar graphs.
This fact permits to collect all divergent parts to counterterms
\begin {equation} 
                                                          \label {5.01}
{\cal L}_{ct}(\psi^+, \psi^-, \phi )
=\sum _{H}{\cal L}_n(H_n),
\end   {equation} 
where the summation is over all renormalization parts. Therefore
\begin {equation} 
                                                          \label {5.01a}
{\cal L}_{ct}(\psi^+, \psi^-, \phi )
=( Z_{\psi}-1)(\partial \psi _{\mu})^2
+\delta M \psi ^2+
\delta g
 \psi :\phi \phi :  \psi +
 \delta \lambda
 :\psi \psi: \psi ^- \psi ^-  .
\end   {equation} 

We add  counterterms (\ref {5.01a})
to the interaction Lagrangian and consider the
renormalized correlation functions
$$ F_{n}(x_1,...,x_n)=
$$
\begin {equation} 
                                                          \label {5.00}
\langle 0|\psi (x_1)\phi (x_2)...\phi (x_{n-1})\psi (x_n)
 \frac{1}{1+\int d^{D}x[g
 \psi :\phi \phi :  \psi
+{\cal L}_{ct}(\psi^+, \psi^-, \phi )]}
|0\rangle .
\end   {equation} 

It is easy to see that the counterterms remove all divergencies in the graphs
corresponding to (\ref {5.00}).
Indeed, let
us consider some half-planar graph
$H_n$.
If $n > 4$ then $H_n$ is  1P reducible and it does not
require any subtraction.
If $H_2$  and  $H_4$ are overall
divergent graphs then these divergences
are subtracted by the graph created by the counterterm vertices
$ (Z_{\psi}-1)(\partial \psi _{\mu})^2$,
$\delta M \psi ^2$ and
$\delta g
 \psi :\phi \phi :  \psi $
 respectively.
Now let us supposed that $H _n$  has overall divergent  1PI subgraphs
$\gamma _1,...,\gamma _k$.
Since the graphs $\gamma_1,...,\gamma_k$ are overall divergent  there are
corresponding counterterms in  (\ref {5.01}).
Divergent subgraphs may be halp-planar ones,
that corresponds to the first three terms in
(\ref{5.01a}), or $\Pi $-type graphs, that corresponds to the last 
counterterm (\ref{5.01a}).  These counterterms produce the graphs in 
which one or more of overall divergent subgraphs are replaced by 
their counterterms and there are the corresponding terms in the 
forest formula.  The counterterms  (\ref {5.01a}) do not generated 
new divergences which are not cancelled with divergences of graphs 
generated by the basic Lagrangian since there is one to one 
correspondence between the terms of 
the forest formula and the terms generated
by the counterterms (\ref {5.01a}).

The distinctive feature of the Boltzmann theory
is that  there are the  wave function counterterms
$(\delta Z -1)(\partial \psi )^2$
and the mass counterterms  $\delta m ^2 \psi ^2$
only for $\psi $-field (see Fig. 15).

To associate  counterterm
(\ref {psict}) with a change of
parameter of the theory it is convenient to add the
corresponding term to the interaction Lagrangian,
\begin{equation}
                                  \label{re9}
{\cal L}_{int}(\psi ^+, \psi ^-, \phi)=
 g\psi :\phi \phi :  \psi+
\lambda :\psi \psi :  \psi ^- \psi ^-.
\end{equation}
Therefore the theory (\ref {re9}) is specifyed by  four parameters,
two masses $M$ and  $m$ and two coupling constants $g$ and $\lambda $.

Let us for completeness write down the Boltzmann Schwinger-Dyson equations
for the interaction (\ref{re9})
$$
(-\bigtriangleup +M^{2})_{x}F_{2}(x,y)=
gF_{2}(x,y) D_M(x,x) -gF_{4}(x,x,x,y)-
$$
 \begin {equation} 
                                                          \label {r2p}
\lambda F_2(x,y)F_2(x,x)+\lambda F_2(x,y)D_M(x,x)
+\delta (x-y) ,
\end   {equation} 
\begin {equation} 
                                                          \label {r4p}
(-\bigtriangleup +m^{2})_{y}F_{4}(x,y,z,t)=
-gF_{2}(x,y)F_{4}(y,y,z,t)
+\delta (y-z)F_{2}(x,t).
\end   {equation} 
We see that equations (\ref{r4p}) is just
equation (\ref{42i}) for the connected four-point Green's 
function.
 However equation (\ref{r2p})
for the two-point function
contains  new terms. From these equations it follows that equations
(\ref{bsl}) and (\ref{si4}) are not changed and instead of
(\ref{si}) we have
\begin {equation} 
\label{si4r}
F_2=\frac{1}{p^2+M_{\lambda}^2+\Sigma _2},
\end   {equation} 
where
\begin {equation} 
\label{si4rr}
M_{\lambda}^2=M^2+\lambda (F_2(0)-D_M(0)).
\end   {equation} 

Let us assume dimensional regularization and the minimal subtraction scheme
with a scale $\mu$. We have
\begin{equation}
\label{re9'}
{\cal L}_{int}= \mu ^{4-D}g\psi :\phi \phi : \psi +
\mu ^{4-D} \lambda :\psi \psi :  \psi ^- \psi ^-,
\end{equation}
\begin{equation}
\label{ct10}
{\cal L}_{ct}=
(Z_{\psi}-1)
(\partial  _{\mu} \psi )^{2}
+\delta M^2\psi ^2
+\mu ^{4-D}\delta g\psi :\phi \phi :  \psi+
\mu ^{4-D}\delta \lambda :\psi \psi:  \psi ^- \psi ^-,
\end{equation}
\begin {equation} 
\label {5.0}
 F_{n}(x_1,...,x_n;m,M,g,\lambda,\mu)=
\langle 0|\psi (x_1)\phi (x_2)...
\phi (x_{n-1})\psi (x_n)
 (1+\int d^{D}x ({\cal L}_{int} +{\cal L}_{ct}))^{-1}
|0\rangle .
\end{equation}
Now we can include the wave function renormalization $Z_{\psi} -1 $
in the rescaling of the field $\psi$  and
the renormalized correlations functions up to the renormalization factor
are equal to unrenormalized
correlation functions in the theory with bare  parameters,
$$
F_n(x_1,...,x_n;  m, M,g,\lambda ,\mu)=
Z_{\psi}^{-1}F_{n}^{unren}(x_1,...,x_n;m,M_0,g_0,\lambda _0)=
$$
\begin {equation} 
                                                          \label {re105}
Z_{\psi}^{-1}\langle 0|\psi (x_1)\phi (x_2)...\phi (x_{n-1})\psi (x_n)
\frac{1}{1+\int d^{D}x
{\cal L}_{int}(m, M_0,g_0,\lambda _0)}
|0\rangle ,
\end   {equation} 
where
\begin{equation}
\label{re106}
M_0=Z_{\psi}^{-1} (M^{2}+\delta M^{2} ),~~
g_0=Z_{\psi}^{-1} \mu ^{4-D}(g+\delta g),~~
\lambda_0=Z_{\psi}^{-2} \mu ^{4-D}(\lambda+\delta \lambda).
\end{equation}

It is evident that
the renormalized correlations functions
being multiplied on $Z_{\psi} $ do not depend on $\mu$
since $F_{n}^{unren}$  depends
only on bare parameters $M_0,$, $g_0$ and $\lambda _0$
which do not depend on the scale parameter
specifying the renormalization
prescription. This implies  the following
renormalization group equation
\begin{equation}
\label{5.23}
(\mu \frac{\partial}{\partial \mu}+
\beta _{g}\frac{\partial}{\partial g}+
\beta _{\lambda}\frac{\partial}{\partial \lambda}-
\gamma _M  \frac{\partial}{\partial \ln M^2}+
 \gamma )
F_n(x_1,...,x_n;m,M,g,\lambda ,\mu)=
0,
\end{equation}
where
\begin{equation}
\beta _{g}= \mu \frac{\partial g(\mu )}{\partial \mu},~~
\beta _{\lambda}= \mu \frac{\partial \lambda (\mu )}{\partial \mu},~~
\gamma _M = -\frac{\mu}{M^2} \frac{\partial M^2(\mu )}{\partial 
\mu},~~ 
\gamma = \mu \frac{\partial \ln Z_{\psi }(\mu )}{\partial \mu}.
\label{re15}
\end{equation}

The difference between  (\ref{5.23}) and the usual renormalization group
equation is that the anomalous dimension
$\gamma $ in (\ref{5.23}) is not multiplied on $n/2$.
There is also a difference in the expression of the $\beta $ and
$\gamma$ functions in terms of the counterterms
$\delta c$ and $\delta g$. 
In dimensional regularization the counterterms are
poles in $(D-4)$
\begin{equation}
\label{5.24}
Z_{\psi}=1+\sum_{i=1}^{\infty}\frac{c_i(g, \lambda)}{(4-D)^i},
~~~\delta g=\sum_{i=1}^{\infty}\frac{a_i(g, \lambda)}{(4-D)^i}
~~~\delta\lambda=\sum_{i=1}^{\infty}\frac{b_i(g,\lambda)}{(4-D)^i}.
\end   {equation}

As in the usual case one can get the low-order expression
for the $\beta$-function in terms of $c_{1}$ and $a_{1}$.
The difference with the usual case comes from the fact that now there is a
new expression for $g_{0}$ in terms of $\delta g$  and $Z_{\psi} $.
We have
\begin {equation} 
                                                          \label {5.25'}
g_0=\mu ^{4-D}(g+\delta g)Z_{\psi} ^{-1},~~~
\lambda_0=\mu ^{4-D}(\lambda+\delta \lambda)Z_{\psi} ^{-2}.
\end   {equation} 
Substituting (\ref {5.24}) in (\ref {5.25'}) gives
$$
g_0=\mu ^{4-D}[g+\frac{a_1(g, \lambda)-gc_1(g, \lambda)}{4-D}~+~~{\rm
poles~ of~ order~ more~ them~ 2}],
$$
\begin{equation}
\label{5.25}
\lambda_0=\mu ^{4-D}[\lambda+\frac{b_1(g,\lambda)-\lambda
c_1(g, \lambda)}{4-D}~+~~{\rm
poles~ of~ order~ more~ them~ 2}].
\end   {equation}
Applying $\mu \frac{d}{d\mu}$ to both sides of (\ref {5.25}) yields
$$
\beta _g(g)=(1-\frac{\partial}{\partial \ln g})(g c_1(g, \lambda)-
a_1(g, \lambda)),
$$
\begin{equation}
\label{5.26}
\beta _{\lambda}(\lambda)=(1-\frac{\partial}{\partial \ln \lambda})
(2\lambda c_1(g,\lambda)-b_1(g,\lambda)).
\end   {equation}
Taking into account the explicit form of $c_1(g,\lambda)$, $a_1(g, \lambda)$
and $b_1(g,\lambda )$
we get
\begin {equation} 
                                                          \label {ren.f}
 \beta _g(g, \lambda)=
\frac{g^2}{8 \pi ^2}+
 \frac{g^3}{(16\pi ^2)^2}+O( g^4, \lambda ^4),~~~
\beta _{\lambda}(g, \lambda)=
\frac{\lambda ^2}{8 \pi ^2}+
O( g^4, \lambda ^4).
\end   {equation} 

We see that
the sign of the half-planar beta function
$\beta _{g}$ is the same as the sign
of the standard beta function in $\varphi ^4$ theory
but there is  a  numerical difference of the half-planar
beta function with the standard beta  function.

A non-standard  wave function renormalization in the Boltzmann 
theory (\ref {5.1})  
implies that a dependence of the renormalized  correlation
function on the scale parameter $\mu$ is given by the formula
$$
F_n(x_1,...,x_n;m,M(\mu),g(\mu),\lambda (\mu),\mu)=
\exp [-\int _{\mu ^{\prime}}^{\mu}
\frac{d \mu ^{\prime \prime}}{\mu ^{\prime\prime}}
\gamma (g(\mu ^{\prime\prime}), \lambda (\mu^{\prime \prime}))]
\times
$$
\begin{equation}
\label{5.27}
F_n(x_1,...,x_n;m,M(\mu ^{\prime \prime}),
g(\mu ^{\prime \prime}),
\lambda(\mu ^{\prime \prime}),
\mu ^{\prime \prime}).
\end{equation}

Since due to a special form
of $\lambda$-interaction  $\lambda$-vertices contribute only in the mass
renormalization of $\psi $-lines and it is natural to expect that the
dependence on $\lambda$  in the 
high-energy asymptotic of correlation functions
 may be neglected together with dependence on $M(\mu)$  and one get
 the following asymptotic formula
\begin {equation} 
                                                          \label {as}
F_n(\kappa x_1,...,\kappa x_n;m,M(\mu),g(\mu),\lambda (\mu),\mu)
\sim 
\kappa ^{d}\xi ^{-1}(\kappa \mu  ) F_n(x_1,...,x_n;0,0,g(\mu),0,\mu),
\end   {equation} 
for  ${\kappa \to 0}$.
Here $d$ is a dimension of correlation function and
\begin {equation} 
                                                          \label {xi}
\xi ^{-1}(\kappa \mu  )=
\exp [-\int _{\mu ^{\prime}}^{\mu}
\frac{d \mu ^{\prime \prime}}{\mu ^{\prime\prime}}
\gamma (g(\mu ^{\prime\prime}),0) ].
\end   {equation} 


$$~$$
{\bf ACKNOWLEDGMENT}

\vspace{5mm}

Both authors are supported by RFFR
grant 96-01-00608.
We are grate\-ful to
G.E.Aru\-tyu\-nov, P.B.Med\-ve\-dev and 
I.V.Vo\-lo\-vich for use\-ful dis\-cussions.
$$~$$

\vspace{10mm}

{\large \bf APPENDIX }

\appendix

\section{A Note on the Combinatoric of Planar Graphs
of a  Matrix Model}

In this Appendix we  prove the following statement.
The planar correlation functions without
vacuum insertions in all orders of perturbation theory are represented by
the sum of all topologically non-equivalent graphs without
any combinatiric factors.

The proof of this statement follows from the following considerations.
Let us consider an arbitrary planar graph with
$n$ external lines
and with $m$ four-point vertices.
We will consider the external lines
of graph corresponding to global invariant Green
functions as the lines corresponding to an $n$-point generalized 
vertex.  We suppose that $k_1$ from $m$ four-point vertices are 
connected with the generalized vertex at least by one line (see Fig. 
16, where  all double lines are drawn by single lines for 
simplicity).

The combinatoric factor associated with all possible contractions
of  $k_1$  vertices with the $n$-point generalized vertex
is equal to
$$
\frac{m!}{(m-k_1)!}
4^{k_1}.
$$
Indeed, to produce topologically equivalent graphs
$k_1$ vertices can be chosen from $m$ vertices by
$\frac{m!}{(m-k_1)!k_1!}$ ways,
$k_1$ vertices can be permutated by $k_1!$ ways and each vertex can be
attached to the generalized vertex by 4  ways.

After  connecting  the generalized vertex by lines with $k_1$
vertices we get the graph  which can be considered as
some set of generalized vertices which are connected with $m-k_1$
remaining four-point vertices.
For example, on  Fig. 16 there are at least
two new generalized vertices. They  are
formed by lines outcoming from 4-vertices labeled by 1, 2, 3, 6, ...
and 4, 5.
Let us supposed that
there are $r$ new generalized vertices and that
$k_2$ from $m-k_1$ remaining four-point vertices are connected
with  new generalized vertices at least by one line
so that the first
generalized vertex is connected with $k^{(1)}_2$
4-vertices, the second
generalized vertex is connected with $k^{(2)}_2$
4-vertices and so on.
There are
$$
\frac{(m-k_1)!}{(m-k_1-k_2)!k_2!}
4^{k_2}
\frac{k_2!}{k^{(1)}_2!...k^{(r)}_2!}
k^{(1)}_2!...k^{(r)}_2!
=
\frac{(m-k_1)!}{(m-k_1-k_2)!}4^{k_2}
$$
possibilities to do this.
Again we get the graph that can be considered as  some
set of generalized vertices which are connected in some way
with $m-k_1-k_2$ remaining four-point vertices.

Analyzing the construction of a full graph  by  steps 
that consist in attaching some vertices to generalized
vertices and assuming that it is need to do $l$ steps
to get the graph one gets that the total 
combinatoric factor will be equal to $$ \frac{m!}{(m-k_1)!}4^{k_1} 
\frac{(m-k_1)!}{(m-k_1-k_2)!}4^{k_2}
...
\frac{(m-k_1-k_2-...-k_{l-1})!}{(m-k_1-k_2-...-k_l)!}4^{k_l}=
m!4^m.
$$
This factor is just canceled by the factor $1/(m!4^m)$ appearing from
the expansion of the exponent of the interaction Lagrangian,
that gives the prove of the statement.


\newpage

\begin{figure}
\begin{center}
\unitlength=1.00mm
\special{em:linewidth 0.4pt}
\linethickness{0.4pt}
\begin{picture}(131.00,36.00)
\put(40.00,20.00){\circle{2.00}}
\put(50.00,20.00){\circle{2.00}}
\put(70.00,20.00){\circle{2.00}}
\put(80.00,20.00){\circle{2.00}}
\put(90.00,20.00){\circle{2.00}}
\put(110.00,20.00){\circle{2.00}}
\put(120.00,20.00){\circle{2.00}}
\put(130.00,20.00){\circle{2.00}}
\put(30.00,20.00){\circle{2.00}}
\put(104.00,20.00){\circle*{0.67}}
\put(100.33,20.00){\circle*{0.67}}
\put(96.67,20.00){\circle*{0.67}}
\put(63.67,20.00){\circle*{0.67}}
\put(59.67,20.00){\circle*{0.67}}
\put(55.67,20.00){\circle*{0.67}}
\put(30.00,16.00){\makebox(0,0)[cc]{{\small 1}}}
\put(40.00,16.00){\makebox(0,0)[cc]{{\small 2}}}
\put(50.00,16.00){\makebox(0,0)[cc]{{\small 3}}}
\put(68.67,16.00){\makebox(0,0)[cc]{{\small $2m-1$}}}
\put(80.00,16.00){\makebox(0,0)[cc]{{\small $2m$}}}
\put(91.00,16.00){\makebox(0,0)[cc]{{\small $2m+1$}}}
\put(118.67,16.00){\makebox(0,0)[cc]{{\small $2n-1$}}}
\put(130.00,16.00){\makebox(0,0)[cc]{{\small $2n$}}}
\put(55.00,20.00){\oval(50.00,16.00)[t]}
\put(109.67,10.00){\makebox(0,0)[cc]{$\underbrace{~~~~~~~~~~~~~~~~~~~~~~~~~~~~~~}_{2n-2m}$}}
\put(55.67,10.00){\makebox(0,0)[cc]{$\underbrace{~~~~~~~~~~~~~~~~~~~~~~~}_{2n-2}$}}
\end{picture}
\end{center}
\caption{Derivation of the  Schwinger-Dyson equations
for free Green's functions
in the
Boltzmannian Fock space. }
\label{fig1}
\end{figure}
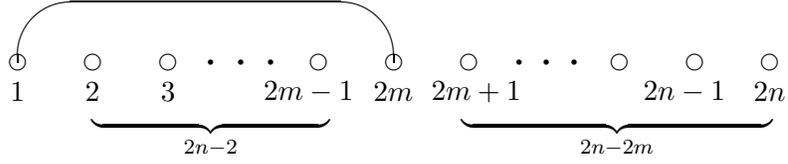


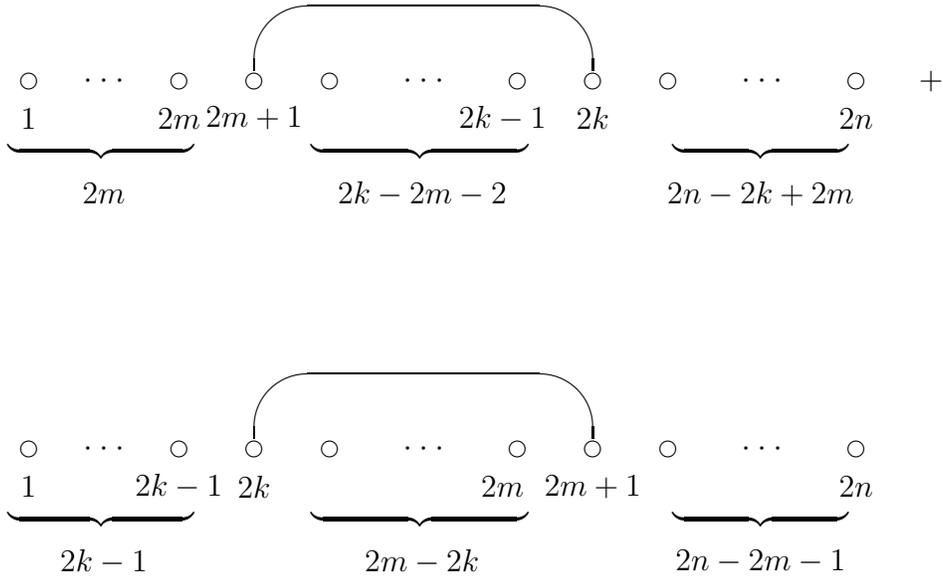
\begin{figure}
\begin{center} 
\unitlength=1.00mm
\special{em:linewidth 0.4pt}
\linethickness{0.4pt}
\begin{picture}(126.00,29.99)
\put(15.00,20.00){\circle{2.00}}
\put(35.00,20.00){\circle{2.00}}
\put(45.00,20.00){\circle{2.00}}
\put(55.00,20.00){\circle{2.00}}
\put(80.00,20.00){\circle{2.00}}
\put(90.00,20.00){\circle{2.00}}
\put(100.00,20.00){\circle{2.00}}
\put(25.00,20.00){\makebox(0,0)[cc]{$\cdot \cdot \cdot$}}
\put(67.50,20.00){\makebox(0,0)[cc]{$\cdot \cdot \cdot$}}
\put(112.50,20.00){\makebox(0,0)[cc]{$\cdot \cdot \cdot$}}
\put(125.00,20.00){\circle{2.00}}
\put(67.50,21.33){\oval(45.00,17.33)[t]}
\put(15.00,15.00){\makebox(0,0)[cc]{$1$}}
\put(35.00,15.00){\makebox(0,0)[cc]{$2m$}}
\put(45.00,15.00){\makebox(0,0)[cc]{$2m+1$}}
\put(78.00,15.00){\makebox(0,0)[cc]{$2k-1$}}
\put(90.00,15.00){\makebox(0,0)[cc]{$2k$}}
\put(125.00,15.00){\makebox(0,0)[cc]{$2n$}}
\put(135.00,20.00){\makebox(0,0)[cc]{$+$}}
\put(67.33,5.00){\makebox(0,0)[cc]{$2k-2m-2$}}
\put(25.00,5.00){\makebox(0,0)[cc]{$2m$}}
\put(112.33,5.00){\makebox(0,0)[cc]{$2n-2k+2m$}}
\put(112.33,10.67){\makebox(0,0)[cc]{$\underbrace{~~~~~~~~~~~~~~~~~}$}}
\put(67.00,10.67){\makebox(0,0)[cc]{$\underbrace{~~~~~~~~~~~~~~~~~~~~~}$}}
\put(24.67,10.67){\makebox(0,0)[cc]{$\underbrace{~~~~~~~~~~~~~~~~~~}$}}
\end{picture}

$$~$$

\unitlength=1.00mm
\special{em:linewidth 0.4pt}
\linethickness{0.4pt}
\begin{picture}(126.00,29.99)
\put(15.00,20.00){\circle{2.00}}
\put(35.00,20.00){\circle{2.00}}      
\put(45.00,20.00){\circle{2.00}}
\put(55.00,20.00){\circle{2.00}}
\put(80.00,20.00){\circle{2.00}}
\put(90.00,20.00){\circle{2.00}}
\put(100.00,20.00){\circle{2.00}}
\put(25.00,20.00){\makebox(0,0)[cc]{$\cdot \cdot \cdot$}}
\put(67.50,20.00){\makebox(0,0)[cc]{$\cdot \cdot \cdot$}}
\put(112.50,20.00){\makebox(0,0)[cc]{$\cdot \cdot \cdot$}}
\put(125.00,20.00){\circle{2.00}}
\put(67.50,21.33){\oval(45.00,17.33)[t]}
\put(15.00,15.00){\makebox(0,0)[cc]{$1$}}
\put(35.00,15.00){\makebox(0,0)[cc]{$2k-1$}}
\put(45.00,15.00){\makebox(0,0)[cc]{$2k$}}
\put(78.00,15.00){\makebox(0,0)[cc]{$2m$}}
\put(90.00,15.00){\makebox(0,0)[cc]{$2m+1$}}
\put(125.00,15.00){\makebox(0,0)[cc]{$2n$}}
\put(67.33,5.00){\makebox(0,0)[cc]{$2m-2k$}}
\put(25.00,5.00){\makebox(0,0)[cc]{$2k-1$}}
\put(112.33,5.00){\makebox(0,0)[cc]{$2n-2m-1$}}
\put(112.33,10.67){\makebox(0,0)[cc]{$\underbrace{~~~~~~~~~~~~~~~~~}$}}
\put(67.00,10.67){\makebox(0,0)[cc]{$\underbrace{~~~~~~~~~~~~~~~~~~~~~}$}}
\put(24.67,10.67){\makebox(0,0)[cc]{$\underbrace{~~~~~~~~~~~~~~~~~~}$}}
\end{picture}
\end{center}
\caption{Derivation of the  Schwinger-Dyson equations
for free Green's functions
in the
Boltzmannian Fock space (the $2k$-th marked operator). }
\label{fig2}
\end{figure}

          

\begin{figure}
\begin{center} 
\unitlength=1.00mm
\special{em:linewidth 0.4pt}
\linethickness{0.4pt}
\begin{picture}(126.00,29.99)
\put(15.00,20.00){\circle{2.00}}
\put(35.00,20.00){\circle{2.00}}
\put(45.00,20.00){\circle{2.00}}
\put(55.00,20.00){\circle{2.00}}
\put(80.00,20.00){\circle{2.00}}
\put(90.00,20.00){\circle{2.00}}
\put(100.00,20.00){\circle{2.00}}
\put(125.00,20.00){\circle{2.00}}
\put(67.50,21.33){\oval(45.00,17.33)[t]}
\put(15.00,15.00){\makebox(0,0)[cc]{$1$}}
\put(32.00,15.00){\makebox(0,0)[cc]{$2m+1$}}
\put(48.00,15.00){\makebox(0,0)[cc]{$2m+2$}}
\put(75.00,15.00){\makebox(0,0)[cc]{$2k$}}
\put(90.00,15.00){\makebox(0,0)[cc]{$2k+1$}}
\put(125.00,15.00){\makebox(0,0)[cc]{$2n$}}
\put(135.00,20.00){\makebox(0,0)[cc]{$+$}}
\put(67.33,5.00){\makebox(0,0)[cc]{$2k-2m-2$}}
\put(25.00,5.00){\makebox(0,0)[cc]{$2m+1$}}
\put(112.33,5.00){\makebox(0,0)[cc]{$2n-2k+2m$}}
\put(112.33,10.67){\makebox(0,0)[cc]{$\underbrace{~~~~~~~~~~~~~~~~~}$}}
\put(67.00,10.67){\makebox(0,0)[cc]{$\underbrace{~~~~~~~~~~~~~~~~~~~~~}$}}
\put(24.67,10.67){\makebox(0,0)[cc]{$\underbrace{~~~~~~~~~~~~~~~~~~}$}}
\put(25.00,20.00){\makebox(0,0)[cc]{$\cdot \cdot \cdot$}}
\put(67.50,20.00){\makebox(0,0)[cc]{$\cdot \cdot \cdot$}}
\put(112.50,20.00){\makebox(0,0)[cc]{$\cdot \cdot \cdot$}}
\end{picture}

$$~$$

\unitlength=1.00mm
\special{em:linewidth 0.4pt}
\linethickness{0.4pt}
\begin{picture}(126.00,29.99)
\put(15.00,20.00){\circle{2.00}}
\put(35.00,20.00){\circle{2.00}}
\put(45.00,20.00){\circle{2.00}}
\put(55.00,20.00){\circle{2.00}}
\put(80.00,20.00){\circle{2.00}}
\put(90.00,20.00){\circle{2.00}}
\put(100.00,20.00){\circle{2.00}}
\put(125.00,20.00){\circle{2.00}}
\put(67.50,21.33){\oval(45.00,17.33)[t]}
\put(15.00,15.00){\makebox(0,0)[cc]{$1$}}
\put(35.00,15.00){\makebox(0,0)[cc]{$2k$}}
\put(45.00,15.00){\makebox(0,0)[cc]{$2k+1$}}
\put(75.00,15.00){\makebox(0,0)[cc]{$2m+1$}}
\put(90.00,15.00){\makebox(0,0)[cc]{$2m+2$}}
\put(125.00,15.00){\makebox(0,0)[cc]{$2n$}}
\put(67.33,5.00){\makebox(0,0)[cc]{$2m-2k$}}
\put(25.00,5.00){\makebox(0,0)[cc]{$2k$}}
\put(112.33,5.00){\makebox(0,0)[cc]{$2n-2m-2$}}
\put(112.33,10.67){\makebox(0,0)[cc]{$\underbrace{~~~~~~~~~~~~~~~~~}$}}
\put(67.00,10.67){\makebox(0,0)[cc]{$\underbrace{~~~~~~~~~~~~~~~~~~~~~}$}}
\put(24.67,10.67){\makebox(0,0)[cc]{$\underbrace{~~~~~~~~~~~~~~~~~~}$}}
\put(25.00,20.00){\makebox(0,0)[cc]{$\cdot \cdot \cdot$}}
\put(67.50,20.00){\makebox(0,0)[cc]{$\cdot \cdot \cdot$}}
\put(112.50,20.00){\makebox(0,0)[cc]{$\cdot \cdot \cdot$}}
\end{picture}
\end{center}
\caption{Derivation of the  Schwinger-Dyson equations
for free Green's functions
in the
Boltzmannian Fock space (the $2k+1$-th marked operator). }
\label{fig3}
\end{figure}
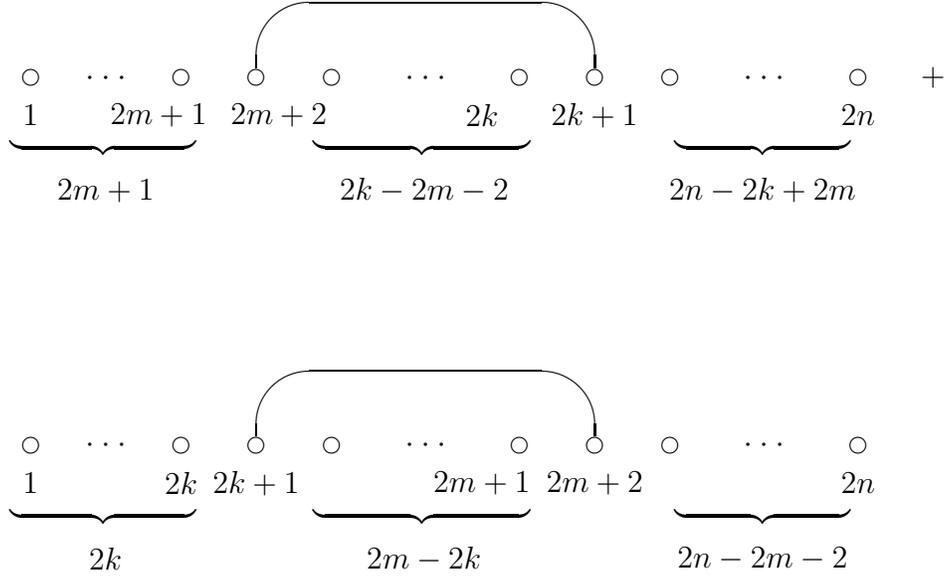


 \begin{figure}
\begin{center}
\unitlength=1.00mm
\special{em:linewidth 0.4pt}
\linethickness{0.4pt}
\begin{picture}(145.66,34.00)
\put(10.00,20.00){\circle{2.00}}
\put(20.00,20.00){\circle{2.00}}
\put(25.00,20.00){\circle{2.00}}
\put(30.00,20.00){\circle{2.00}}
\put(40.00,20.00){\circle{2.00}}
\put(60.00,20.00){\circle*{1.33}}
\put(63.00,20.00){\circle*{1.33}}
\put(70.00,20.00){\circle*{1.33}}
\put(145.00,20.00){\circle*{1.33}}
\put(138.00,20.00){\circle*{1.33}}
\put(135.00,20.00){\circle*{1.33}}
\put(103.00,20.00){\circle*{1.33}}
\put(106.00,20.00){\circle*{1.33}}
\put(113.00,20.00){\circle*{1.33}}
\put(100.00,20.00){\circle*{1.33}}
\put(93.00,20.00){\circle*{1.33}}
\put(15.00,20.00){\circle*{0.33}}
\put(17.00,20.00){\circle*{0.33}}
\put(13.00,20.00){\circle*{0.33}}
\put(35.00,20.00){\circle*{0.33}}
\put(37.00,20.00){\circle*{0.33}}
\put(33.00,20.00){\circle*{0.33}}
\put(66.33,20.00){\circle*{0.33}}
\put(68.00,20.00){\circle*{0.33}}
\put(64.67,20.00){\circle*{0.33}}
\put(96.67,20.00){\circle*{0.33}}
\put(98.33,20.00){\circle*{0.33}}
\put(95.00,20.00){\circle*{0.33}}
\put(109.33,20.00){\circle*{0.33}}
\put(111.00,20.00){\circle*{0.33}}
\put(107.67,20.00){\circle*{0.33}}
\put(141.33,20.00){\circle*{0.33}}
\put(143.00,20.00){\circle*{0.33}}
\put(139.67,20.00){\circle*{0.33}}
\put(124.00,20.00){\circle*{0.33}}
\put(127.00,20.00){\circle*{0.33}}
\put(121.00,20.00){\circle*{0.33}}
\put(81.00,20.00){\circle*{0.33}}
\put(84.00,20.00){\circle*{0.33}}
\put(78.00,20.00){\circle*{0.33}}
\put(64.00,20.50){\oval(78.00,27.00)[t]}
\put(10.00,16.00){\makebox(0,0)[cc]{$_1$}}
\put(20.00,16.00){\makebox(0,0)[cc]{$_{k-1}$}}
\put(25.00,16.00){\makebox(0,0)[cc]{$_k$}}
\put(30.00,16.00){\makebox(0,0)[cc]{$_{k+1}$}}
\put(40.00,16.00){\makebox(0,0)[cc]{$_n$}}
\put(65.33,16.00){\makebox(0,0)[cc]{$\underbrace{~~~~~~~~~}_{r}$}}
\put(96.67,16.00){\makebox(0,0)[cc]{$\underbrace{~~~~~~}_{s}$}}
\put(103.00,12.00){\makebox(0,0)[cc]{$\underbrace{~~~~~~~~~~~~~~~~~}_{r}$}}
\put(140.33,16.00){\makebox(0,0)[cc]{$\underbrace{~~~~~~~~~}_{r}$}}
\end{picture}
\end{center}
\caption{Derivation of the  Schwinger-Dyson equations in the
Boltzmannian Fock space for the interaction $\phi ^r$.}
\label{fig4}
\end{figure}
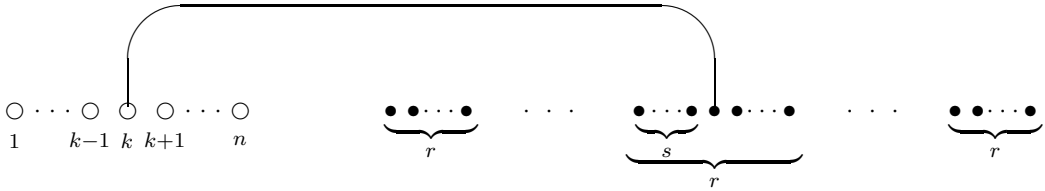

\begin{figure}
\begin{center}
\unitlength=1.00mm
\special{em:linewidth 0.4pt}
\linethickness{0.4pt}
\begin{picture}(140.00,150.00)
\put(10.00,150.00){\line(0,-1){145.00}}
\put(10.00,5.00){\line(1,0){130.00}}
\put(140.00,5.00){\line(0,1){145.00}}
\put(140.00,150.00){\line(-1,0){130.00}}
\put(10.00,140.00){\line(1,0){130.00}}
\put(40.00,150.00){\line(0,-1){145.00}}
\put(70.00,150.00){\line(0,-1){145.00}}
\put(25.00,145.00){\makebox(0,0)[cc]{$g$}}
\put(55.00,145.00){\makebox(0,0)[cc]{$g^2$}}
\put(105.00,145.00){\makebox(0,0)[cc]{$g^3$}}
\put(55.00,131.00){\circle{6.00}}
\put(85.00,128.50){\oval(10.00,13.00)[t]}
\put(120.00,134.00){\oval(10.00,12.00)[b]}
\put(110.00,134.00){\line(1,0){20.00}}
\put(120.00,128.00){\line(-5,6){5.00}}
\put(120.00,128.00){\line(5,6){5.00}}
\put(25.00,128.00){\circle{6.00}}
\put(17.00,125.00){\line(1,0){16.00}}
\put(25.00,110.00){\circle{6.00}}
\put(17.00,113.00){\line(1,0){16.00}}
\put(55.00,119.00){\circle{4.00}}
\put(55.00,115.00){\circle{4.00}}
\put(47.00,113.00){\line(1,0){17.00}}
\put(85.00,118.00){\circle{8.00}}
\put(120.00,118.00){\circle{8.00}}
\put(120.00,102.00){\circle{6.00}}
\put(85.00,100.00){\circle{6.00}}
\put(85.00,105.00){\circle{4.00}}
\put(120.00,97.00){\circle{4.00}}
\put(75.00,100.00){\line(1,0){20.00}}
\put(110.00,102.00){\line(1,0){20.00}}
\put(55.00,99.00){\circle{4.00}}
\put(55.00,103.00){\circle{4.00}}
\put(47.00,105.00){\line(1,0){17.00}}
\put(55.00,86.00){\circle{8.00}}
\put(55.00,71.00){\circle{8.00}}
\put(85.00,86.00){\circle{8.00}}
\put(85.00,71.00){\circle{8.00}}
\put(120.00,86.00){\circle{8.00}}
\put(120.00,71.00){\circle{8.00}}
\put(75.00,86.00){\line(1,0){20.00}}
\put(95.00,71.00){\line(-1,0){20.00}}
\put(110.00,71.00){\line(1,0){20.00}}
\put(130.00,86.00){\line(-1,0){20.00}}
\put(55.00,88.00){\circle{4.00}}
\put(55.00,69.00){\circle{4.00}}
\put(75.00,113.70){\line(1,0){20.00}}
\put(110.00,122.30){\line(1,0){20.00}}
\put(125.00,57.33){\circle{8.00}}
\put(83.00,57.00){\circle{8.00}}
\put(83.00,42.00){\circle{8.00}}
\put(125.00,42.00){\circle{8.00}}
\put(135.00,42.00){\line(-1,0){25.00}}
\put(100.00,42.00){\line(-1,0){25.00}}
\put(75.00,57.00){\line(1,0){24.67}}
\put(110.33,57.00){\line(1,0){24.33}}
\put(116.00,40.33){\circle{3.33}}
\put(116.00,58.67){\circle{3.33}}
\put(93.33,58.67){\circle{3.33}}
\put(93.33,40.33){\circle{3.33}}
\put(47.00,43.67){\line(1,0){17.00}}
\put(64.00,55.33){\line(-1,0){16.67}}
\put(51.33,57.33){\circle{4.00}}
\put(59.00,57.33){\circle{4.00}}
\put(59.00,41.67){\circle{4.00}}
\put(51.33,41.67){\circle{4.00}}
\put(47.00,26.00){\line(1,0){17.00}}
\put(64.00,11.00){\line(-1,0){17.00}}
\put(51.33,9.00){\circle{4.00}}
\put(59.00,13.00){\circle{4.00}}
\put(59.00,24.00){\circle{4.00}}
\put(51.33,28.00){\circle{4.00}}
\put(91.00,26.00){\circle{8.00}}
\put(100.00,21.70){\line(-1,0){18.00}}
\put(95.67,30.33){\circle{4.67}}
\put(86.33,30.33){\circle{4.67}}
\put(109.67,24.67){\line(1,0){25.33}}
\put(121.67,26.33){\circle{3.33}}
\put(129.33,26.33){\circle{3.33}}
\put(114.67,26.33){\circle{3.33}}
\put(123.67,8.33){\circle{3.33}}
\put(123.67,11.67){\circle{3.33}}
\put(123.67,15.00){\circle{3.33}}
\put(134.00,6.67){\line(-1,0){20.00}}
\put(100.33,8.33){\line(-1,0){21.67}}
\put(84.00,10.00){\circle{3.33}}
\put(93.33,10.00){\circle{3.33}}
\put(93.33,13.33){\circle{3.33}}
\put(76.00,25.33){\makebox(0,0)[cc]{$8$}}
\put(74.00,10.67){\makebox(0,0)[cc]{$16$}}
\put(109.67,10.67){\makebox(0,0)[cc]{$8$}}
\put(107.00,26.33){\makebox(0,0)[cc]{$8$}}
\put(47.00,131.00){\line(1,0){17.00}}
\put(81.67,120.67){\line(1,0){6.33}}
\put(117.00,115.17){\line(1,0){6.33}}
\put(120.17,115.33){\oval(6.33,5.00)[t]}
\put(46.67,81.70){\line(1,0){17.67}}
\put(46.67,75.30){\line(1,0){18.00}}
\put(85.17,120.50){\oval(6.33,5.00)[b]}
\put(120.00,69.67){\circle{2.67}}
\put(85.00,72.33){\circle{2.67}}
\put(85.00,88.67){\circle{2.67}}
\put(120.00,83.33){\circle{2.67}}
\put(80.00,128.67){\line(4,5){5.00}}
\put(85.00,135.00){\line(4,-5){5.00}}
\put(95.00,128.67){\line(-1,0){20.00}}
\put(47.00,11.00){\dashbox{1.00}(17.00,6.33)[cc]{}}
\put(47.00,26.00){\dashbox{0.67}(17.00,7.00)[cc]{}}
\put(47.00,43.67){\dashbox{0.67}(17.00,5.33)[cc]{}}
\put(47.33,55.33){\dashbox{0.67}(16.67,7.67)[cc]{}}
\put(46.67,75.33){\dashbox{0.67}(18.00,3.33)[cc]{}}
\put(46.33,81.67){\dashbox{0.67}(18.00,11.00)[cc]{}}
\put(46.67,105.00){\dashbox{0.67}(17.33,4.33)[cc]{}}
\put(46.67,113.00){\dashbox{0.67}(17.33,10.33)[cc]{}}
\put(47.00,131.00){\dashbox{0.67}(17.00,5.67)[cc]{}}
\put(17.00,125.00){\dashbox{0.67}(16.00,9.00)[cc]{}}
\put(16.67,113.00){\dashbox{0.67}(16.33,5.00)[cc]{}}
\put(74.67,42.00){\dashbox{0.67}(25.33,7.33)[cc]{}}
\put(75.00,57.00){\dashbox{0.67}(25.00,7.00)[cc]{}}
\put(109.67,42.00){\dashbox{0.67}(25.33,7.00)[cc]{}}
\put(110.33,57.00){\dashbox{0.67}(24.67,7.00)[cc]{}}
\put(75.00,71.00){\dashbox{0.67}(20.00,7.00)[cc]{}}
\put(110.00,71.00){\dashbox{0.67}(20.00,7.00)[cc]{}}
\put(110.00,86.00){\dashbox{0.67}(20.00,6.67)[cc]{}}
\put(110.00,102.00){\dashbox{0.67}(20.00,6.67)[cc]{}}
\put(110.00,122.33){\dashbox{0.67}(20.00,3.67)[cc]{}}
\put(110.00,134.00){\dashbox{0.67}(20.00,3.67)[cc]{}}
\put(75.00,86.00){\dashbox{0.67}(20.00,6.67)[cc]{}}
\put(75.00,100.00){\dashbox{0.67}(20.00,9.33)[cc]{}}
\put(75.00,113.67){\dashbox{0.67}(20.00,10.67)[cc]{}}
\put(75.00,128.67){\dashbox{0.67}(20.00,8.67)[cc]{}}
\end{picture}
\end{center}
\caption{Planar graphs in the matrix theory with the quartic interaction
up to order $g^3$.}
\label{fig5}
\end{figure}

\begin{figure}
\begin{center}
\unitlength=1.00mm
\special{em:linewidth 0.4pt}
\linethickness{0.4pt}
\begin{picture}(140.00,150.00)
\put(10.00,150.00){\line(0,-1){145.00}}
\put(10.00,5.00){\line(1,0){130.00}}
\put(140.00,5.00){\line(0,1){145.00}}
\put(140.00,150.00){\line(-1,0){130.00}}
\put(10.00,140.00){\line(1,0){130.00}}
\put(45.00,150.00){\line(0,-1){145.00}}
\put(92.33,145.00){\makebox(0,0)[cc]{$g^2$}}
\put(27.00,145.00){\makebox(0,0)[cc]{$g$}}
\emline{15.00}{126.00}{1}{15.00}{137.00}{2}
\emline{15.00}{137.00}{3}{33.00}{137.00}{4}
\emline{33.00}{137.00}{5}{33.00}{126.00}{6}
\emline{33.00}{126.00}{7}{38.00}{133.00}{8}
\emline{38.00}{133.00}{9}{41.00}{129.00}{10}
\emline{41.00}{129.00}{11}{33.00}{126.00}{12}
\emline{33.00}{126.00}{13}{31.00}{132.00}{14}
\emline{31.00}{132.00}{15}{19.00}{132.00}{16}
\emline{19.00}{132.00}{17}{19.00}{126.00}{18}
\emline{15.00}{109.00}{19}{15.00}{121.00}{20}
\emline{15.00}{121.00}{21}{40.00}{121.00}{22}
\emline{25.00}{117.00}{23}{19.00}{117.00}{24}
\emline{19.00}{117.00}{25}{19.00}{109.00}{26}
\emline{15.00}{91.00}{27}{15.00}{103.00}{28}
\emline{15.00}{103.00}{29}{40.00}{103.00}{30}
\emline{40.00}{103.00}{31}{33.00}{91.00}{32}
\emline{33.00}{91.00}{33}{33.00}{99.00}{34}
\emline{33.00}{99.00}{35}{19.00}{99.00}{36}
\emline{19.00}{99.00}{37}{19.00}{91.00}{38}
\emline{33.00}{91.00}{39}{29.33}{97.00}{40}
\emline{29.33}{97.00}{41}{26.67}{93.00}{42}
\emline{26.67}{93.00}{43}{33.00}{91.00}{44}
\emline{50.00}{127.00}{45}{50.00}{137.00}{46}
\emline{50.00}{137.00}{47}{88.00}{137.00}{48}
\emline{88.00}{137.00}{49}{83.00}{127.00}{50}
\emline{83.00}{127.00}{51}{83.00}{135.00}{52}
\emline{83.00}{135.00}{53}{65.00}{135.00}{54}
\emline{65.00}{135.00}{55}{65.00}{127.00}{56}
\emline{65.00}{127.00}{57}{60.00}{135.00}{58}
\emline{60.00}{135.00}{59}{54.00}{135.00}{60}
\emline{54.00}{135.00}{61}{54.00}{127.00}{62}
\emline{65.00}{127.00}{63}{68.00}{133.00}{64}
\emline{68.00}{133.00}{65}{80.00}{133.00}{66}
\emline{80.00}{133.00}{67}{83.00}{127.00}{68}
\emline{83.00}{127.00}{69}{78.00}{131.00}{70}
\emline{78.00}{131.00}{71}{70.00}{131.00}{72}
\emline{70.00}{131.00}{73}{65.00}{127.00}{74}
\emline{100.00}{127.00}{75}{100.00}{137.00}{76}
\emline{100.00}{137.00}{77}{113.00}{137.00}{78}
\emline{113.00}{137.00}{79}{117.00}{127.00}{80}
\emline{117.00}{127.00}{81}{117.00}{135.00}{82}
\emline{117.00}{135.00}{83}{131.00}{135.00}{84}
\emline{131.00}{135.00}{85}{131.00}{127.00}{86}
\emline{131.00}{127.00}{87}{138.00}{130.00}{88}
\emline{138.00}{130.00}{89}{135.00}{134.00}{90}
\emline{135.00}{134.00}{91}{131.00}{127.00}{92}
\emline{131.00}{127.00}{93}{128.00}{133.00}{94}
\emline{128.00}{133.00}{95}{120.00}{133.00}{96}
\emline{120.00}{133.00}{97}{117.00}{127.00}{98}
\emline{117.00}{127.00}{99}{112.00}{133.00}{100}
\emline{112.00}{133.00}{101}{104.00}{133.00}{102}
\emline{104.00}{133.00}{103}{104.00}{127.00}{104}
\emline{50.00}{112.00}{105}{50.00}{122.00}{106}
\emline{50.00}{122.00}{107}{64.00}{122.00}{108}
\emline{64.00}{122.00}{109}{68.00}{112.00}{110}
\emline{68.00}{112.00}{111}{68.00}{122.00}{112}
\emline{68.00}{122.00}{113}{88.00}{122.00}{114}
\emline{88.00}{122.00}{115}{85.00}{112.00}{116}
\emline{85.00}{112.00}{117}{85.00}{119.00}{118}
\emline{85.00}{119.00}{119}{81.00}{118.00}{120}
\emline{81.00}{118.00}{121}{85.00}{112.00}{122}
\emline{85.00}{112.00}{123}{77.00}{118.00}{124}
\emline{77.00}{118.00}{125}{71.00}{118.00}{126}
\emline{71.00}{118.00}{127}{68.00}{112.00}{128}
\emline{68.00}{112.00}{129}{61.33}{119.00}{130}
\emline{61.33}{119.00}{131}{54.00}{119.00}{132}
\emline{54.00}{119.00}{133}{54.00}{112.00}{134}
\emline{100.00}{112.00}{135}{100.00}{122.00}{136}
\emline{100.00}{122.00}{137}{113.00}{122.00}{138}
\emline{113.00}{122.00}{139}{117.00}{112.00}{140}
\emline{117.00}{112.00}{141}{117.00}{122.00}{142}
\emline{117.00}{122.00}{143}{137.00}{122.00}{144}
\emline{137.00}{122.00}{145}{134.00}{112.00}{146}
\emline{134.00}{112.00}{147}{134.00}{119.00}{148}
\emline{134.00}{119.00}{149}{120.00}{119.00}{150}
\emline{120.00}{119.00}{151}{117.00}{112.00}{152}
\emline{117.00}{112.00}{153}{110.00}{119.00}{154}
\emline{110.00}{119.00}{155}{104.00}{119.00}{156}
\emline{104.00}{119.00}{157}{104.00}{112.00}{158}
\emline{134.00}{112.00}{159}{130.00}{117.00}{160}
\emline{130.00}{117.00}{161}{128.00}{113.67}{162}
\emline{128.00}{113.67}{163}{134.00}{112.00}{164}
\emline{50.00}{97.00}{165}{50.00}{107.00}{166}
\emline{50.00}{107.00}{167}{88.00}{107.00}{168}
\emline{88.00}{107.00}{169}{85.00}{97.00}{170}
\emline{85.00}{97.00}{171}{82.00}{103.00}{172}
\emline{82.00}{103.00}{173}{71.00}{103.00}{174}
\emline{71.00}{103.00}{175}{71.00}{97.00}{176}
\emline{71.00}{97.00}{177}{68.00}{102.33}{178}
\emline{68.00}{102.33}{179}{65.67}{98.67}{180}
\emline{65.67}{98.67}{181}{71.00}{97.00}{182}
\emline{71.00}{97.00}{183}{73.00}{101.00}{184}
\emline{73.00}{101.00}{185}{80.00}{101.00}{186}
\emline{85.00}{105.00}{187}{54.00}{105.00}{188}
\emline{54.00}{105.00}{189}{54.00}{97.00}{190}
\emline{100.00}{97.00}{191}{100.00}{107.00}{192}
\emline{100.00}{107.00}{193}{137.00}{107.00}{194}
\emline{137.00}{107.00}{195}{134.00}{97.00}{196}
\emline{134.00}{97.00}{197}{131.00}{103.00}{198}
\emline{131.00}{103.00}{199}{113.00}{103.00}{200}
\emline{113.00}{103.00}{201}{117.00}{97.00}{202}
\emline{117.00}{97.00}{203}{117.00}{101.00}{204}
\emline{117.00}{101.00}{205}{120.67}{100.00}{206}
\emline{120.67}{100.00}{207}{117.00}{97.00}{208}
\emline{117.00}{97.00}{209}{125.00}{100.00}{210}
\emline{125.00}{100.00}{211}{129.00}{100.00}{212}
\emline{129.00}{100.00}{213}{134.00}{97.00}{214}
\emline{134.00}{97.00}{215}{134.00}{105.00}{216}
\emline{134.00}{105.00}{217}{104.00}{105.00}{218}
\emline{104.00}{105.00}{219}{104.00}{97.00}{220}
\emline{50.00}{82.00}{221}{50.00}{92.00}{222}
\emline{50.00}{92.00}{223}{88.00}{92.00}{224}
\emline{88.00}{92.00}{225}{85.00}{82.00}{226}
\emline{85.00}{82.00}{227}{82.33}{88.00}{228}
\emline{82.33}{88.00}{229}{68.00}{88.00}{230}
\emline{68.00}{88.00}{231}{71.00}{82.00}{232}
\emline{71.00}{82.00}{233}{71.00}{86.00}{234}
\emline{71.00}{86.00}{235}{80.00}{86.00}{236}
\emline{80.00}{86.00}{237}{85.00}{82.00}{238}
\emline{85.00}{82.00}{239}{85.00}{90.00}{240}
\emline{85.00}{90.00}{241}{54.00}{90.00}{242}
\emline{54.00}{90.00}{243}{54.00}{82.00}{244}
\emline{71.00}{82.00}{245}{76.00}{85.00}{246}
\emline{76.00}{85.00}{247}{76.67}{82.67}{248}
\emline{76.67}{82.67}{249}{71.00}{82.00}{250}
\emline{100.00}{82.00}{251}{100.00}{92.00}{252}
\emline{100.00}{92.00}{253}{137.00}{92.00}{254}
\emline{137.00}{92.00}{255}{134.00}{82.00}{256}
\emline{134.00}{82.00}{257}{134.00}{88.00}{258}
\emline{134.00}{88.00}{259}{130.67}{87.00}{260}
\emline{130.67}{87.00}{261}{134.00}{82.00}{262}
\emline{134.00}{82.00}{263}{127.00}{88.00}{264}
\emline{127.00}{88.00}{265}{122.00}{88.00}{266}
\emline{122.00}{88.00}{267}{117.00}{82.00}{268}
\emline{117.00}{82.00}{269}{118.67}{87.33}{270}
\emline{118.67}{87.33}{271}{114.67}{87.33}{272}
\emline{114.67}{87.33}{273}{117.00}{82.00}{274}
\emline{117.00}{82.00}{275}{111.00}{88.00}{276}
\emline{111.00}{88.00}{277}{104.00}{88.00}{278}
\emline{104.00}{88.00}{279}{104.00}{82.00}{280}
\emline{100.00}{67.00}{281}{100.00}{77.00}{282}
\emline{100.00}{77.00}{283}{131.00}{77.00}{284}
\emline{131.00}{77.00}{285}{131.00}{67.00}{286}
\emline{131.00}{67.00}{287}{127.33}{73.00}{288}
\emline{127.33}{73.00}{289}{122.00}{73.00}{290}
\emline{122.00}{73.00}{291}{117.00}{67.00}{292}
\emline{117.00}{67.00}{293}{118.67}{72.00}{294}
\emline{118.67}{72.00}{295}{114.67}{72.00}{296}
\emline{114.67}{72.00}{297}{117.00}{67.00}{298}
\emline{117.00}{67.00}{299}{111.00}{73.00}{300}
\emline{111.00}{73.00}{301}{104.00}{73.00}{302}
\emline{104.00}{73.00}{303}{104.00}{67.00}{304}
\emline{131.00}{67.00}{305}{134.67}{72.67}{306}
\emline{134.67}{72.67}{307}{137.33}{69.00}{308}
\emline{137.33}{69.00}{309}{131.00}{67.00}{310}
\emline{50.00}{67.00}{311}{50.00}{77.00}{312}
\emline{50.00}{77.00}{313}{88.00}{77.00}{314}
\emline{88.00}{77.00}{315}{85.00}{67.00}{316}
\emline{85.00}{67.00}{317}{85.00}{75.00}{318}
\emline{85.00}{75.00}{319}{74.00}{75.00}{320}
\emline{74.00}{75.00}{321}{71.00}{67.00}{322}
\emline{71.00}{67.00}{323}{71.00}{73.00}{324}
\emline{71.00}{73.00}{325}{67.33}{71.00}{326}
\emline{67.33}{71.00}{327}{71.00}{67.00}{328}
\emline{71.00}{67.00}{329}{60.67}{73.00}{330}
\emline{60.67}{73.00}{331}{54.00}{73.00}{332}
\emline{54.00}{73.00}{333}{54.00}{67.00}{334}
\emline{85.00}{67.00}{335}{82.33}{72.00}{336}
\emline{82.33}{72.00}{337}{80.00}{68.67}{338}
\emline{80.00}{68.67}{339}{85.00}{67.00}{340}
\emline{50.00}{52.00}{341}{50.00}{62.00}{342}
\emline{50.00}{62.00}{343}{88.00}{62.00}{344}
\emline{88.00}{62.00}{345}{85.00}{52.00}{346}
\emline{85.00}{52.00}{347}{85.00}{57.33}{348}
\emline{85.00}{57.33}{349}{81.33}{56.00}{350}
\emline{81.33}{56.00}{351}{85.00}{52.00}{352}
\emline{85.00}{52.00}{353}{76.67}{58.00}{354}
\emline{76.67}{58.00}{355}{73.00}{58.00}{356}
\emline{73.00}{58.00}{357}{71.00}{52.00}{358}
\emline{71.00}{52.00}{359}{71.00}{59.00}{360}
\emline{71.00}{59.00}{361}{54.00}{59.00}{362}
\emline{54.00}{59.00}{363}{54.00}{52.00}{364}
\emline{71.00}{52.00}{365}{68.33}{57.00}{366}
\emline{68.33}{57.00}{367}{66.00}{53.67}{368}
\emline{66.00}{53.67}{369}{71.00}{52.00}{370}
\emline{100.00}{52.00}{371}{100.00}{62.00}{372}
\emline{100.00}{62.00}{373}{137.00}{62.00}{374}
\emline{137.00}{62.00}{375}{134.00}{52.00}{376}
\emline{134.00}{52.00}{377}{134.00}{60.00}{378}
\emline{134.00}{60.00}{379}{122.00}{60.00}{380}
\emline{122.00}{60.00}{381}{117.00}{52.00}{382}
\emline{117.00}{52.00}{383}{117.00}{60.00}{384}
\emline{117.00}{60.00}{385}{104.00}{60.00}{386}
\emline{104.00}{60.00}{387}{104.00}{52.00}{388}
\emline{117.00}{52.00}{389}{113.33}{57.00}{390}
\emline{113.33}{57.00}{391}{111.67}{53.00}{392}
\emline{111.67}{53.00}{393}{117.00}{52.00}{394}
\emline{134.00}{52.00}{395}{130.67}{56.67}{396}
\emline{130.67}{56.67}{397}{128.67}{53.33}{398}
\emline{128.67}{53.33}{399}{134.00}{52.00}{400}
\emline{50.00}{37.00}{401}{50.00}{47.00}{402}
\emline{50.00}{47.00}{403}{83.00}{47.00}{404}
\emline{83.00}{47.00}{405}{83.00}{37.00}{406}
\emline{83.00}{37.00}{407}{79.00}{43.33}{408}
\emline{79.00}{43.33}{409}{74.00}{43.33}{410}
\emline{74.00}{43.33}{411}{71.00}{37.00}{412}
\emline{71.00}{37.00}{413}{68.00}{42.00}{414}
\emline{68.00}{42.00}{415}{65.67}{38.33}{416}
\emline{65.67}{38.33}{417}{71.00}{37.00}{418}
\emline{71.00}{37.00}{419}{71.00}{45.00}{420}
\emline{71.00}{45.00}{421}{54.00}{45.00}{422}
\emline{54.00}{45.00}{423}{54.00}{37.00}{424}
\emline{83.00}{37.00}{425}{86.33}{42.67}{426}
\emline{86.33}{42.67}{427}{89.33}{39.00}{428}
\emline{89.33}{39.00}{429}{83.00}{37.00}{430}
\emline{100.00}{37.00}{431}{100.00}{47.00}{432}
\emline{100.00}{47.00}{433}{137.00}{47.00}{434}
\emline{137.00}{47.00}{435}{134.00}{37.00}{436}
\emline{134.00}{37.00}{437}{134.00}{43.00}{438}
\emline{134.00}{43.00}{439}{130.00}{41.33}{440}
\emline{130.00}{41.33}{441}{134.00}{37.00}{442}
\emline{134.00}{37.00}{443}{123.00}{45.00}{444}
\emline{123.00}{45.00}{445}{117.00}{45.00}{446}
\emline{117.00}{45.00}{447}{117.00}{37.00}{448}
\emline{117.00}{37.00}{449}{120.00}{42.33}{450}
\emline{120.00}{42.33}{451}{122.67}{39.00}{452}
\emline{122.67}{39.00}{453}{117.00}{37.00}{454}
\emline{117.00}{37.00}{455}{112.00}{45.00}{456}
\emline{112.00}{45.00}{457}{104.00}{45.00}{458}
\emline{104.00}{45.00}{459}{104.00}{37.00}{460}
\emline{100.00}{22.00}{461}{100.00}{32.00}{462}
\emline{100.00}{32.00}{463}{131.00}{32.00}{464}
\emline{131.00}{32.00}{465}{131.00}{22.00}{466}
\emline{131.00}{22.00}{467}{134.33}{27.33}{468}
\emline{134.33}{27.33}{469}{137.00}{24.33}{470}
\emline{137.00}{24.33}{471}{131.00}{22.00}{472}
\emline{131.00}{22.00}{473}{127.67}{30.00}{474}
\emline{127.67}{30.00}{475}{117.00}{30.00}{476}
\emline{117.00}{30.00}{477}{117.00}{22.00}{478}
\emline{117.00}{22.00}{479}{120.00}{27.33}{480}
\emline{120.00}{27.33}{481}{123.00}{24.00}{482}
\emline{123.00}{24.00}{483}{117.00}{22.00}{484}
\emline{117.00}{22.00}{485}{112.00}{30.00}{486}
\emline{112.00}{30.00}{487}{104.00}{30.00}{488}
\emline{104.00}{30.00}{489}{104.00}{22.00}{490}
\emline{50.00}{22.00}{491}{50.00}{32.00}{492}
\emline{50.00}{32.00}{493}{88.00}{32.00}{494}
\emline{88.00}{32.00}{495}{85.00}{22.00}{496}
\emline{85.00}{22.00}{497}{85.00}{30.00}{498}
\emline{85.00}{30.00}{499}{64.33}{30.00}{500}
\emline{64.33}{30.00}{501}{64.33}{22.00}{502}
\emline{64.33}{22.00}{503}{61.00}{30.00}{504}
\emline{61.00}{30.00}{505}{54.00}{30.00}{506}
\emline{54.00}{30.00}{507}{54.00}{22.00}{508}
\emline{64.33}{22.00}{509}{67.00}{27.67}{510}
\emline{67.00}{27.67}{511}{70.00}{24.33}{512}
\emline{70.00}{24.33}{513}{64.33}{22.00}{514}
\emline{85.00}{22.00}{515}{82.33}{27.67}{516}
\emline{82.33}{27.67}{517}{80.00}{24.00}{518}
\emline{80.00}{24.00}{519}{85.00}{22.00}{520}
\emline{25.00}{117.00}{521}{33.00}{109.00}{522}
\emline{33.00}{109.00}{523}{31.00}{115.33}{524}
\emline{31.00}{115.33}{525}{35.00}{115.67}{526}
\emline{35.00}{115.67}{527}{33.00}{109.00}{528}
\emline{33.00}{109.00}{529}{40.00}{121.00}{530}
\emline{80.00}{101.00}{531}{85.00}{97.00}{532}
\emline{85.00}{97.00}{533}{85.00}{105.00}{534}
\end{picture}
\end{center}
\caption{Half-planar graphs in the Boltzmann theory with the quartic 
interaction up to order $g^2$.} 
\label{fig6} 
\end{figure}
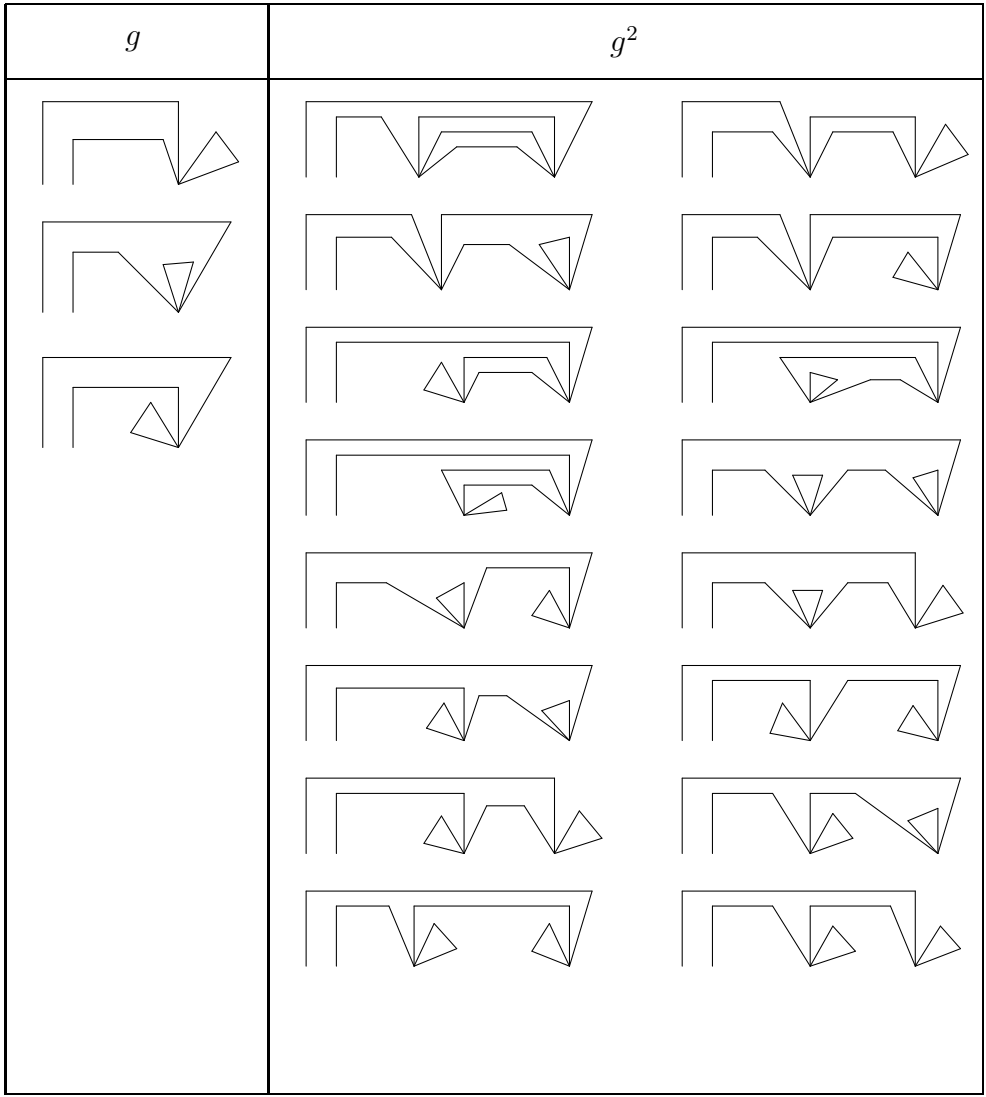

          
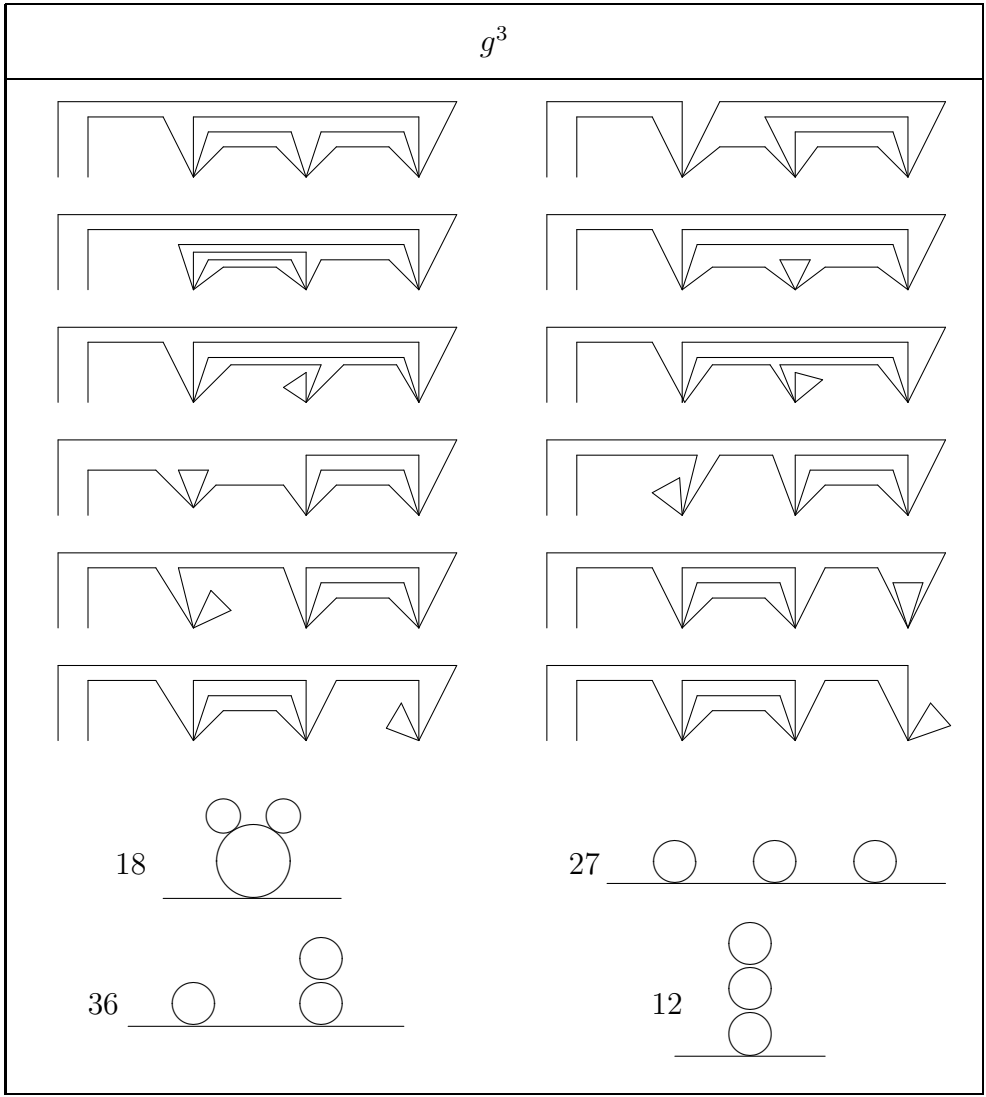
\begin{figure}
\begin{center}
\unitlength=1.00mm
\special{em:linewidth 0.4pt}
\linethickness{0.4pt}
\begin{picture}(140.00,150.00)
\put(10.00,150.00){\line(0,-1){145.00}}
\put(10.00,5.00){\line(1,0){130.00}}
\put(140.00,5.00){\line(0,1){145.00}}
\put(140.00,150.00){\line(-1,0){130.00}}
\put(10.00,140.00){\line(1,0){130.00}}
\put(75.00,145.00){\makebox(0,0)[cc]{$g^3$}}
\emline{17.00}{127.00}{1}{17.00}{137.00}{2}
\emline{17.00}{137.00}{3}{70.00}{137.00}{4}
\emline{70.00}{137.00}{5}{65.00}{127.00}{6}
\emline{65.00}{127.00}{7}{65.00}{135.00}{8}
\emline{65.00}{135.00}{9}{35.00}{135.00}{10}
\emline{35.00}{135.00}{11}{35.00}{127.00}{12}
\emline{35.00}{127.00}{13}{37.00}{133.00}{14}
\emline{37.00}{133.00}{15}{48.00}{133.00}{16}
\emline{48.00}{133.00}{17}{50.00}{127.00}{18}
\emline{50.00}{127.00}{19}{52.00}{133.00}{20}
\emline{52.00}{133.00}{21}{63.00}{133.00}{22}
\emline{63.00}{133.00}{23}{65.00}{127.00}{24}
\emline{65.00}{127.00}{25}{61.00}{131.00}{26}
\emline{61.00}{131.00}{27}{54.00}{131.00}{28}
\emline{54.00}{131.00}{29}{50.00}{127.00}{30}
\emline{50.00}{127.00}{31}{46.00}{131.00}{32}
\emline{46.00}{131.00}{33}{39.00}{131.00}{34}
\emline{39.00}{131.00}{35}{35.00}{127.00}{36}
\emline{35.00}{127.00}{37}{31.00}{135.00}{38}
\emline{31.00}{135.00}{39}{21.00}{135.00}{40}
\emline{21.00}{135.00}{41}{21.00}{127.00}{42}
\emline{17.00}{112.00}{43}{17.00}{122.00}{44}
\emline{17.00}{122.00}{45}{70.00}{122.00}{46}
\emline{70.00}{122.00}{47}{65.00}{112.00}{48}
\emline{65.00}{112.00}{49}{65.00}{120.00}{50}
\emline{65.00}{120.00}{51}{21.00}{120.00}{52}
\emline{21.00}{120.00}{53}{21.00}{112.00}{54}
\emline{65.00}{112.00}{55}{63.00}{118.00}{56}
\emline{63.00}{118.00}{57}{33.00}{118.00}{58}
\emline{33.00}{118.00}{59}{35.00}{112.00}{60}
\emline{35.00}{112.00}{63}{35.00}{117.00}{64}
\emline{35.00}{117.00}{65}{50.00}{117.00}{66}
\emline{50.00}{117.00}{67}{50.00}{112.00}{68}
\emline{50.00}{112.00}{69}{48.00}{116.00}{70}
\emline{48.00}{116.00}{71}{37.00}{116.00}{72}
\emline{37.00}{116.00}{73}{35.00}{112.00}{74}
\emline{35.00}{112.00}{75}{39.00}{115.00}{76}
\emline{39.00}{115.00}{77}{46.00}{115.00}{78}
\emline{46.00}{115.00}{79}{50.00}{112.00}{80}
\emline{50.00}{112.00}{81}{52.00}{116.00}{82}
\emline{52.00}{116.00}{83}{61.00}{116.00}{84}
\emline{61.00}{116.00}{85}{65.00}{112.00}{86}
\emline{82.00}{127.00}{87}{82.00}{137.00}{88}
\emline{82.00}{137.00}{89}{100.00}{137.00}{90}
\emline{100.00}{137.00}{91}{100.00}{127.00}{92}
\emline{100.00}{127.00}{93}{105.00}{137.00}{94}
\emline{105.00}{137.00}{95}{135.00}{137.00}{96}
\emline{135.00}{137.00}{97}{130.00}{127.00}{98}
\emline{130.00}{127.00}{99}{130.00}{135.00}{100}
\emline{130.00}{135.00}{101}{111.00}{135.00}{102}
\emline{111.00}{135.00}{103}{115.00}{127.00}{104}
\emline{115.00}{127.00}{105}{115.00}{133.00}{106}
\emline{115.00}{133.00}{107}{128.00}{133.00}{108}
\emline{128.00}{133.00}{109}{130.00}{127.00}{110}
\emline{130.00}{127.00}{111}{126.00}{131.00}{112}
\emline{126.00}{131.00}{113}{118.00}{131.00}{114}
\emline{118.00}{131.00}{115}{115.00}{127.00}{116}
\emline{115.00}{127.00}{117}{111.00}{131.00}{118}
\emline{111.00}{131.00}{119}{105.00}{131.00}{120}
\emline{105.00}{131.00}{121}{100.00}{127.00}{122}
\emline{100.00}{127.00}{123}{96.00}{135.00}{124}
\emline{96.00}{135.00}{125}{86.00}{135.00}{126}
\emline{86.00}{135.00}{127}{86.00}{127.00}{128}
\emline{82.00}{112.00}{129}{82.00}{122.00}{130}
\emline{82.00}{122.00}{131}{135.00}{122.00}{132}
\emline{135.00}{122.00}{133}{130.00}{112.00}{134}
\emline{130.00}{112.00}{135}{130.00}{120.00}{136}
\emline{130.00}{120.00}{137}{100.00}{120.00}{138}
\emline{100.00}{120.00}{139}{100.00}{112.00}{140}
\emline{100.00}{112.00}{141}{102.00}{118.00}{142}
\emline{102.00}{118.00}{143}{128.00}{118.00}{144}
\emline{128.00}{118.00}{145}{130.00}{112.00}{146}
\emline{130.00}{112.00}{147}{126.00}{115.00}{148}
\emline{126.00}{115.00}{149}{119.00}{115.00}{150}
\emline{119.00}{115.00}{151}{115.00}{112.00}{152}
\emline{115.00}{112.00}{153}{111.00}{115.00}{154}
\emline{111.00}{115.00}{155}{104.00}{115.00}{156}
\emline{104.00}{115.00}{157}{100.00}{112.00}{158}
\emline{100.00}{112.00}{159}{96.00}{120.00}{160}
\emline{96.00}{120.00}{161}{86.00}{120.00}{162}
\emline{86.00}{120.00}{163}{86.00}{112.00}{164}
\emline{115.00}{112.00}{165}{117.00}{116.00}{166}
\emline{117.00}{116.00}{167}{113.00}{116.00}{168}
\emline{113.00}{116.00}{169}{115.00}{112.00}{170}
\emline{82.00}{97.00}{171}{82.00}{107.00}{172}
\emline{82.00}{107.00}{173}{135.00}{107.00}{174}
\emline{135.00}{107.00}{175}{130.00}{97.00}{176}
\emline{130.00}{97.00}{177}{130.00}{105.00}{178}
\emline{130.00}{105.00}{179}{100.00}{105.00}{180}
\emline{100.00}{105.00}{181}{100.00}{97.00}{182}
\emline{100.00}{97.00}{183}{102.00}{103.00}{184}
\emline{102.00}{103.00}{185}{128.00}{103.00}{186}
\emline{128.00}{103.00}{187}{130.00}{97.00}{188}
\emline{130.00}{97.00}{189}{126.00}{102.00}{190}
\emline{126.00}{102.00}{191}{113.00}{102.00}{192}
\emline{113.00}{102.00}{193}{115.00}{97.00}{194}
\emline{115.00}{97.00}{195}{115.00}{101.00}{196}
\emline{115.00}{101.00}{197}{118.67}{100.00}{198}
\emline{118.67}{100.00}{199}{115.00}{97.00}{200}
\emline{115.00}{97.00}{201}{111.67}{102.00}{202}
\emline{111.67}{102.00}{203}{104.00}{102.00}{204}
\emline{104.00}{102.00}{205}{100.33}{97.00}{206}
\emline{100.33}{97.00}{207}{96.00}{105.00}{208}
\emline{96.00}{105.00}{209}{86.00}{105.00}{210}
\emline{86.00}{105.00}{211}{86.00}{97.00}{212}
\emline{17.00}{97.00}{213}{17.00}{107.00}{214}
\emline{17.00}{107.00}{215}{70.00}{107.00}{216}
\emline{70.00}{107.00}{217}{65.00}{97.00}{218}
\emline{65.00}{97.00}{219}{65.00}{105.00}{220}
\emline{65.00}{105.00}{221}{35.00}{105.00}{222}
\emline{35.00}{105.00}{223}{35.00}{97.00}{224}
\emline{35.00}{97.00}{225}{37.00}{103.00}{226}
\emline{37.00}{103.00}{227}{63.00}{103.00}{228}
\emline{63.00}{103.00}{229}{65.00}{97.00}{230}
\emline{65.00}{97.00}{231}{62.00}{102.00}{232}
\emline{62.00}{102.00}{233}{55.00}{102.00}{234}
\emline{55.00}{102.00}{235}{50.00}{97.00}{236}
\emline{50.00}{97.00}{237}{52.00}{102.00}{238}
\emline{52.00}{102.00}{239}{40.00}{102.00}{240}
\emline{40.00}{102.00}{241}{35.00}{97.00}{242}
\emline{35.00}{97.00}{243}{31.00}{105.00}{244}
\emline{31.00}{105.00}{245}{21.00}{105.00}{246}
\emline{21.00}{105.00}{247}{21.00}{97.00}{248}
\emline{50.00}{97.00}{249}{50.00}{101.00}{250}
\emline{50.00}{101.00}{251}{47.00}{99.00}{252}
\emline{47.00}{99.00}{253}{50.00}{97.00}{254}
\emline{17.00}{82.00}{255}{17.00}{92.00}{256}
\emline{17.00}{92.00}{257}{70.00}{92.00}{258}
\emline{70.00}{92.00}{259}{65.00}{82.00}{260}
\emline{65.00}{82.00}{261}{65.00}{90.00}{262}
\emline{65.00}{90.00}{263}{50.00}{90.00}{264}
\emline{50.00}{90.00}{265}{50.00}{82.00}{266}
\emline{50.00}{82.00}{267}{52.00}{88.00}{268}
\emline{52.00}{88.00}{269}{63.00}{88.00}{270}
\emline{63.00}{88.00}{271}{65.00}{82.00}{272}
\emline{65.00}{82.00}{273}{61.00}{86.00}{274}
\emline{61.00}{86.00}{275}{54.00}{86.00}{276}
\emline{54.00}{86.00}{277}{50.00}{82.00}{278}
\emline{50.00}{82.00}{279}{47.00}{86.00}{280}
\emline{47.00}{86.00}{281}{38.00}{86.00}{282}
\emline{38.00}{86.00}{283}{35.00}{83.00}{284}
\emline{35.00}{83.00}{285}{30.00}{88.00}{286}
\emline{30.00}{88.00}{287}{21.00}{88.00}{288}
\emline{21.00}{88.00}{289}{21.00}{82.00}{290}
\emline{35.00}{83.00}{291}{37.00}{88.00}{292}
\emline{37.00}{88.00}{293}{33.00}{88.00}{294}
\emline{33.00}{88.00}{295}{35.00}{83.00}{296}
\emline{82.00}{82.00}{297}{82.00}{92.00}{298}
\emline{82.00}{92.00}{299}{135.00}{92.00}{300}
\emline{135.00}{92.00}{301}{130.00}{82.00}{302}
\emline{130.00}{82.00}{303}{130.00}{90.00}{304}
\emline{130.00}{90.00}{305}{115.00}{90.00}{306}
\emline{115.00}{90.00}{307}{115.00}{82.00}{308}
\emline{115.00}{82.00}{309}{117.00}{88.00}{310}
\emline{117.00}{88.00}{311}{128.00}{88.00}{312}
\emline{128.00}{88.00}{313}{130.00}{82.00}{314}
\emline{130.00}{82.00}{315}{126.00}{86.00}{316}
\emline{126.00}{86.00}{317}{119.00}{86.00}{318}
\emline{119.00}{86.00}{319}{115.00}{82.00}{320}
\emline{115.00}{82.00}{321}{112.00}{90.00}{322}
\emline{112.00}{90.00}{323}{105.00}{90.00}{324}
\emline{105.00}{90.00}{325}{100.00}{82.00}{326}
\emline{100.00}{82.00}{327}{102.00}{90.00}{328}
\emline{102.00}{90.00}{329}{86.00}{90.00}{330}
\emline{86.00}{90.00}{331}{86.00}{82.00}{332}
\emline{100.00}{82.00}{333}{99.67}{87.00}{334}
\emline{99.67}{87.00}{335}{96.00}{85.00}{336}
\emline{96.00}{85.00}{337}{100.00}{82.00}{338}
\emline{17.00}{67.00}{339}{17.00}{77.00}{340}
\emline{17.00}{77.00}{341}{70.00}{77.00}{342}
\emline{70.00}{77.00}{343}{65.00}{67.00}{344}
\emline{65.00}{67.00}{345}{65.00}{75.00}{346}
\emline{65.00}{75.00}{347}{50.00}{75.00}{348}
\emline{50.00}{75.00}{349}{50.00}{67.00}{350}
\emline{50.00}{67.00}{351}{52.00}{73.00}{352}
\emline{52.00}{73.00}{353}{63.00}{73.00}{354}
\emline{63.00}{73.00}{355}{65.00}{67.00}{356}
\emline{65.00}{67.00}{357}{61.00}{71.00}{358}
\emline{61.00}{71.00}{359}{54.00}{71.00}{360}
\emline{54.00}{71.00}{361}{50.00}{67.00}{362}
\emline{50.00}{67.00}{363}{47.00}{75.00}{364}
\emline{47.00}{75.00}{365}{33.00}{75.00}{366}
\emline{33.00}{75.00}{367}{35.00}{67.00}{368}
\emline{35.00}{67.00}{369}{37.33}{72.00}{370}
\emline{37.33}{72.00}{371}{40.00}{69.33}{372}
\emline{40.00}{69.33}{373}{35.00}{67.00}{374}
\emline{35.00}{67.00}{375}{30.00}{75.00}{376}
\emline{30.00}{75.00}{377}{21.00}{75.00}{378}
\emline{21.00}{75.00}{379}{21.00}{67.00}{380}
\emline{82.00}{67.00}{381}{82.00}{77.00}{382}
\emline{82.00}{77.00}{383}{135.00}{77.00}{384}
\emline{135.00}{77.00}{385}{130.00}{67.00}{386}
\emline{130.00}{67.00}{387}{126.00}{75.00}{388}
\emline{126.00}{75.00}{389}{119.00}{75.00}{390}
\emline{119.00}{75.00}{391}{115.00}{67.00}{392}
\emline{115.00}{67.00}{393}{115.00}{75.00}{394}
\emline{115.00}{75.00}{395}{100.00}{75.00}{396}
\emline{100.00}{75.00}{397}{100.00}{67.00}{398}
\emline{100.00}{67.00}{399}{102.00}{73.00}{400}
\emline{102.00}{73.00}{401}{113.00}{73.00}{402}
\emline{113.00}{73.00}{403}{115.00}{67.00}{404}
\emline{115.00}{67.00}{405}{111.00}{71.00}{406}
\emline{111.00}{71.00}{407}{104.00}{71.00}{408}
\emline{104.00}{71.00}{409}{100.00}{67.00}{410}
\emline{100.00}{67.00}{411}{96.00}{75.00}{412}
\emline{96.00}{75.00}{413}{86.00}{75.00}{414}
\emline{86.00}{75.00}{415}{86.00}{67.00}{416}
\emline{130.00}{67.00}{417}{132.00}{73.00}{418}
\emline{132.00}{73.00}{419}{128.00}{73.00}{420}
\emline{128.00}{73.00}{421}{130.00}{67.00}{422}
\emline{82.00}{52.00}{423}{82.00}{62.00}{424}
\emline{82.00}{62.00}{425}{130.00}{62.00}{426}
\emline{130.00}{62.00}{427}{130.00}{52.00}{428}
\emline{130.00}{52.00}{429}{133.00}{57.00}{430}
\emline{133.00}{57.00}{431}{135.67}{54.00}{432}
\emline{135.67}{54.00}{433}{130.00}{52.00}{434}
\emline{130.00}{52.00}{435}{126.00}{60.00}{436}
\emline{126.00}{60.00}{437}{119.00}{60.00}{438}
\emline{119.00}{60.00}{439}{115.00}{52.00}{440}
\emline{115.00}{52.00}{441}{115.00}{60.00}{442}
\emline{115.00}{60.00}{443}{100.00}{60.00}{444}
\emline{100.00}{60.00}{445}{100.00}{52.00}{446}
\emline{100.00}{52.00}{447}{102.00}{58.00}{448}
\emline{102.00}{58.00}{449}{113.00}{58.00}{450}
\emline{113.00}{58.00}{451}{115.00}{52.00}{452}
\emline{115.00}{52.00}{453}{111.00}{56.00}{454}
\emline{111.00}{56.00}{455}{104.00}{56.00}{456}
\emline{104.00}{56.00}{457}{100.00}{52.00}{458}
\emline{100.00}{52.00}{459}{96.00}{60.00}{460}
\emline{96.00}{60.00}{461}{86.00}{60.00}{462}
\emline{86.00}{60.00}{463}{86.00}{52.00}{464}
\emline{17.00}{52.00}{465}{17.00}{62.00}{466}
\emline{17.00}{62.00}{467}{70.00}{62.00}{468}
\emline{70.00}{62.00}{469}{65.00}{52.00}{470}
\emline{65.00}{52.00}{471}{65.00}{60.00}{472}
\emline{65.00}{60.00}{473}{54.00}{60.00}{474}
\emline{54.00}{60.00}{475}{50.00}{52.00}{476}
\emline{50.00}{52.00}{477}{50.00}{60.00}{478}
\emline{50.00}{60.00}{479}{35.00}{60.00}{480}
\emline{35.00}{60.00}{481}{35.00}{52.00}{482}
\emline{35.00}{52.00}{483}{37.00}{58.00}{484}
\emline{37.00}{58.00}{485}{48.00}{58.00}{486}
\emline{48.00}{58.00}{487}{50.00}{52.00}{488}
\emline{50.00}{52.00}{489}{46.00}{56.00}{490}
\emline{46.00}{56.00}{491}{39.00}{56.00}{492}
\emline{39.00}{56.00}{493}{35.00}{52.00}{494}
\emline{35.00}{52.00}{495}{30.00}{60.00}{496}
\emline{30.00}{60.00}{497}{21.00}{60.00}{498}
\emline{21.00}{60.00}{499}{21.00}{52.00}{500}
\emline{65.00}{52.00}{501}{62.67}{57.00}{502}
\emline{62.67}{57.00}{503}{60.67}{53.67}{504}
\emline{60.67}{53.67}{505}{65.00}{52.00}{506}
\put(43.00,36.00){\circle{10.00}}
\put(39.00,42.00){\circle{4.67}}
\put(47.00,42.00){\circle{4.67}}
\put(35.00,17.00){\circle{6.00}}
\put(99.00,36.00){\circle{6.00}}
\put(112.33,36.00){\circle{6.00}}
\put(125.67,36.00){\circle{6.00}}
\put(109.00,13.00){\circle{6.00}}
\emline{90.00}{33.00}{507}{135.00}{33.00}{508}
\emline{99.00}{10.00}{509}{119.00}{10.00}{510}
\emline{26.33}{14.00}{511}{63.00}{14.00}{512}
\emline{54.67}{31.00}{513}{31.00}{31.00}{514}
\put(26.67,35.67){\makebox(0,0)[cc]{$18$}}
\put(87.00,35.67){\makebox(0,0)[cc]{$27$}}
\put(23.00,17.00){\makebox(0,0)[cc]{$36$}}
\put(98.00,17.00){\makebox(0,0)[cc]{$12$}}
\put(52.00,17.00){\circle{6.00}}
\put(52.00,23.00){\circle{6.00}}
\put(109.00,19.00){\circle{6.00}}
\put(109.00,25.00){\circle{6.00}}
\end{picture}
\end{center}
\caption{Half-planar graphs in the Boltzmann theory with the quartic 
interaction in the order $g^3$.} 
\label{fig7}
\end{figure}

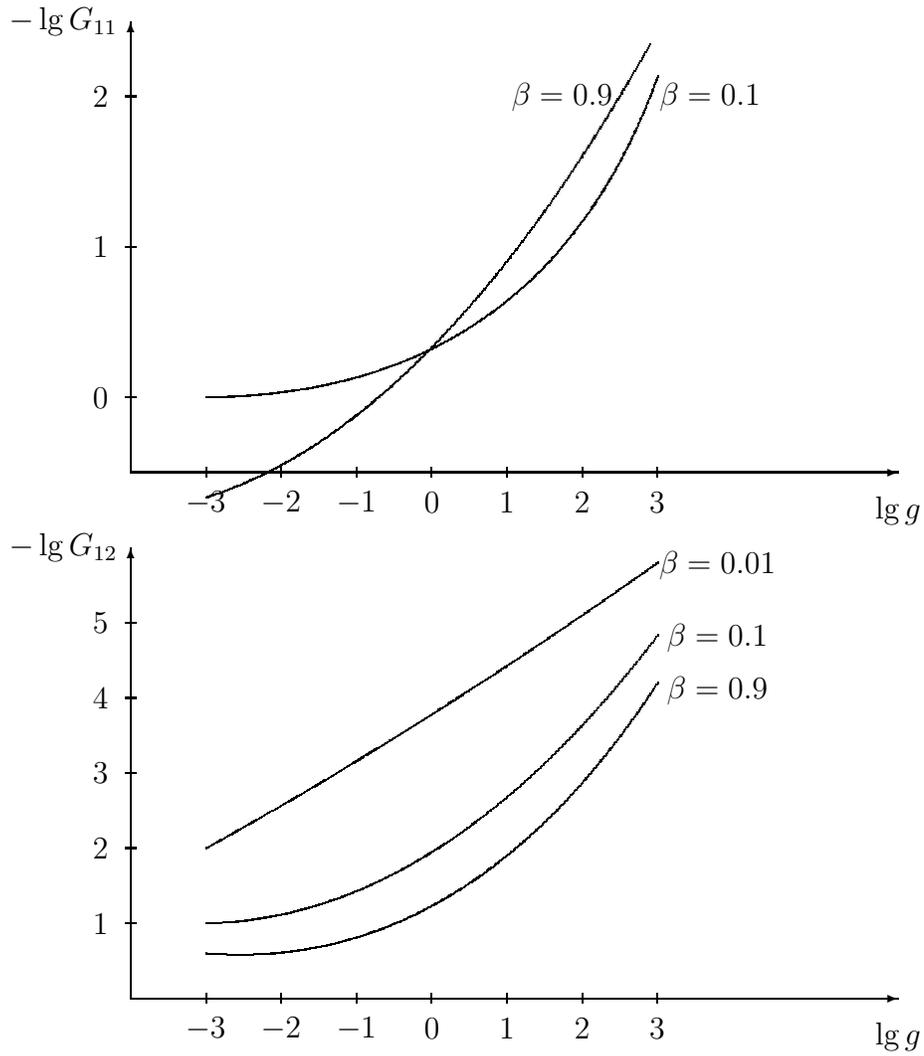
\begin{figure}
\begin{center}
\unitlength=1.00mm
\special{em:linewidth 0.4pt}
\linethickness{0.4pt}
\begin{picture}(132.00,140.00)
\put(30.00,10.00){\vector(1,0){102.00}}
\put(30.00,10.00){\vector(0,1){60.00}}
\put(30.00,80.00){\vector(0,1){60.00}}
\put(30.00,80.00){\vector(1,0){102.00}}
\put(40.00,80.67){\line(0,-1){1.33}}
\put(50.00,80.67){\line(0,-1){1.33}}
\put(60.00,80.67){\line(0,-1){1.33}}
\put(70.00,80.67){\line(0,-1){1.33}}
\put(80.00,80.67){\line(0,-1){1.33}}
\put(90.00,80.67){\line(0,-1){1.33}}
\put(100.00,80.67){\line(0,-1){1.33}}
\put(40.00,10.67){\line(0,-1){1.33}}
\put(50.00,10.67){\line(0,-1){1.33}}
\put(60.00,10.67){\line(0,-1){1.33}}
\put(70.00,10.67){\line(0,-1){1.33}}
\put(80.00,10.67){\line(0,-1){1.33}}
\put(90.00,10.67){\line(0,-1){1.33}}
\put(100.00,10.67){\line(0,-1){1.33}}
\put(29.33,20.00){\line(1,0){1.33}}
\put(29.33,30.00){\line(1,0){1.33}}
\put(29.33,40.00){\line(1,0){1.33}}
\put(29.33,50.00){\line(1,0){1.33}}
\put(29.33,60.00){\line(1,0){1.33}}
\put(21.00,70.00){\makebox(0,0)[cc]{$-\lg G_{12}$}}
\put(21.00,140.00){\makebox(0,0)[cc]{$-\lg G_{11}$}}
\put(26.00,20.00){\makebox(0,0)[cc]{$1$}}
\put(26.00,30.00){\makebox(0,0)[cc]{$2$}}
\put(26.00,40.00){\makebox(0,0)[cc]{$3$}}
\put(26.00,50.00){\makebox(0,0)[cc]{$4$}}
\put(26.00,60.00){\makebox(0,0)[cc]{$5$}}
\put(132.00,75.00){\makebox(0,0)[cc]{$\lg g$}}
\put(132.00,5.00){\makebox(0,0)[cc]{$\lg g$}}
\put(40.00,76.00){\makebox(0,0)[cc]{$-3$}}
\put(50.00,76.00){\makebox(0,0)[cc]{$-2$}}
\put(60.00,76.00){\makebox(0,0)[cc]{$-1$}}
\put(70.00,76.00){\makebox(0,0)[cc]{$0$}}
\put(80.00,76.00){\makebox(0,0)[cc]{$1$}}
\put(90.00,76.00){\makebox(0,0)[cc]{$2$}}
\put(100.00,76.00){\makebox(0,0)[cc]{$3$}}
\put(40.00,6.00){\makebox(0,0)[cc]{$-3$}}
\put(50.00,6.00){\makebox(0,0)[cc]{$-2$}}
\put(60.00,6.00){\makebox(0,0)[cc]{$-1$}}
\put(70.00,6.00){\makebox(0,0)[cc]{$0$}}
\put(80.00,6.00){\makebox(0,0)[cc]{$1$}}
\put(90.00,6.00){\makebox(0,0)[cc]{$2$}}
\put(100.00,6.00){\makebox(0,0)[cc]{$3$}}
\put(29.33,90.00){\line(1,0){1.33}}
\put(29.33,110.00){\line(1,0){1.33}}
\put(29.33,130.00){\line(1,0){1.33}}
\put(26.00,90.00){\makebox(0,0)[cc]{$0$}}
\put(26.00,110.00){\makebox(0,0)[cc]{$1$}}
\put(26.00,130.00){\makebox(0,0)[cc]{$2$}}
\bezier{360}(40.00,90.00)(84.67,90.33)(100.00,132.67)
\bezier{360}(40.00,76.67)(70.33,86.33)(99.00,137.00)
\put(87.33,130.00){\makebox(0,0)[cc]{$\beta =0.9$}}
\put(107.00,130.00){\makebox(0,0)[cc]{$\beta =0.1$}}
\bezier{284}(40.00,30.00)(70.00,46.67)(100.00,68.00)
\bezier{316}(40.00,20.00)(71.33,20.67)(100.00,58.33)
\bezier{324}(40.00,16.00)(74.67,13.33)(100.00,52.00)
\put(108.00,67.67){\makebox(0,0)[cc]{$\beta =0.01$}}
\put(108.00,58.00){\makebox(0,0)[cc]{$\beta =0.1$}}
\put(108.00,51.00){\makebox(0,0)[cc]{$\beta =0.9$}}
\end{picture}
\end{center}
\caption{The dependences of the two-point Green's functions $G_{11}$
and $G_{12}$ on the coupling constant $g$ for different values of 
$\beta$.}
\label{fig8}
\end{figure}

          
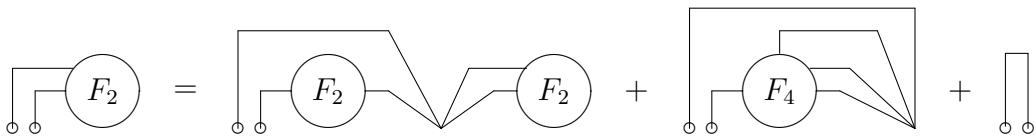
\begin{figure}
\begin{center}
\unitlength=1mm
\special{em:linewidth 0.4pt}
\linethickness{0.4pt}
\begin{picture}(143.67,26.00)
\put(20.00,15.00){\circle{10.00}}
\put(50.00,15.00){\circle{10.00}}
\put(80.00,15.00){\circle{10.00}}
\put(110.00,15.00){\circle{10.00}}
\emline{11.00}{10.00}{1}{11.00}{15.00}{2}
\emline{11.00}{15.00}{3}{15.00}{15.00}{4}
\emline{8.00}{10.00}{5}{8.00}{18.00}{6}
\emline{8.00}{18.00}{7}{16.00}{18.00}{8}
\emline{41.00}{10.00}{9}{41.00}{15.00}{10}
\emline{41.00}{15.00}{11}{45.00}{15.00}{12}
\emline{38.00}{10.00}{13}{38.00}{23.00}{14}
\emline{38.00}{23.00}{15}{58.00}{23.00}{16}
\emline{58.00}{23.00}{17}{65.00}{10.00}{18}
\emline{65.00}{10.00}{19}{58.00}{15.00}{20}
\emline{58.00}{15.00}{21}{55.00}{15.00}{22}
\emline{65.00}{10.00}{23}{72.00}{15.00}{24}
\emline{72.00}{15.00}{25}{75.00}{15.00}{26}
\emline{65.00}{10.00}{27}{68.67}{18.00}{28}
\emline{68.67}{18.00}{29}{76.33}{18.00}{30}
\emline{101.00}{10.00}{31}{101.00}{15.00}{32}
\emline{101.00}{15.00}{33}{105.00}{15.00}{34}
\emline{98.00}{10.00}{35}{98.00}{26.00}{36}
\emline{98.00}{26.00}{37}{128.00}{26.00}{38}
\emline{128.00}{26.00}{39}{128.00}{10.00}{40}
\emline{128.00}{10.00}{41}{123.00}{23.00}{42}
\emline{123.00}{23.00}{43}{110.00}{23.00}{44}
\emline{110.00}{23.00}{45}{110.00}{20.00}{46}
\emline{115.00}{15.00}{47}{118.00}{15.00}{48}
\emline{118.00}{15.00}{49}{128.00}{10.00}{50}
\emline{128.00}{10.00}{51}{119.33}{18.00}{52}
\emline{119.33}{18.00}{53}{114.00}{18.00}{54}
\emline{140.00}{10.33}{55}{140.00}{20.00}{56}
\emline{140.00}{20.00}{57}{143.00}{20.00}{58}
\emline{143.00}{20.00}{59}{143.00}{10.00}{60}
\put(8.00,10.00){\circle{1.33}}
\put(11.00,10.00){\circle{1.33}}
\put(38.00,10.00){\circle{1.33}}
\put(41.00,10.00){\circle{1.33}}
\put(98.00,10.00){\circle{1.33}}
\put(101.00,10.00){\circle{1.33}}
\put(140.00,10.00){\circle{1.33}}
\put(143.00,10.00){\circle{1.33}}
\put(20.00,15.00){\makebox(0,0)[cc]{$F_2$}}
\put(50.00,15.00){\makebox(0,0)[cc]{$F_2$}}
\put(80.00,15.00){\makebox(0,0)[cc]{$F_2$}}
\put(110.00,15.00){\makebox(0,0)[cc]{$F_4$}}
\put(31.00,15.00){\makebox(0,0)[cc]{$=$}}
\put(91.00,15.00){\makebox(0,0)[cc]{$+$}}
\put(134.00,15.00){\makebox(0,0)[cc]{$+$}}
\end{picture}
\end{center}
\caption{Graphical representation  of equation (4.21).}
\label{fig9}
\end{figure}


\begin{figure}
\begin{center}
\unitlength=1mm
\special{em:linewidth 0.4pt}
\linethickness{0.4pt}
\begin{picture}(145.00,71.33)
\put(40.00,60.00){\circle{10.00}}
\put(90.00,60.00){\circle{10.00}}
\put(120.00,60.00){\circle{10.00}}
\put(30.00,25.00){\circle{10.00}}
\put(60.00,25.00){\circle{10.00}}
\put(140.00,25.00){\circle{10.00}}
\put(101.00,25.00){\circle{10.00}}
\emline{29.00}{55.00}{1}{29.00}{60.00}{2}
\emline{29.00}{60.00}{3}{35.00}{60.00}{4}
\emline{26.00}{55.00}{5}{26.00}{63.00}{6}
\emline{26.00}{63.00}{7}{36.00}{63.00}{8}
\emline{23.00}{67.00}{9}{35.33}{67.00}{10}
\emline{35.33}{67.00}{11}{37.33}{64.33}{12}
\emline{23.00}{55.00}{13}{23.00}{67.00}{14}
\emline{20.00}{55.00}{15}{20.00}{71.00}{16}
\emline{20.00}{71.00}{17}{40.00}{71.00}{18}
\emline{40.00}{71.00}{19}{40.00}{65.00}{20}
\emline{79.00}{55.00}{21}{79.00}{60.00}{22}
\emline{79.00}{60.00}{23}{85.00}{60.00}{24}
\emline{76.00}{55.00}{25}{76.00}{63.00}{26}
\emline{76.00}{63.00}{27}{86.00}{63.00}{28}
\emline{73.00}{55.00}{29}{73.00}{68.00}{30}
\emline{73.00}{68.00}{31}{93.00}{68.00}{32}
\emline{93.00}{68.00}{33}{105.00}{55.00}{34}
\emline{105.00}{55.00}{35}{110.00}{60.00}{36}
\emline{110.00}{60.00}{37}{115.00}{60.00}{38}
\emline{116.00}{63.00}{39}{110.00}{63.00}{40}
\emline{110.00}{63.00}{41}{105.00}{55.00}{42}
\emline{105.00}{55.00}{43}{105.00}{68.00}{44}
\emline{105.00}{68.00}{45}{114.33}{68.00}{46}
\emline{114.33}{68.00}{47}{118.00}{64.33}{48}
\emline{120.00}{65.00}{49}{120.00}{71.33}{50}
\emline{120.00}{71.33}{51}{70.00}{71.33}{52}
\emline{70.00}{71.33}{53}{70.00}{55.00}{54}
\emline{19.00}{20.00}{55}{19.00}{25.00}{56}
\emline{19.00}{25.00}{57}{25.00}{25.00}{58}
\emline{16.00}{20.00}{59}{16.00}{28.00}{60}
\emline{16.00}{28.00}{61}{26.00}{28.00}{62}
\emline{13.00}{20.00}{63}{13.00}{35.00}{64}
\emline{13.00}{35.00}{65}{47.00}{35.00}{66}
\emline{47.00}{35.00}{67}{47.00}{20.00}{68}
\emline{47.00}{20.00}{69}{51.67}{25.00}{70}
\emline{51.67}{25.00}{71}{55.00}{25.00}{72}
\emline{60.00}{30.00}{73}{60.00}{39.00}{74}
\emline{60.00}{39.00}{75}{10.00}{39.00}{76}
\emline{10.00}{39.00}{77}{10.00}{20.00}{78}
\emline{35.00}{25.00}{79}{38.00}{25.00}{80}
\emline{38.00}{25.00}{81}{47.00}{20.00}{82}
\emline{47.00}{20.00}{83}{40.00}{28.00}{84}
\emline{40.00}{28.00}{85}{33.67}{28.00}{86}
\emline{90.00}{20.00}{87}{90.00}{25.00}{88}
\emline{90.00}{25.00}{89}{96.00}{25.00}{90}
\emline{87.00}{20.00}{91}{87.00}{28.00}{92}
\emline{87.00}{28.00}{93}{97.00}{28.00}{94}
\emline{84.00}{20.00}{95}{84.00}{28.00}{96}
\emline{84.00}{28.00}{97}{81.00}{28.00}{98}
\emline{81.00}{28.00}{99}{81.00}{20.00}{100}
\emline{129.00}{20.00}{101}{129.00}{25.00}{102}
\emline{129.00}{25.00}{103}{135.00}{25.00}{104}
\emline{126.00}{20.00}{105}{126.00}{25.00}{106}
\emline{126.00}{25.00}{107}{123.00}{25.00}{108}
\emline{123.00}{25.00}{109}{123.00}{20.00}{110}
\emline{120.00}{20.00}{113}{120.00}{28.00}{114}
\emline{120.00}{28.00}{115}{136.00}{28.00}{116}
\put(20.00,55.00){\circle{1.33}}
\put(23.00,55.00){\circle{1.33}}
\put(26.00,55.00){\circle{1.33}}
\put(29.00,55.00){\circle{1.33}}
\put(70.00,55.00){\circle{1.33}}
\put(73.00,55.00){\circle{1.33}}
\put(76.00,55.00){\circle{1.33}}
\put(79.00,55.00){\circle{1.33}}
\put(10.00,20.00){\circle{1.33}}
\put(13.00,20.00){\circle{1.33}}
\put(16.00,20.00){\circle{1.33}}
\put(19.00,20.00){\circle{1.33}}
\put(81.00,20.00){\circle{1.33}}
\put(84.00,20.00){\circle{1.33}}
\put(87.00,20.00){\circle{1.33}}
\put(90.00,20.00){\circle{1.33}}
\put(120.00,20.00){\circle{1.33}}
\put(123.00,20.00){\circle{1.33}}
\put(126.00,20.00){\circle{1.33}}
\put(129.00,20.00){\circle{1.33}}
\put(40.00,60.00){\makebox(0,0)[cc]{$F_4$}}
\put(90.00,60.00){\makebox(0,0)[cc]{$F_2$}}
\put(120.00,60.00){\makebox(0,0)[cc]{$F_4$}}
\put(30.00,25.00){\makebox(0,0)[cc]{$F_4$}}
\put(60.00,25.00){\makebox(0,0)[cc]{$F_2$}}
\put(101.00,25.00){\makebox(0,0)[cc]{$F_2$}}
\put(140.00,25.00){\makebox(0,0)[cc]{$F_2$}}
\put(56.00,59.67){\makebox(0,0)[cc]{$=$}}
\put(133.00,60.00){\makebox(0,0)[cc]{$+$}}
\put(72.00,25.00){\makebox(0,0)[cc]{$+$}}
\put(112.00,25.00){\makebox(0,0)[cc]{$+$}}
\end{picture}
\end{center}
\caption{Graphical representation  of equation (4.20).}
\label{fig10}
\end{figure}
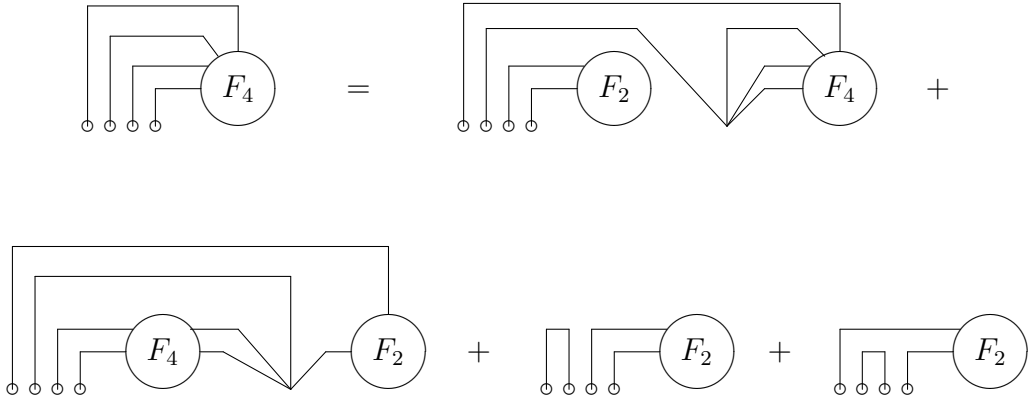

          
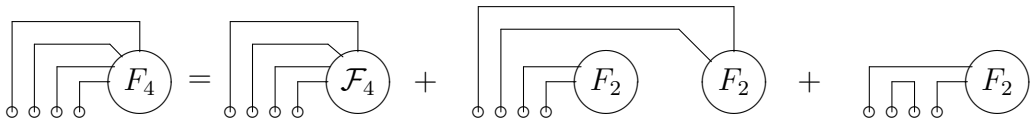
\begin{figure}
\begin{center}
\unitlength=1.00mm
\special{em:linewidth 0.4pt}
\linethickness{0.4pt}
\begin{picture}(143.00,30.00)
\put(25.00,20.00){\circle{8.00}}
\put(54.00,20.00){\circle{8.00}}
\put(87.00,20.00){\circle{8.00}}
\put(104.00,20.00){\circle{8.00}}
\put(139.00,20.00){\circle{8.00}}
\emline{17.00}{16.00}{1}{17.00}{20.00}{2}
\emline{17.00}{20.00}{3}{21.00}{20.00}{4}
\emline{14.00}{16.00}{5}{14.00}{22.00}{6}
\emline{11.00}{16.00}{9}{11.00}{25.00}{10}
\emline{11.00}{25.00}{11}{21.00}{25.00}{12}
\emline{21.00}{25.00}{13}{22.67}{23.33}{14}
\emline{8.00}{16.00}{15}{8.00}{28.00}{16}
\emline{8.00}{28.00}{17}{25.00}{28.00}{18}
\emline{25.00}{28.00}{19}{25.00}{24.00}{20}
\emline{46.00}{16.00}{21}{46.00}{20.00}{22}
\emline{46.00}{20.00}{23}{50.00}{20.00}{24}
\emline{43.00}{16.00}{25}{43.00}{22.00}{26}
\emline{40.00}{16.00}{27}{40.00}{25.00}{28}
\emline{37.00}{16.00}{29}{37.00}{28.00}{30}
\emline{37.00}{28.00}{31}{54.00}{28.00}{32}
\emline{54.00}{28.00}{33}{54.00}{24.00}{34}
\emline{79.00}{16.00}{35}{79.00}{20.00}{36}
\emline{79.00}{20.00}{37}{83.00}{20.00}{38}
\emline{76.00}{16.00}{39}{76.00}{22.00}{40}
\emline{73.00}{16.00}{41}{73.00}{27.00}{42}
\emline{73.00}{27.00}{43}{96.67}{27.00}{44}
\emline{96.67}{27.00}{45}{101.00}{22.67}{46}
\emline{70.00}{16.00}{47}{70.00}{30.00}{48}
\emline{70.00}{30.00}{49}{104.00}{30.00}{50}
\emline{104.00}{30.00}{51}{104.00}{24.00}{52}
\emline{131.00}{16.00}{53}{131.00}{20.00}{54}
\emline{131.00}{20.00}{55}{135.00}{20.00}{56}
\emline{128.00}{16.00}{57}{128.00}{20.00}{58}
\emline{128.00}{20.00}{59}{125.00}{20.00}{60}
\emline{125.00}{20.00}{61}{125.00}{16.00}{62}
\emline{122.00}{16.00}{63}{122.00}{22.00}{64}
\put(8.00,16.00){\circle{1.33}}
\put(11.00,16.00){\circle{1.33}}
\put(14.00,16.00){\circle{1.33}}
\put(17.00,16.00){\circle{1.33}}
\put(37.00,16.00){\circle{1.33}}
\put(40.00,16.00){\circle{1.33}}
\put(43.00,16.00){\circle{1.33}}
\put(46.00,16.00){\circle{1.33}}
\put(70.00,16.00){\circle{1.33}}
\put(73.00,16.00){\circle{1.33}}
\put(76.00,16.00){\circle{1.33}}
\put(79.00,16.00){\circle{1.33}}
\put(122.00,16.00){\circle{1.33}}
\put(125.00,16.00){\circle{1.33}}
\put(128.00,16.00){\circle{1.33}}
\put(131.00,16.00){\circle{1.33}}
\put(25.00,20.00){\makebox(0,0)[cc]{$F_4$}}
\put(33.00,20.00){\makebox(0,0)[cc]{$=$}}
\put(54.00,20.00){\makebox(0,0)[cc]{${\cal F}_4$}}
\put(63.00,20.00){\makebox(0,0)[cc]{$+$}}
\put(87.00,20.00){\makebox(0,0)[cc]{$F_2$}}
\put(104.00,20.00){\makebox(0,0)[cc]{$F_2$}}
\put(139.00,20.00){\makebox(0,0)[cc]{$F_2$}}
\put(114.00,20.00){\makebox(0,0)[cc]{$+$}}
\emline{14.00}{22.00}{65}{21.67}{22.00}{66}
\emline{43.00}{22.00}{67}{50.33}{22.00}{68}
\emline{40.00}{25.00}{69}{50.00}{25.00}{70}
\emline{50.00}{25.00}{71}{52.00}{23.33}{72}
\emline{76.00}{22.00}{73}{83.67}{22.00}{74}
\emline{122.00}{22.00}{75}{135.67}{22.00}{76}
\end{picture}
\end{center}
\caption{Graphical representation  of equation (4.23).}
\label{fig11}
\end{figure}


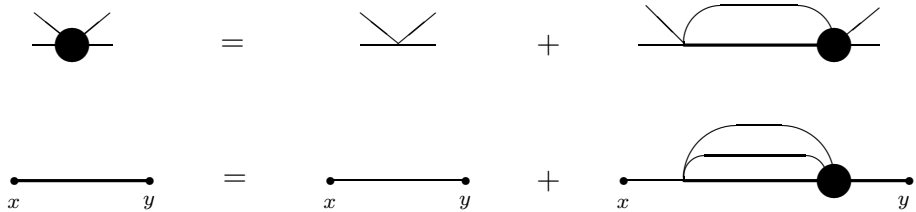
\begin{figure}
\begin{center}
\unitlength=1.00mm
\linethickness{0.4pt}
\begin{picture}(143.50,34.00)
\put(123.00,10.84){\oval(20.00,14.33)[t]}
\put(133.00,10.67){\circle*{4.67}}
\put(31.67,28.67){\circle*{4.67}}
\put(123.00,28.84){\oval(20.00,10.33)[t]}
\put(133.00,28.67){\circle*{4.67}}
\put(53.33,10.67){\makebox(0,0)[cc]{$=$}}
\put(95.00,10.67){\makebox(0,0)[cc]{$+$}}
\put(95.00,28.67){\makebox(0,0)[cc]{$+$}}
\put(53.00,28.67){\makebox(0,0)[cc]{$=$}}
\put(24.00,10.67){\circle*{1.00}}
\put(42.00,10.67){\circle*{1.00}}
\put(66.00,10.67){\circle*{1.00}}
\put(84.00,10.67){\circle*{1.00}}
\put(105.00,10.67){\circle*{1.00}}
\put(143.00,10.67){\circle*{1.00}}
\put(142.33,7.67){\makebox(0,0)[cc]{$_y$}}
\put(105.00,7.67){\makebox(0,0)[cc]{$_x$}}
\put(84.00,7.67){\makebox(0,0)[cc]{$_y$}}
\put(66.00,7.67){\makebox(0,0)[cc]{$_x$}}
\put(42.00,7.67){\makebox(0,0)[cc]{$_y$}}
\put(24.00,7.67){\makebox(0,0)[cc]{$_x$}}
\put(113.00,10.67){\line(0,0){0.00}}
\put(122.50,10.50){\oval(19.00,7.00)[t]}
\put(66.00,10.67){\line(1,0){18.00}}
\put(105.00,10.67){\line(1,0){8.00}}
\put(133.00,28.67){\line(1,0){6.00}}
\put(113.00,28.67){\line(-1,0){6.00}}
\put(80.00,28.67){\line(-1,0){10.00}}
\put(37.00,28.67){\line(-1,0){10.67}}
\put(80.00,33.00){\line(-6,-5){5.00}}
\put(75.00,29.00){\line(-5,4){5.00}}
\put(36.67,33.00){\line(-6,-5){5.00}}
\put(31.67,29.00){\line(-5,4){5.00}}
\put(108.00,34.00){\line(1,-1){5.00}}
\put(133.00,28.67){\line(6,5){6.00}}
\special{em:linewidth 1.2pt}
\linethickness{1.2pt}
\emline{113.00}{28.67}{1}{133.00}{28.67}{2}
\emline{24.00}{10.67}{3}{42.00}{10.67}{4}
\emline{113.00}{10.67}{5}{143.00}{10.67}{6}
\end{picture}
\end{center}
\caption{Graphical representation  of equations (4.29)
and (4.30)}
\label{fig12}
\end{figure}

          
\begin{figure}
\begin{center}
\unitlength=1.00mm
\special{em:linewidth 0.4pt}
\linethickness{0.4pt}
\begin{picture}(145.00,139.67)
\put(120.00,25.17){\oval(10.00,10.33)[t]}
\put(130.00,25.17){\oval(10.00,10.33)[t]}
\put(110.00,25.17){\oval(10.00,10.33)[t]}
\emline{135.00}{25.00}{1}{143.00}{33.00}{2}
\emline{105.00}{25.00}{3}{97.33}{32.67}{4}
\put(120.00,25.00){\oval(30.00,28.00)[t]}
\put(110.00,20.00){\dashbox{0.67}(20.00,14.00)[cc]{}}
\put(37.50,25.17){\oval(15.00,10.33)[t]}
\put(37.50,25.33){\oval(15.00,18.67)[t]}
\put(37.50,25.17){\oval(35.00,32.33)[t]}
\emline{20.00}{25.00}{5}{12.33}{32.67}{6}
\emline{55.00}{25.00}{7}{63.00}{33.00}{8}
\put(25.00,20.00){\dashbox{0.67}(25.00,17.00)[cc]{}}
\bezier{224}(60.00,66.00)(75.00,89.67)(90.00,66.00)
\bezier{288}(60.00,66.00)(75.00,99.00)(90.00,66.00)
\bezier{552}(40.00,66.00)(75.00,125.33)(110.00,66.00)
\bezier{652}(40.00,66.00)(75.00,139.67)(110.00,66.00)
\put(66.00,66.00){\dashbox{0.67}(18.00,7.00)[cc]{}}
\put(58.00,66.00){\dashbox{0.67}(34.00,19.00)[cc]{}}
\put(35.00,66.00){\dashbox{0.67}(80.00,42.00)[cc]{}}
\put(84.00,62.00){\dashbox{0.67}(8.00,4.00)[cc]{}}
\put(58.00,62.00){\dashbox{0.67}(8.00,4.00)[cc]{}}
\put(35.00,57.00){\dashbox{0.67}(33.00,9.00)[cc]{}}
\put(82.00,57.00){\dashbox{0.67}(33.00,9.00)[cc]{}}
\put(75.00,50.00){\makebox(0,0)[cc]{$a)$}}
\put(38.00,11.00){\makebox(0,0)[cc]{$b)$}}
\put(120.00,11.00){\makebox(0,0)[cc]{$c)$}}
\put(68.00,66.00){\dashbox{0.67}(14.00,5.00)[cc]{}}
\bezier{72}(70.00,66.00)(75.00,73.33)(80.00,66.00)
\bezier{52}(70.00,66.00)(75.00,70.00)(80.00,66.00)
\linethickness{1.2pt}
\put(30.00,66.00){\line(1,0){91.00}}
\put(10.00,25.00){\line(1,0){55.00}}
\put(95.00,25.00){\line(1,0){50.00}}
\end{picture}
\end{center}
\caption{
a) Example of divergent
parts of $\Pi $-type of the two-point half-planar
graph.
b), c) Examples of    half-planar divergent subgraphs.}
\label{fig13}
\end{figure}
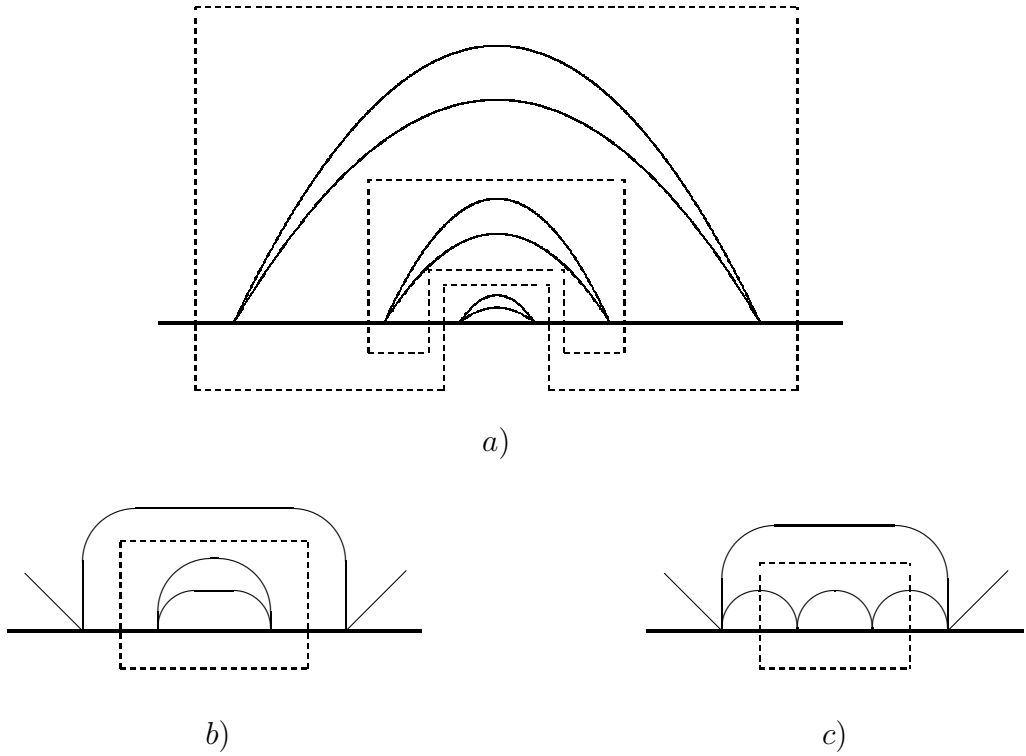

          
\begin{figure}
\begin{center}
\unitlength=1.00mm
\special{em:linewidth 0.4pt}
\linethickness{0.4pt}
\begin{picture}(135.00,95.00)
\put(40.00,50.00){\line(1,0){70.00}}
\put(135.00,10.00){\line(-1,0){45.00}}
\put(60.00,10.00){\line(-1,0){45.00}}
\bezier{104}(65.00,50.00)(75.00,58.00)(85.00,50.00)
\bezier{144}(65.00,50.00)(74.67,65.00)(85.00,50.00)
\bezier{328}(50.00,50.00)(75.00,82.67)(100.00,50.00)
\bezier{412}(50.00,50.00)(75.00,95.00)(100.00,50.00)
\bezier{284}(25.00,10.00)(42.67,41.00)(60.00,10.00)
\bezier{284}(90.00,10.00)(107.00,41.00)(125.00,10.00)
\bezier{136}(40.00,10.00)(50.00,24.00)(60.00,10.00)
\bezier{104}(40.00,10.00)(50.00,18.00)(60.00,10.00)
\bezier{136}(90.00,10.00)(100.00,24.00)(110.00,10.00)
\bezier{104}(90.00,10.00)(100.00,18.00)(110.00,10.00)
\put(125.00,10.00){\line(2,3){6.67}}
\put(125.00,10.00){\circle*{2.00}}
\put(25.00,10.00){\line(-2,3){6.67}}
\put(25.00,10.00){\circle*{2.00}}
\put(25.00,5.00){\makebox(0,0)[cc]{$A$}}
\put(125.00,5.00){\makebox(0,0)[cc]{$B$}}
\put(100.00,45.00){\makebox(0,0)[cc]{$B$}}
\put(50.00,45.00){\makebox(0,0)[cc]{$A$}}
\put(65.00,40.00){\vector(-3,-4){8.33}}
\put(85.00,40.00){\vector(2,-3){7.33}}
\put(46.00,50.00){\dashbox{0.67}(58.00,25.00)[cc]{}}
\put(62.00,50.00){\dashbox{0.67}(26.00,9.33)[cc]{}}
\put(46.00,47.00){\dashbox{0.67}(16.00,3.00)[cc]{}}
\put(88.00,47.00){\dashbox{0.67}(16.00,3.00)[cc]{}}
\end{picture}
\end{center}
\caption{
Two possibilities making
a contraction of $\Pi$-type subgraph to a point.}
\label{fig14} 
\end{figure}
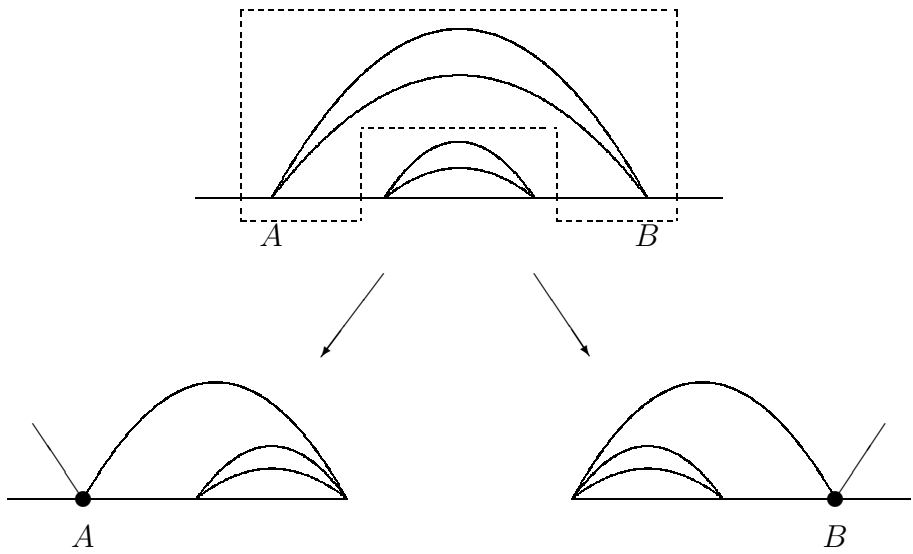


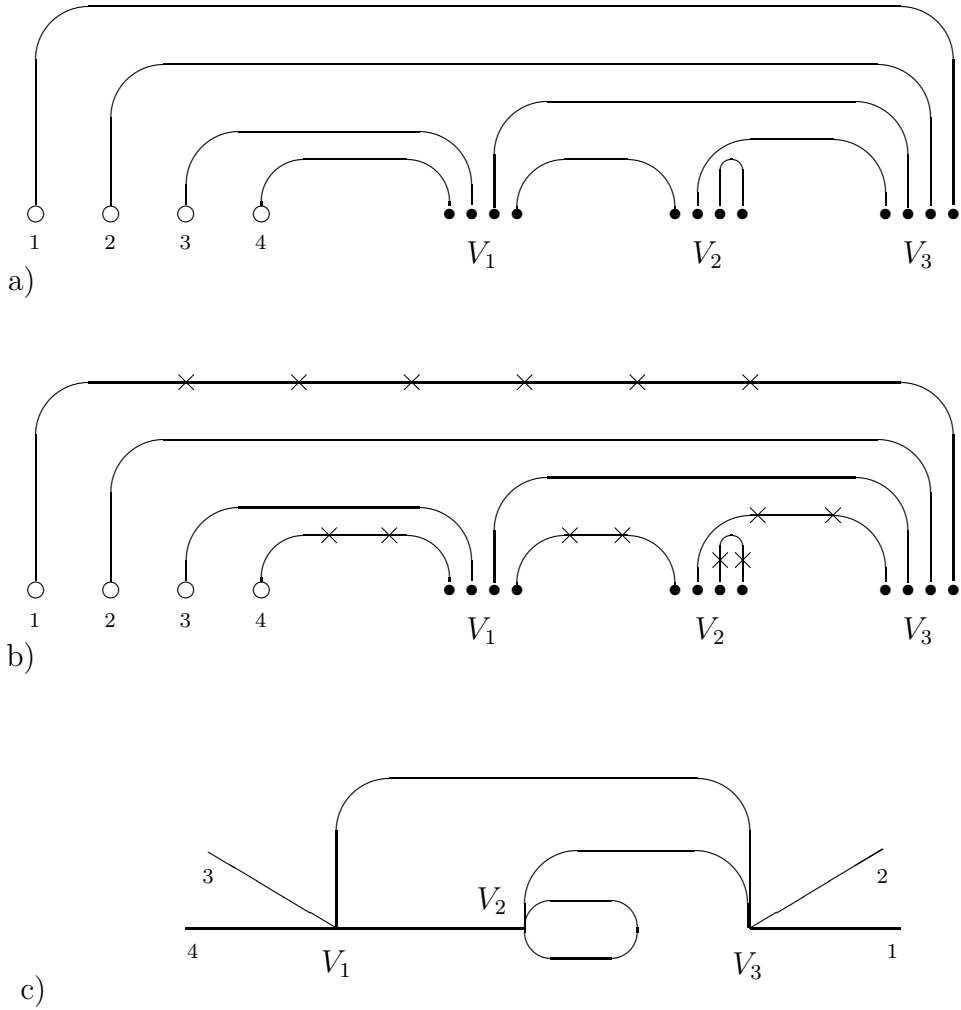
\begin{figure}
\begin{center}
\unitlength=1.00mm
\special{em:linewidth 0.4pt}
\linethickness{0.4pt}
\begin{picture}(137.66,142.66)
\put(15.00,65.00){\circle{2.00}}
\put(25.00,65.00){\circle{2.00}}
\put(35.00,65.00){\circle{2.00}}
\put(45.00,65.00){\circle{2.00}}
\put(70.00,65.00){\circle*{1.33}}
\put(73.00,65.00){\circle*{1.33}}
\put(76.00,65.00){\circle*{1.33}}
\put(79.00,65.00){\circle*{1.33}}
\put(100.00,65.00){\circle*{1.33}}
\put(103.00,65.00){\circle*{1.33}}
\put(106.00,65.00){\circle*{1.33}}
\put(109.00,65.00){\circle*{1.33}}
\put(128.00,65.00){\circle*{1.33}}
\put(131.00,65.00){\circle*{1.33}}
\put(134.00,65.00){\circle*{1.33}}
\put(137.00,65.00){\circle*{1.33}}
\put(76.00,66.33){\oval(122.00,52.67)[t]}
\put(79.50,66.17){\oval(109.00,37.67)[t]}
\put(107.50,66.17){\oval(3.00,12.33)[t]}
\put(115.50,66.17){\oval(25.00,17.67)[t]}
\put(54.00,66.33){\oval(38.00,19.33)[t]}
\put(57.50,66.17){\oval(25.00,12.33)[t]}
\put(103.50,66.00){\oval(55.00,28.00)[t]}
\put(89.50,65.67){\oval(21.00,13.33)[t]}
\put(15.00,115.00){\circle{2.00}}
\put(25.00,115.00){\circle{2.00}}
\put(35.00,115.00){\circle{2.00}}
\put(45.00,115.00){\circle{2.00}}
\put(70.00,115.00){\circle*{1.33}}
\put(73.00,115.00){\circle*{1.33}}
\put(76.00,115.00){\circle*{1.33}}
\put(79.00,115.00){\circle*{1.33}}
\put(100.00,115.00){\circle*{1.33}}
\put(103.00,115.00){\circle*{1.33}}
\put(106.00,115.00){\circle*{1.33}}
\put(109.00,115.00){\circle*{1.33}}
\put(128.00,115.00){\circle*{1.33}}
\put(131.00,115.00){\circle*{1.33}}
\put(134.00,115.00){\circle*{1.33}}
\put(137.00,115.00){\circle*{1.33}}
\put(76.00,116.33){\oval(122.00,52.67)[t]}
\put(79.50,116.17){\oval(109.00,37.67)[t]}
\put(107.50,116.17){\oval(3.00,12.33)[t]}
\put(115.50,116.17){\oval(25.00,17.67)[t]}
\put(54.00,116.33){\oval(38.00,19.33)[t]}
\put(57.50,116.17){\oval(25.00,12.33)[t]}
\put(103.50,116.00){\oval(55.00,28.00)[t]}
\put(89.50,115.67){\oval(21.00,13.33)[t]}
\put(15.00,106.00){\makebox(0,0)[rc]{a)}}
\put(15.00,55.00){\makebox(0,0)[rb]{b)}}
\put(15.00,111.33){\makebox(0,0)[cc]{$_1$}}
\put(25.00,111.33){\makebox(0,0)[cc]{$_2$}}
\put(35.00,111.33){\makebox(0,0)[cc]{$_3$}}
\put(45.00,111.33){\makebox(0,0)[cc]{$_4$}}
\put(74.33,109.67){\makebox(0,0)[cc]{$V_1$}}
\put(104.33,109.67){\makebox(0,0)[cc]{$V_2$}}
\put(132.33,109.67){\makebox(0,0)[cc]{$V_3$}}
\put(132.33,59.67){\makebox(0,0)[cc]{$V_3$}}
\put(104.67,59.67){\makebox(0,0)[cc]{$V_2$}}
\put(74.33,59.67){\makebox(0,0)[cc]{$V_1$}}
\put(45.00,61.00){\makebox(0,0)[cc]{$_4$}}
\put(35.00,61.00){\makebox(0,0)[cc]{$_3$}}
\put(25.00,61.00){\makebox(0,0)[cc]{$_2$}}
\put(15.00,61.00){\makebox(0,0)[cc]{$_1$}}
\put(35.00,20.00){\line(1,0){20.00}}
\put(55.00,20.00){\line(-5,3){17.00}}
\put(54.67,20.00){\line(1,0){25.33}}
\put(82.50,20.00){\oval(55.00,40.00)[t]}
\put(130.00,20.00){\line(-1,0){20.00}}
\put(110.00,20.00){\line(5,3){17.67}}
\put(94.83,20.17){\oval(29.67,20.33)[t]}
\put(129.00,17.00){\makebox(0,0)[cc]{$_1$}}
\put(127.67,27.00){\makebox(0,0)[cc]{$_2$}}
\put(38.00,27.00){\makebox(0,0)[cc]{$_3$}}
\put(36.00,17.00){\makebox(0,0)[cc]{$_4$}}
\put(55.00,15.33){\makebox(0,0)[cc]{$V_1$}}
\put(75.67,23.67){\makebox(0,0)[cc]{$V_2$}}
\put(109.67,15.00){\makebox(0,0)[cc]{$V_3$}}
\put(14.67,11.67){\makebox(0,0)[cc]{c)}}
\put(87.50,19.83){\oval(15.00,7.67)[]}
\emline{34.00}{93.67}{1}{36.00}{91.67}{2}
\emline{36.00}{93.67}{3}{34.00}{91.67}{4}
\emline{49.00}{93.67}{5}{51.00}{91.67}{6}
\emline{51.00}{93.67}{7}{49.00}{91.67}{8}
\emline{64.00}{93.67}{9}{66.00}{91.67}{10}
\emline{66.00}{93.67}{11}{64.00}{91.67}{12}
\emline{79.00}{93.67}{13}{81.00}{91.67}{14}
\emline{81.00}{93.67}{15}{79.00}{91.67}{16}
\emline{94.00}{93.67}{17}{96.00}{91.67}{18}
\emline{96.00}{93.67}{19}{94.00}{91.67}{20}
\emline{109.00}{93.67}{21}{111.00}{91.67}{22}
\emline{111.00}{93.67}{23}{109.00}{91.67}{24}
\emline{55.00}{71.33}{25}{53.00}{73.33}{26}
\emline{55.00}{73.33}{27}{53.00}{71.33}{28}
\emline{63.00}{73.33}{29}{61.00}{71.33}{30}
\emline{61.00}{73.33}{31}{63.00}{71.33}{32}
\emline{85.00}{73.33}{33}{87.00}{71.33}{34}
\emline{87.00}{73.33}{35}{85.00}{71.33}{36}
\emline{94.00}{73.33}{37}{92.00}{71.33}{38}
\emline{92.00}{73.33}{39}{94.00}{71.33}{40}
\emline{110.00}{76.00}{41}{112.00}{74.00}{42}
\emline{112.00}{76.00}{43}{110.00}{74.00}{44}
\emline{122.00}{76.00}{45}{120.00}{74.00}{46}
\emline{120.00}{76.00}{47}{122.00}{74.00}{48}
\emline{105.00}{70.00}{49}{107.00}{68.00}{50}
\emline{107.00}{70.00}{51}{105.00}{68.00}{52}
\emline{110.00}{70.00}{53}{108.00}{68.00}{54}
\emline{108.00}{70.00}{55}{110.00}{68.00}{56}
\end{picture}
\end{center}
\label{fig15}
\caption{
Graph for 4-point correlation function in the Boltzman field theory.
 Some lines of graph
 a) are marked by crosses.
 Insertions of $\delta M^2 \phi ^2$ and 
$(Z_{\psi}-1)(\partial _{\mu}\phi )^2 $
 are admissible only on the marked lines.}
\end{figure}

\begin{figure}
\begin{center}
\unitlength=1.00mm
\special{em:linewidth 0.4pt}
\linethickness{0.4pt}
\begin{picture}(143.00,65.67)
\put(20.00,15.00){\line(1,0){2.00}}
\put(24.00,15.00){\line(1,0){2.00}}
\put(28.00,15.00){\line(1,0){2.00}}
\put(32.00,15.00){\line(1,0){2.00}}
\put(36.00,15.00){\line(1,0){2.00}}
\put(40.00,15.00){\line(1,0){2.00}}
\put(44.00,15.00){\line(1,0){2.00}}
\put(48.00,15.00){\line(1,0){2.00}}
\put(52.00,15.00){\line(1,0){2.00}}
\put(56.00,15.00){\line(1,0){2.00}}
\put(60.00,15.00){\line(1,0){2.00}}
\put(64.00,15.00){\line(1,0){2.00}}
\put(68.00,15.00){\line(1,0){2.00}}
\put(72.00,15.00){\line(1,0){2.00}}
\put(76.00,15.00){\line(1,0){2.00}}
\put(80.00,15.00){\line(1,0){2.00}}
\put(84.00,15.00){\line(1,0){2.00}}
\put(88.00,15.00){\line(1,0){2.00}}
\put(92.00,15.00){\line(1,0){2.00}}
\put(96.00,15.00){\line(1,0){2.00}}
\put(100.00,15.00){\line(1,0){2.00}}
\put(104.00,15.00){\line(1,0){2.00}}
\put(108.00,15.00){\line(1,0){2.00}}
\put(112.00,15.00){\line(1,0){2.00}}
\put(116.00,15.00){\line(1,0){2.00}}
\put(120.00,15.00){\line(1,0){2.00}}
\put(124.00,15.00){\line(1,0){2.00}}
\put(128.00,15.00){\line(1,0){2.00}}
\put(20.33,15.00){\line(0,1){25.00}}
\put(20.33,40.00){\line(-1,1){5.33}}
\put(20.33,40.00){\line(6,5){6.33}}
\put(20.33,40.33){\line(0,1){7.33}}
\put(32.00,48.67){\line(4,-5){6.00}}
\put(38.00,41.00){\line(-1,-3){8.67}}
\put(44.33,47.67){\line(-1,-1){6.67}}
\put(37.67,41.00){\line(1,-5){5.33}}
\put(66.00,15.00){\line(0,1){20.00}}
\put(70.00,34.67){\line(-3,-4){4.00}}
\put(66.00,29.33){\line(-3,4){3.67}}
\put(76.33,34.67){\line(0,-1){19.67}}
\put(80.67,34.33){\line(-1,-1){4.33}}
\put(76.33,30.00){\line(-5,6){3.67}}
\put(52.67,15.00){\line(2,5){17.67}}
\put(70.33,59.33){\line(1,-2){22.00}}
\put(75.00,65.67){\line(-2,-3){4.33}}
\put(70.67,59.00){\line(-1,2){3.33}}
\put(101.33,15.00){\line(1,2){8.00}}
\put(109.33,31.00){\line(0,-1){16.00}}
\put(109.33,39.00){\line(0,-1){8.00}}
\put(109.33,31.00){\line(1,-2){8.00}}
\put(132.00,15.00){\line(1,0){2.00}}
\put(136.00,15.00){\line(1,0){2.00}}
\put(138.00,15.00){\line(0,1){23.67}}
\put(143.00,38.67){\line(-4,-5){5.00}}
\put(138.00,32.33){\line(-5,6){4.67}}
\put(125.33,22.00){\makebox(0,0)[cc]{$.~.~.~.~.~.$}}
\put(138.00,9.00){\makebox(0,0)[cc]{$x_n$}}
\put(20.33,9.00){\makebox(0,0)[cc]{$x_1$}}
\put(29.33,9.00){\makebox(0,0)[cc]{$x_2$}}
\put(42.67,9.00){\makebox(0,0)[cc]{$x_3$}}
\put(53.00,9.00){\makebox(0,0)[cc]{$x_4$}}
\put(96.00,9.00){\makebox(0,0)[cc]{$.~.~.~.~.~.~.~.~.$}}
\put(22.67,38.67){\makebox(0,0)[cc]{1}}
\put(40.67,40.33){\makebox(0,0)[cc]{2}}
\put(73.00,59.00){\makebox(0,0)[cc]{3}}
\put(68.67,29.00){\makebox(0,0)[cc]{4}}
\put(79.00,29.00){\makebox(0,0)[cc]{5}}
\put(112.33,30.33){\makebox(0,0)[cc]{6}}
\put(141.67,31.00){\makebox(0,0)[cc]{$k_1$}}
\end{picture}
\end{center}
\label{fig16}
\caption{Construction of planar graph.
$k_1$ of $m$ four-point vertices
are connected with the generalized
vertex at least by one line.
Two new generalized vertices are
formed by lines outcoming from 4-vertices labeled by 1, 2, 3, 6, ...
and 4, 5.}
\end{figure}
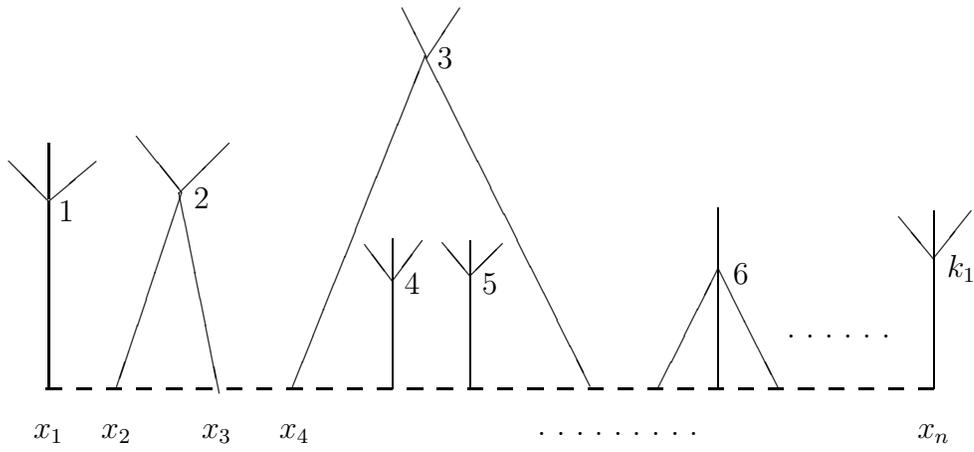

\end{document}